\documentclass[letterpaper,onecolumn,journal,final]{IEEEtran}
\usepackage{amsmath,amsfonts}
\usepackage[ruled,vlined,linesnumbered]{algorithm2e}
\usepackage{array}
\usepackage{amsmath}

\DeclareMathOperator*{\argmin}{arg\,min}
\usepackage{textcomp}
\usepackage{url}
\usepackage{acro}
\usepackage{verbatim}
\usepackage{graphicx}
\usepackage{cite}
\usepackage{pgfplots}
\usepackage{centernot}
\usepackage{booktabs}
\usepackage{xcolor}
\newlength{\commentWidth}
\usepackage{multirow}
\usepgfplotslibrary{groupplots}
\usepackage{amsmath}
\usepackage{pgfplotstable}
\definecolor{Paired-1}{RGB}{31,120,180}
\definecolor{Paired-2}{RGB}{166,206,227}
\definecolor{Paired-3}{RGB}{51,160,44}
\definecolor{Paired-4}{RGB}{178,223,138}
\definecolor{Paired-5}{RGB}{227,26,28}
\definecolor{Paired-6}{RGB}{251,154,153}
\definecolor{Paired-7}{RGB}{255,127,0}
\definecolor{Paired-8}{RGB}{253,191,111}
\definecolor{Paired-9}{RGB}{106,61,154}
\definecolor{Paired-10}{RGB}{202,178,214}
\definecolor{Paired-11}{RGB}{177,89,40}
\definecolor{Paired-12}{RGB}{105,105,105}
\definecolor{Paired-13}{RGB}{80,80,80}
\usetikzlibrary{matrix, positioning, patterns, shapes, arrows}
\pgfplotsset{compat=1.18}
\newtheorem{definition}{Definition}

\DeclareAcronym{ML}{
  short = ML ,
  long  = Maximum-likelihood
}
\DeclareAcronym{WHD}{
  short = WHD ,
  long  = Weighted Hamming Distance
}
\DeclareAcronym{pdf}{
  short = PDF ,
  long  = Probability Density Function
}
\DeclareAcronym{AWGN}{
  short = AWGN ,
  long  = additive white Gaussian noise
}

\DeclareAcronym{BPSK}{
  short = BPSK ,
  long  = Binary Phase-Shift Keying
}

\DeclareAcronym{OSD}{
  short = OSD ,
  long  = Ordered Statistics Decoding
}

\DeclareAcronym{GRAND}{
  short = GRAND ,
  long  = Guessing Random Additive Noise decoding
}
\DeclareAcronym{SGRAND}{
  short = SGRAND ,
  long  = Soft Guessing Random Additive Noise decoding
}
\DeclareAcronym{POSD}{
  short = POSD ,
  long  = Partial Ordered Statistics Decoding
}

\DeclareAcronym{GCD}{
  short = GCD ,
  long  = Guessing Codeword Decoding
}
\DeclareAcronym{GE}{
  short = GE ,
  long  = Gaussian elimination
}
\DeclareAcronym{CSI}{
  short = CSI ,
  long  = Channel State Information
}
\DeclareAcronym{PDF}{
  short = PDF ,
  long  = Probability Density Function
}

\DeclareAcronym{CDF}{
  short = CDF ,
  long  = Cumulative Density Function
}
\DeclareAcronym{LW}{
  short = LW ,
  long  = Logistic Weight
}
\DeclareAcronym{ILW}{
  short = ILW ,
  long  = Improved Logistic Weight
}
\DeclareAcronym{EW}{
  short = EW ,
  long  = Ellipsoidal Weight
}
\DeclareAcronym{HW}{
  short = HW ,
  long  = Hamming Weight
}
\DeclareAcronym{ISTEP}{
  short = ISTEP ,
  long  = Improved Step GRAND
}
\DeclareAcronym{TEP}{
  short = TEP ,
  long  = Test Error Pattern
}
\DeclareAcronym{LLR}{
  short = LLR ,
  long  = Log Likelihood Ratio
}
\DeclareAcronym{CA}{
  short = CA ,
  long  = Cyclic Redundancy Check Aided
}
\DeclareAcronym{RLC}{  
short = RLC ,
  long  = Random Linear Code
  }

  \DeclareAcronym{BCH}{  
short = BCH ,
  long  = Bose-Chaudhuri-Hocquenghem
  }
    \DeclareAcronym{SNR}{  
short = SNR ,
  long  = Signal to Noise Ratio
  }
  \DeclareAcronym{NCSI}{
  short = NCSI ,
  long  = no perfect knowledge of Channel State Information
}

  \DeclareAcronym{SOTA}{
  short = SOTA ,
  long  = state of the art
}
\begin{document}

\title{A Universal Systematic Method to Generate Error Patterns on Memoryless Channels}

\author{Marwan Jalaleddine, Jiajie Li, Syed Mohsin Abbas,  Warren J. Gross
\thanks{ Marwan Jalaleddine, Jiajie Li and Warren J. Gross are with the Department of Electrical and Computer Engineering at McGill university, Montreal, Quebec, Canada.  Their emails are: marwan.jalaleddine@mail.mcgill.ca, jiajie.li@mail.mcgill.ca,   warren.gross@mcgill.ca. Syed Mohsin Abbas is with Tampere University, Finland. His email is: mohsin.abbas@tuni.fi}}



\maketitle

\begin{abstract}
The high computational cost of approaching the performance of \ac{ML} decoding has limited its practical use for decades. Because the complexity grows exponentially with the message length, researchers have spent years developing algorithms like \ac{OSD}, \ac{POSD} and \ac{GRAND} which try to  {approach} ML performance. \ac{OSD}, \ac{POSD} and \ac{GRAND} work by trying to guess the error patterns affecting the received signals. However, there does not exist a systematic  {method} to extend the error pattern guesses to novel channels. This work introduces a systematic  {method} that uses the \ac{pdf} of a memoryless channel to generate a set of error patterns that can be applied on any future received signal on this channel. Simulation results show that our proposed method applied on \ac{GRAND}, \ac{OSD} and \ac{POSD} generally matches or outperforms current pre-generated error patterns on \ac{AWGN} channel, mixture of Gaussian distribution channels, Rayleigh fading channel with perfect knowledge of \ac{CSI} and Rayleigh fading channel with \ac{NCSI}.
\end{abstract}

\section{Introduction}

In $1922$, the statistical principle of  {\acf{ML}} estimation established a benchmark to the optimality of the estimate \cite{fisher_mathematical_1922}. This notion became an important concept in coding theory, which emerged in 1948 \cite{shannon_mathematical_1948}, as the ML estimate  {provides} the optimal performance target for any decoder. \ac{ML} decoding identifies the most likely codeword given a received channel signal; however, \ac{ML} decoding is an NP complete problem \cite{berlekamp_inherent_1978} as it require {s} evaluating $2^k$ candidates  {for binary codes} with $k$ being the message length.  {The infeasible computational complexity of ML decoding has driven the search for practical approximate \ac{ML} decoding algorithms.} 

In the 1950s, one of the first attempts at approximate \ac{ML} decoding emerged through sequential decoding \cite{wozencraft_sequential_1957}. The sequential decoder was the first practical algorithm capable of approaching ML decoding performance on low-rate convolutional codes. In the 1960s, a simplified sequential decoding algorithm with low memory requirements \cite{r_fano_heuristic_1963} refined the original approach. Due to the sequential decoder's effectiveness in the low signal-to-noise conditions, NASA deployed the sequential decoder on the Pioneer 9 mission and used it as the standard coding system for deep space \cite{massey_deep-space_1992,heller_performance_1968}. Sequential decoders face a fundamental limitation known as the cutoff rate. Exceeding this rate causes the number of states to grow exponentially, leading to high computational and time complexities. In a separate line of work, in the 1960s, an efficient exact ML Viterbi decoder for convolutional codes \cite{a_viterbi_error_1967} emerged that had Ohigher memory and computational requirements than the previously proposed sequential decoders. Unlike the previously proposed sequential decoders, the Viterbi decoder was amenable for hardware implementations using shift registers as it did not require complex control logic to lead the search for the codeword. Additionally, the Viterbi decoder provided deterministic latency in contrast to the previously proposed sequential decoders' unpredictable latency.

For block codes, the exhaustive ML search remained intractable since using the ML Viterbi decoder on the equivalent trellis results in high computational complexity \cite{kschischang_trellis_1995}.  
Among the earliest near-\ac{ML} decoding approaches on linear block codes, a scheme proposed in 1987 \cite{battail_decodage_1983,battail_decodage_1986,battail_we_1993,fang_decodage_1987} generates candidate codewords by applying error patterns to the message bits in order of increasing likelihood. This method effectively approximates the \ac{ML} search by examining error patterns starting from the most probable ones. However, enumerating error patterns in this order is not well suited to hardware implementation and is difficult to parallelize, limiting the practicality of the approach in modern communication systems.

\ac{OSD}, introduced in 1995  \cite{fossorier_soft-decision_1995}, leverages the most reliable $k$ received bits to generate near-\ac{ML} codewords. \ac{OSD} first reorders the received bits by decreasing reliability (so that the first $k$ positions correspond to the most reliable bits) and then applies  \ac{GE}  to the permuted generator matrix to obtain a systematic form with these $k$ bits as the information basis. OSD generates candidate codewords by flipping combinations of these most reliable bits according to a predetermined set of error patterns.  The analysis shows that OSD can achieve a decoding performance equivalent to \ac{ML} decoding by querying all error patterns with Hamming weight up to $\lceil d_H/4 - 1 \rceil$, where $d_H$ denotes the minimum distance of the code \cite{fossorier_soft-decision_1995}. Many works have proposed complexity-reduction techniques for \ac{OSD}, such as the box-and-match method \cite{valembois_box_2004}, the linear-equation technique \cite{yue_linear-equation_2022}, and the segmentation discarding technique \cite{yue_segmentation-discarding_2019}, all of which aim to discard unpromising error patterns. Alternatively, the approach in \cite{yang_efficient_2025} exploits the structure of Reed Solomon codes to avert the use of \ac{GE} for \ac{BCH} codes \cite{yang_efficient_2025} at the cost of limiting OSD's use for BCH codes.  Although fixed error patterns and error-pattern elimination methods reduce OSD's complexity, the required \ac{GE} for all codes except BCH codes impedes efficient hardware implementation \cite{fossorier_modified_2024}. 

In 2001, a \ac{ML} method \cite{valembois_improved_2001} that directly enumerates likely noise (error) sequences affecting the codeword emerged. Instead of guessing codewords directly, this method generates error patterns in order of decreasing likelihood and subtracts each pattern from the received signal.  {The first valid codeword produced by this method corresponds to the \ac{ML} codeword, albeit at the cost of a hardware-inefficient error-pattern scheduling}. {Since 2019, the underlying technique for guessing error patterns has been referred to as \ac{GRAND} \cite{duffy_guessing_2019}. The \ac{ML} error pattern generation technique, which produces the error patterns in the same order to those produced by \cite{valembois_improved_2001}, is referred to as \ac{SGRAND} \cite{solomon_soft_2020}. Numerous hardware-efficient error-pattern ordering strategies have been developed for GRAND, most notably the logistic-weight error patterns used in ORBGRAND in 2021 \cite{duffy_ordered_2021, duffy_ordered_2022}. These error patterns were the first to approximate the ML decoding behavior of GRAND with only minor performance degradation.
Logistic weight error patterns and other hardware-friendly error patterns \cite{abbas_improved_2025,condo_fixed_2022}, have driven the development of efficient hardware implementations for \ac{GRAND} \cite{abbas_improved_2025,abbas_guessing_2023,riaz_sub-08pjb_2023,abbas_high-throughput_2022,condo_fixed_2022}. While such patterns mitigate implementation complexity, they do not address a core algorithmic limitation of GRAND: it typically must guess the entire error vector affecting the codeword, which becomes impractical for codes with large error-correcting capability.} 

To decode codes of lower rates without the need of \ac{GE}, Partial Ordered Statistics Decoding (POSD) \cite{alnawayseh_ordered_2012,alnawayseh_low-complexity_2009,jalaleddine_partial_2023}, initially proposed in 2009 can be used as a simplification of the ML approach proposed in \cite{battail_decodage_1983,battail_decodage_1986,battail_we_1993,fang_decodage_1987}. The original POSD \cite{alnawayseh_ordered_2012,alnawayseh_low-complexity_2009} employed fixed error patterns inspired from the \ac{OSD} framework, but the choice of error patterns led to relatively poor decoding performance as the patterns used did not map correctly the error patterns seen on the channel. Subsequently in 2023, improved error-patterns inspired by \cite{duffy_ordered_2022,condo_high_2021} achieved enhanced decoding performance with POSD while retaining the hardware-friendly, predetermined error patterns approach. In a separate line of work, Guessing Codeword Decoding (GCD) \cite{ma_guessing_2024}, introduced in 2024, is effectively a rediscovery of the original 1986 method \cite{battail_decodage_1983,battail_decodage_1986,battail_we_1993,fang_decodage_1987} upon which POSD is based on. In fact, a summary of this algorithm proposed in 1986 can be found in the earlier literature (see, e.g., p.~361-362 of \cite{battail_we_1993}) in english and the equivalency of the distance metrics is established in Appendix \ref{appdx:1} in this work.

While \ac{GRAND}, \ac{POSD}, and \ac{OSD} algorithms can be extended to novel channel models, a systematic framework for such adaptations, without the use of Monte Carlo simulations {or piecewise-linear approximation techniques}, remains elusive. Although Monte Carlo methods effectively characterize error patterns in low Signal-to-Noise Ratio (SNR) regimes, their utility diminishes at higher SNRs. In these regions, the scarcity of observed test error patterns prevents a statistically significant ranking of test error patterns. Moreover, the piecewise-linear approximation technique proposed in ORBGRAND \cite{duffy_ordered_2022} can cause a large degradation in performance compared to the \ac{ML} error patterns when the channel's log-likelihood ratio distribution is non-smooth.

In this work, we introduce a novel method to generate channel adaptive error patterns for OSD, POSD and GRAND on memoryless channels. We first determine the \ac{PDF} of the sorted reliabilities of the received channel signal, the bits GRAND, OSD and POSD operate on. Then, we generate error patterns in increasing expected value of \ac{WHD} based on the fact that the codeword that minimizes the \ac{WHD} is the ML codeword. These generated error patterns can be loaded into the decoder and the same error patterns can be used on any received channel signal.  We validate our proposed method on AWGN, Rayleigh Fading with no knowledge of \ac{CSI}, Rayleigh Fading with perfect knowledge of \ac{CSI} and two mixture of Gaussian  channels with novel additive noise profiles{. We  consistently show that our proposed predetermined error patterns result in a decoding performance equivalent or better than the previous state-of-the-art predetermined error patterns.  We also note that our proposed error patterns can be used to improve the error pattern approximations\cite{duffy_ordered_2022} through generating a set of error patterns to be saved in memory for the regions that cannot be easily linearized}.

The remainder of this paper is organized as follows. Section II establishes the preliminaries and the notations used in this work. Section III details existing code-agnostic decoding algorithms, specifically \ac{GRAND}, \ac{OSD}, and \ac{POSD}, and explores various existing error pattern ordering techniques. In Section IV, we propose a theoretical framework that utilizes both Log-Likelihood Ratio (LLR) and received signal distributions to generate optimized error patterns. The subsequent sections evaluate this framework across diverse channel environments: Section V addresses the Additive White Gaussian Noise (AWGN) channel, providing a performance analysis of Test Error Patterns (TEPs); Section VI extends this to Mixture of White Gaussian channels; and Section VII examines uncorrelated fast-fading scenarios. Finally, Section VIII concludes the paper with a summary of our findings.

 

\section{Preliminaries and Notations}
\label{sec:preliminaries}

\subsection{Notations}

Matrices and vectors are denoted by a bold letter symbol ($\mathbf{M}$ or $\mathbf{v}$). The transpose operator is represented by $^\top$.
The number of $k$-combinations from a given set of $n$ elements is noted by $\binom{n}{k}$.
$\mathbf{1}_z$ is the indicator vector where all locations except the $z^{\text{th}}$ are $0$ and the $z^{\text{th}}$ is $1$.
All the indices start at $1$. The $i$'th element of vector $\mathbf{v}$ is represented as $\mathbf{v}[i]$. The subvector composed of elements $i$ to $j$ from vector $\mathbf{v}$ is represented as $\mathbf{v}[i:j]$.
The symbols $\implies$, $\centernot \implies$ denote \textit{implies} and \textit{does not imply} respectively. $\oplus$ denotes the XOR binary operation.
For this work, the Galois field with 2 elements is noted as $\mathbb{F}_2$. Furthermore, we restrict ourselves to $(n,k)$ linear block codes, where $n$ is the code length and $k$ is the message length.
Throughout this analysis and simulations, we assume Binary Phase Shift Keying (BPSK) modulation and that bit $0$ and bit $1$ are equiprobable. 

\subsection{Definitions}

\begin{definition}[Linear Block Code]
     A linear block code is a linear mapping $ \mathbb{F}_2^k \rightarrow \mathbb{F}_2^n$.
 To characterise any linear block code, there exists a $k \times n$ matrix $\mathbf{G}$ called {the} generator matrix.  The generator matrix is in standard form if it is represented as $\mathbf{G}=[I^k|P]$ where $\mathbf{I}^k$ is the identity matrix of size $k\times k$ and $\mathbf{P}$ is an $k\times n-k$ matrix. We use a notation [n,k] to define the codeword length and length of the used code. A codeword $\mathbf{c}$ of length $n$ belonging to the linear block code (represented by codebook $\mathbb{C}$) can be generated by multiplying the message ($\mathbf{u}$) of length $k$ with the generator matrix:

 \begin{equation}
     \mathbf{c}=\mathbf{u}\times \mathbf{G}
 \end{equation}

The corresponding systematic form of the  parity-check matrix is $\mathbf{H}= [\mathbf{P}^T  | I_{n-k}] $. Any codeword belonging to the codebook of generator matrix $\mathbf{G}$ produces a zero syndrome:
\begin{equation}
    \mathbf{s}=\mathbf{cH}^T=\mathbf{0} \hspace{0.5cm}|\hspace{0.5cm} \mathbf{c}\in \mathbb{C}
\end{equation}
\end{definition}

\begin{definition}[\ac{BPSK}]
BPSK is a linear modulation scheme that maps each binary symbol $b \in \{0,1\}$ to a real-valued signal point $x \in \{+1,-1\}$ via the mapping $x = 1 - 2b$.
\end{definition}

\begin{definition}[\ac{LLR}]
The \ac{LLR} ($L$) of an instance of the received signal of the received channel signal is the measure on our confidence that the received coded bit $c$ is $0$ as opposed to $1$ :
\begin{equation}
  L = \mathcal{L}(y)= \ln \left(\frac{P(y\text{ }|\text{   }c=0)}{P(y\text{ }|\text{  }c=1)} \right) .
\end{equation}
A positive LLR, is demodulated to a $0$ bit and a negative LLR is demodulated to a $1$ bit. The absolute value of the LLR $|L|$ represents how confident we are in our demodulated result. The higher the value of $|L|$, the higher the reliability of this received signal. The corresponding vector of LLRs is represented as $\mathbf{L}$.

\end{definition}

\begin{definition}
The hard demodulation function ($\theta$) thresholds the received LLRs :
\begin{align}
    \theta(L)=0 \hspace{0.5cm};\hspace{0.5cm} &  L \geq 0. \\
    \theta(L)=1 \hspace{0.5cm} ;\hspace{0.5cm}  & L<0. 
\end{align}
\end{definition}

\begin{definition}[Weighted Hamming Distance] \label{sec:whd}
The weighted Hamming distance (also known as the ellipsoidal weight, the correlation discrepancy, soft analog weight) represents the soft distance of the current codeword from the received signal:
\begin{equation}
w_H (\mathbf{c},\mathbf{L} )= \sum_{i=1} ^N    \left(\mathbf{c}[i] \oplus \theta(\mathbf{L}[i])\right)\times |\mathbf{L}[i]|.
\end{equation}

Noting that the ($  \mathbf{c}[i]\oplus \theta(\mathbf{L}[i])$) is equivalent to the hard error pattern ($\mathbf{e}[i]$) that was applied by the decoder onto the hard demodulated signals, we can represent the weighted Hamming distance in terms of the error pattern:
\begin{equation}
w_H' (\mathbf{e},\mathbf{L})= \sum_{i=1} ^N    \mathbf{e}[i] \times \left|\mathbf{L}[i]\right|  .
\end{equation}
{If all codewords have equal probability of being transmitted, }the codeword that minimizes the weighted Hamming distance in a memoryless channel is the maximum likelihood codeword  \cite{valembois_comparison_2002}. 
In case the soft channel information are not known, the weighted Hamming distance becomes the Hamming distance which can be calculated as:

\begin{equation}
hw'(\mathbf{e}) = \sum_{i=1} ^N    \mathbf{e}[i] .
\end{equation}
Hence, the maximum likelihood criteria for a hard valued channel is the codeword that minimizes the number of bit-flips from the received signal. 
\end{definition}

\subsection{Computing Lists of Binary
Vectors That Optimize a Given Weight Function} \label{sec:optimizing_binary}

The authors in \cite{valembois_improved_2001} presented an efficient algorithm  for generating binary vectors (patterns) in strictly increasing order of a given weight function $w$, an improvement of an original algorithm proposed by Battail \cite{battail_decodage_1986}. Let $z$ be the number of bit positions under consideration. Each pattern $\mathbf{e}$ is a binary vector of length $N$, and the set $Z$ holds the set of the positions of 1's in $\mathbf{e}$. The weight function $w'(\mathbf{e}) = w(Z)$ satisfies the monotonicity property or all positions $j \notin Z, j \notin Z'$ :
\begin{align}
       w(Z) &\leq w(Z \cup \{j\}) \quad ,  \label{eq:minweight_1}\\
    w(Z) \leq w(Z') &\implies w(Z \cup \{j\})  \leq w(Z' \cup \{j\}) \label{eq:minweight_2} ,
\end{align}
  where $\{j\}$ corresponds to a set with element $j$ ($1$ on position j of the $\mathbf{e}$ vector) and $Z \cup \{j\}$ describes a set $Z$ with an additional element $j$.

\begin{table}[t]
\caption{An example of the execution of weight minimization procedure assuming weights of index $1,2$ and $3$ are $0.3,0.4$ and $0.5$ respectively, and the weight function is additive.}
\label{tab:SGRANDex}
\centering
\resizebox{\columnwidth}{!}{
\renewcommand{\arraystretch}{1.5}
\begin{tabular}{|c|c|c|c|c|c|c|}
\cline{1-7}
$t$ & $Z_{c}$ & $w(Z_{c})$ & $j^*$ & $S$ & $V$ & $A$ \\ \cline{1-7}
1 & \{\} & 0.0 & 0 & \{(0.3,\{1\},1),(0.4,\{2\},1,),(0.5,\{3\},1,)\}& \{1,2,3\} & \{(\{\},0)\} \\ \cline{1-7}
2 & \{1\} & 0.3 & 1 & $\{(0.3,\{1\},1),(0.4,\{2\},1),(0.5,\{3\},1)\}$ &\{1,2,3\} & \{(\{\},0), (\{1\},0.3)\} \\ \cline{1-7}
3 & \{2\} & 0.4 & 2 & \{$(0.3,\{1\},1,),(0.7,\{1,2\},2),(0.5,\{3\},0)$\}  &\{2,3\}& \{(\{\},0), (\{1\},0.3), (\{2\},0.4)\} \\ \cline{1-7}
4 & \{3\} & 0.5 & 3 & \{$(0.3,\{1\},1),(0.7,\{1,2\},2),(0.8,\{1,3\},2)$\} &\{2,3\}& \{(\{\},0), (\{1\},0.3), (\{2\},0.4), (\{3\},0.5)\} \\ \cline{1-7}
5 & \multirow{1}{*}{\{1,2\}} & \multirow{1}{*}{0.7} & \multirow{1}{*}{2} & \multirow{1}{*}{$\{(0.3,\{1\},1),(0.7,\{1,2\},2),(0.8,\{1,3\},2)\}$} &\multirow{1}{*}{\{3\}}& \{(\{\},0), (\{1\},0.3), (\{2\},0.4), (\{3\},0.5),(\{1,2\},0.7)$\}$ \\ \cline{1-7}
6 & \multirow{2}{*}{\{1,3\}} & \multirow{2}{*}{0.8} & \multirow{2}{*}{3} & \multirow{2}{*}{$\{(0.3,\{1\},1),(0.7,\{1,2\},2),(0.9,\{2,3\},3)\}$} &\multirow{2}{*}{\{3\}}& \{(\{\},0), (\{1\},0.3), (\{2\},0.4), (\{3\},0.5),(\{1,2\},0.7), \\ 
  &                      &              &      &                &                                   & $(\{1,3\},0.8)\}$ \\ \cline{1-7}
7 & \multirow{2}{*}{\{2,3\}} & \multirow{2}{*}{0.9} & \multirow{2}{*}{3} & \multirow{2}{*}{$\{(0.3,\{1\},1),(0.7,\{1,2\},2),(1.2,\{1,2,3\},5)\}$} &\multirow{2}{*}{\{3\}}& \{(\{\},0), (\{1\},0.3), (\{2\},0.4), (\{3\},0.5), (\{1,2\},0.7),\\ 
  &                      &                    &                &    &                               & $(\{1,3\},0.8),(\{2,3\},0.9)\}$ \\ \cline{1-7}
8 & \multirow{2}{*}{\{1,2,3\}} & \multirow{2}{*}{1.2} & \multirow{2}{*}{3} & \multirow{2}{*}{$\{(0.3,\{1\},1),(0.7,\{1,2\},2),(1.2,\{1,2,3\},5)\}$} &\multirow{2}{*}{\{\}}& \{(\{\},0), (\{1\},0.3), (\{2\},0.4), (\{3\},0.5),(\{1,2\},0.7), \\ 
  &                          &             &       &                &                                   & $(\{1,3\},0.8),(\{2,3\},0.9),(\{1,2,3\},1.2)\}$ \\ \cline{1-7}
\end{tabular}
}
\end{table}

The algorithm begins by allocating a list of arrays ($S[1:z]$). Each $S[j]$ contains exactly one candidate pattern whose \emph{last} error occurs at position $j$. Each list element is a tuple $(w, Z, p)$ where
$w$ is the weight $w(Z)$ of the pattern,
$Z$ is the set of positions where of $1$'s and
$p$ is pointer to the parent pattern in history list $A$. The history list ($A$) is a sequential record of all patterns that have been generated and processed, stored as $(w, Z)$ pairs. Additionally, the validity set determines whether all the binary vectors originating from that vector have been generated.

The algorithm allocates only
one new pattern $Z_{s}$ each time a new $Z_{c}$ pattern is chosen. If a successor cannot be allocated, the pattern is considered not valid and no further patterns are derived from it. 

{\setlength{\algomargin}{1em}
\SetAlgoNlRelativeSize{-1}
\linespread{0.9}
\begin{algorithm}[t]
    
    \KwIn{Weight values $w[1], w[2], \dots, w[N]$ for each position}
    \KwIn{Maximum number of patterns $M$}
    \KwOut{Sequence of sets of patterns $Z[1], Z[2], \dots , Z[M]$ with $w(Z[1]) \leq w(Z[2]) \leq \cdots \leq w(Z[M])$}
    
    \BlankLine
    \textbf{Initialization:} \\
    $A[1] = (\{\}, 1)$\;
    \For{$j = 1$ \KwTo $k$}{
        $S[j] = (\{j\}, w(\{j\}), 1)$\;
        $V =V \cup \{j\}$;\
    }
    
    \BlankLine
    \textbf{Main Loop:} \\
    \For{$t = 1$ \KwTo $M$}{
       \textbf{Select minimal weight pattern from the valid set:}\\ 
$j^* = \arg\min_{j \in V} S[j].w$\;
$Z_{\text{c}} \leftarrow S[j^*].Z$\;
        
        \BlankLine
        \textbf{Update History:}\\
        Append $(Z_{\text{c}}, w(Z_{\text{c}}))$ to $A$\;
        
        \BlankLine
        \textbf{Find Successor:}\\
        $p_{\text{p}} \leftarrow S[j^*].p$\;
        $p_{\text{n}} \leftarrow p_{\text{p}} + 1$\;
        \While{$p_{\text{n}} < |A|$ \textbf{and} ($A[p_{\text{n}}].w \geq S[j^*].w$ \textbf{or} $\max(A[p_{\text{n}}].Z) \geq j^*$)}{
            $p_{\text{n}} \leftarrow p_{\text{n}} + 1$\;
        }
        
        \BlankLine
        \textbf{Update List:}\\
        \If{$p_{\text{n}} < |A|$}{
            $Z_{\text{s}} \leftarrow A[p_{\text{n}}].Z \cup \{j^*\}$\;
            $S[j^*] \leftarrow (Z_{\text{s}}, w(Z_{\text{s}}), p_{\text{n}})$\;
        }
        \Else{
            $V =V - \{j^*\}$\;
        }
    }
    \caption{Pattern Generation by Increasing Weight}
    \label{alg:pattern-gen}
\end{algorithm}
The algorithm guarantees that the patterns are generated in strictly increasing order of weight ($w(Z_1) \leq w(Z_2) \leq \cdots \leq w(Z_M)$). Additionally, this algorithm guarantees completeness as all patterns can be generated exactly once.
TABLE \ref{tab:SGRANDex} shows an example of the operation of the algorithm on a set of 3 elements. We note that for sequential error generation, all the elements of $A$ with an index less than the minimum $p_{next}$ in set $S$ can be removed. Additionally, the algorithm can be optimized to reduce memory use.

\section{Code Agnostic Decoding Algorithms and Error Patterns}
In this section, we will discuss different code agnostic decoders. Most of the discussed decoders have variants that can achieve close to \ac{ML} performance. Hence, the decoders aim to :
\begin{enumerate}
    \item Generate a result that is a codeword belonging to the codebook.
    \item Minimize the weighted Hamming distance compared to all the other codewords in the codebook.
\end{enumerate}
The maximum-likelihood version of GRAND \cite{solomon_soft_2020,valembois_improved_2001} traverses its error pattern in a manner that minimizes the weighted Hamming distance, but each of its results is not necessary a codeword. Alternatively, \ac{OSD} \cite{fossorier_soft-decision_1995} and POSD \cite{alnawayseh_low-complexity_2009,alnawayseh_ordered_2012,jalaleddine_partial_2023} generate elements that are codewords in the codebook, but in an order that does not necessarily minimize the weighted Hamming distance. This section will discuss the details of these decoding methods.

\subsection{Guessing Random Additive Noise Decoding}
\label{s:B-GRAND}
\begin{algorithm}[t]
\SetKwInOut{Input}{Input}
\SetKwInOut{Output}{Output}
\Input{ Received word LLRs $\mathbf{L}$, parity-check matrix $\mathbf{H}$, maximum query count $M$}
\Output{Decoded codeword $\mathbf{\widehat{x}}$ or $ABANDON$}
\While {$n_b \leq M$}
{
$\mathbf{e} \gets$ NextNoiseSequence ($\phi(\mathbf{L}),n$) \;

$\mathbf{\widehat{x}} = \theta(\mathbf{L}) \oplus \phi_1^{-1}(\mathbf{e})$ \;
\If {$\mathbf{\widehat{x}} \mathbf{H}^T = \mathbf{0}$}{
\Return $\mathbf{\widehat{x}}$\; }
$n_b ++$ \;
}
\caption{GRAND Algorithm}
\label{alg:grand}
\end{algorithm}

{Guessing Random Additive Noise Decoding (GRAND) \cite{duffy_capacity-achieving_2019,valembois_improved_2001,duffy_ordered_2022} has emerged as a universal decoding framework, with certain variants capable of achieving maximum-likelihood performance \cite{valembois_improved_2001,solomon_soft_2020}}. The algorithm operates by estimating the additive noise vector $\mathbf{e}$ that has corrupted the transmitted codeword. Some variations \cite{valembois_improved_2001,solomon_soft_2020} utilize the soft information, in the form of bit reliability metrics $\mathbf{L}[i]$, to order potential noise sequences in descending order of their probability using an algorithm similar to Algorithm \ref{alg:pattern-gen} \cite{valembois_improved_2001}. These ordered noise patterns are sequentially applied to the received word, commencing with the most likely (zero bit-flips) and proceeding to a predetermined maximum number of queries.

To determine whether a candidate vector $\mathbf{\widehat{x}}$ constitutes a valid codeword, GRAND employs the parity-check matrix $\mathbf{H}$. The syndrome $\mathbf{s}$ is computed as:
\begin{equation} \label{eq:appx}
\mathbf{s} = \mathbf{\widehat{x}} \mathbf{H}^T .
\end{equation}
A syndrome vector $\mathbf{s} = \mathbf{0}$ thus indicates that $\mathbf{\widehat{x}} \in \mathbb{C}$.
To avoid the computational complexity associated with computing the optimized weight sequences{, using Algorithm \ref{alg:pattern-gen} on each of the received channel signals \cite{valembois_improved_2001}}, some variations of GRAND aim to create channel specific error patterns that only depend on the ordered bit reliabilities based on the LLRs \cite{duffy_ordered_2022} \cite{abbas_improved_2025} \cite{condo_high_2021}.

\subsection{Ordered Statistics Decoding}
\label{s:B-OSD}

 \ac{OSD} \cite{fossorier_soft-decision_1995} is a decoder capable of achieving \ac{ML} decoding performance at the cost of substantial computational complexity, particularly through its use of  \ac{GE} for generator matrix reordering \cite{fossorier_soft-decision_1995}. Multiple architectures have been proposed to mitigate \ac{OSD} latency through early termination criteria \cite{wu_soft-decision_2007-1} \cite{wu_soft-decision_2007} or by constraining the search space of test patterns \cite{yue_segmentation-discarding_2019}.
Given that the \ac{GE} complexity scales as $\mathcal{O} (n^3)$, implementing an efficient low-latency universal \ac{OSD} decoder remains challenging.

The \ac{OSD} algorithm proceeds as follows.
First an initial processing step involves reordering the received codeword according to decreasing reliability magnitudes. This permutation, denoted $\phi _1 ()$, sorts the absolute values of the log-likelihood ratios (LLRs) $|\mathbf{L}[i]|$ in descending magnitude order. Subsequently, the generator matrix $\mathbf{G}$ undergoes column permutation via $\phi_1$ to yield $\mathbf{G'}$. To ensure linear independence among the first $k$ columns of $\mathbf{G'}$,  \ac{GE}  is applied to obtain a systematic form $\mathbf{G''}$, generating an additional permutation $\phi_2()$:
\begin{align}
    \mathbf{G}'' = \phi_2(\phi_1(\mathbf{G})) 
\end{align}

The received vector LLRs are also mapped to a binary vector via hard decision function $\theta()$. The order-$o$ \ac{OSD} algorithm processes the $k$ most reliable bits according to Algorithm \ref{alg:osd}.
\begin{algorithm}[t]
\DontPrintSemicolon
\SetKwInOut{Input}{Input}
\SetKwInOut{Output}{Output}
\Input{ Received word LLRs $\mathbf{L},$ permuted generator matrix $\mathbf{G''}$, maximum number of queries $M$}
\Output{$\mathbf{\widehat{x}}$}
$n_b = 0$ \tcp*{Current bit-flip count}
$w_H = \infty$ \tcp*{Minimum WHD}
$\mathbf{u_c} = \theta(\phi_2 ( \phi_1 (\mathbf{L}))) [n-k:n]$ \;
\While{$n_b \leq M$}{
$\mathbf{e_s} \gets $NextNoiseSequence$( \phi_2( \phi_1$$(\mathbf{L})) [n-k:n],k )$ \;
$\mathbf{e} = \mathbf{e_s} \times \mathbf{G''}$ \;

$\mathbf{x_c}= (\mathbf{u_c} \times \mathbf{G''}) \oplus \mathbf{e}$ \;
\If {$ \text{WHD}(\mathbf{x_c}, \phi_2 ( \phi_1 (\mathbf{L})) < w_H $}{
$\mathbf{\widehat{x}} = \phi_1 ^{-1} (\phi_2 ^{-1} ( \mathbf{x_c}))$ \; 
$w_H=\text{WHD}(\mathbf{x_c}, \phi_2 ( \phi_1 (\mathbf{L}))$ \; }
$n_b ++$ \;}
\Return $\mathbf{\widehat{x}}$ \;
\caption{OSD-$o$ Algorithm}
\label{alg:osd}
\end{algorithm}

The algorithm initializes by applying the most probable noise sequence to the $k$ most reliable bits, generating candidate vector $\mathbf{\hat{u}}_c$. This candidate is multiplied by $\mathbf{G''}$ to produce codeword $\mathbf{x_c}$. The codeword minimizing this weighted Hamming distance is selected as the output. 

\subsection{Partial Ordered Statistics Decoding}

\begin{algorithm}[t]
\DontPrintSemicolon
\SetKwInOut{Input}{Input}
\SetKwInOut{Output}{Output}
\Input{ Received word LLRs $\mathbf{L},$ generator matrix $\mathbf{G}$, maximum number of queries $M$}
\Output{$\mathbf{\widehat{x}}$}
$n_b = 0$ \tcp*{Current bit-flip count}
$w_H = \infty$ \tcp*{Minimum WHD}
$\mathbf{u_c} =  \theta(\phi_1 (\mathbf{L} [1:k]))$ \;
\While{$n_b \leq M$}{
$\mathbf{e_s} \gets$ NextNoiseSequence   $( \phi_1 (\mathbf{L} [1:k]),k$)\;
$\mathbf{e} = \mathbf{e_s} \times \mathbf{G''}$ \;

$\mathbf{u_c} = \theta(\mathbf{L}) [1:k]$  \;
$\mathbf{x_c}= \mathbf{u_c} \times \mathbf{G}$ $\oplus $ $ \phi_1 ^{-1}(\mathbf{e} )$\;
\If {$ \text{WHD}(\mathbf{x_c}, \mathbf{L}) < w_H $}{
$\mathbf{\widehat{x}} =\mathbf{x_c}$ \; 
$w_H=\text{WHD}(\mathbf{x_c}, \mathbf{L})$ \; }

}
\Return {$\mathbf{\widehat{x}}$} \;
\caption{POSD Algorithm}
\label{alg:seg}
\end{algorithm}

Partial ordered statistics decoding (POSD), detailed in Algorithm \ref{alg:seg}, is a simplified ordered statistics decoder that foregoes the requirement of  \ac{GE}  \cite{alnawayseh_ordered_2012}. Unlike \ac{OSD} which applies error patterns on the most reliable bits of the received signal (which are mostly error free), POSD applies its error patterns on the first $k$ bits of the received signal and tries to generate a codeword which minimizes the \ac{ML} distance criterion. Hence, OSD usually needs to only sort the first $k$ bits of the received signal, reducing the size of the sorting circuitry to be employed with it which is usually the bottleneck in GRAND methods \cite{abbas_hardware_2022}. POSD offers a middle ground between GRAND, which under performs with medium rate codes, and \ac{OSD}, which generally requires the computationally expensive  \ac{GE}  for every decoding attempt. The original version of POSD variant \cite{alnawayseh_low-complexity_2009,alnawayseh_ordered_2012} did not evaluate the channel noise sequences to determine the error patterns to be applied on the first $k$ bits which resulted in a degraded error correction ability. Later work \cite{jalaleddine_partial_2023} solved this issue by suggesting enhanced error patterns, inspired from GRAND \cite{duffy_ordered_2022}.

We note that the base algorithm of the POSD proposed in 2023 \cite{jalaleddine_partial_2023} is identical to themethod proposed in 1986 \cite{battail_decodage_1986,battail_we_1993,fang_decodage_1987}   and the base GCD algorithm \cite{ma_guessing_2024} proposed in 2024. The POSD variation relies on hardware-friendly error patterns that are generated based on the ordered statistics of the LLRs; however the 1986 \cite{battail_decodage_1986,battail_we_1993,fang_decodage_1987} method and GCD proposed using a error geneneration strategy similar to the one discussed in Section \ref{sec:optimizing_binary}. The need for using this optimized error pattern generation method limits its hardware compatibility. 

Additionally due to {the similarities between OSD and POSD, with} POSD being a simplified version of OSD, most of the complexity reduction techniques and modifications for \ac{OSD} could be applied to POSD. For example, the box and match method \cite{valembois_box_2004}, could be applied on POSD. For all the modifications that can be applied to \ac{OSD} and could be later adapted to POSD, we refer the reader to \cite{yue_revisit_2021}.

\subsection{\ac{TEP} Generation} \label{section:TEPs}
We examine various \ac{TEP} generation methods that can be used with the aforementioned decoders.

\subsubsection{\ac{HW} Order}
The \ac{HW} ordered \ac{TEP} generatation method generates test error patterns according to increasing Hamming weight, defined as:
\begin{equation}
HW (\mathbf{e}) = \sum_{i=1}^{n} \mathbf{e}[i]. \label{eq:hw}
\end{equation}
These error patterns { were initially proposed to be used with OSD \cite{fossorier_soft-decision_1995}. Interestingly, these error patterns can operate independent of the soft channel information.}

\subsubsection{\ac{LW} Order}
The logistic weight order, { initially proposed for the GRAND variant Ordered reliability bits GRAND (ORBGRAND) \cite{duffy_ordered_2022},} generates test error patterns according to ascending logistic weight, computed as:
\begin{equation}
LW (\mathbf{e}) = \sum_{i=1}^{n} i \times \mathbf{e}[i]. \label{eq:lw}
\end{equation} 

\subsubsection{\ac{ILW} Order}
The improved logistic weight order, initially proposed with GRAND \cite{condo_high_2021}, prioritizes LW \ac{TEP}s with lower Hamming weight. The objective function that is minimized is:
\begin{equation}
ILW (\mathbf{e})= \sum_{i=1}^{\sum(\mathbf{e}[j])} \text{index}(\mathbf{e}[i]) \times i, \label{eq:ilw}
\end{equation}
where index denotes the index of the ordered vector of indices corresponding to nonzero entries in $\mathbf{e}$:
\begin{equation}
    \mathbf{im}=\mathbf{ind}[\mathbf{e}[i]==1].
\end{equation}

\subsubsection{Maximum Likelihood (ML) Patterns}
Using {Algorithm \ref{alg:pattern-gen}} applied on the currently received LLRs, the error patterns can be produced in increasing weighted Hamming distance\cite{valembois_improved_2001}. This method has been shown to outperform other pre-generated error patterns with {Soft GRAND} and achieve \ac{ML} performance with a large amount of error patterns\cite{solomon_soft_2020}. In \cite{valembois_comparison_2002}, the authors applied this method to generate error patterns for \ac{OSD} and showed that the Hamming weight error patterns have a similar decoding performance to the \ac{ML} patterns. This is due to the fact that :
\begin{equation}
    \argmin_{\mathbf{e}[1:k]} w_H (\mathbf{e}[1:k],\mathbf{L}[1:k]) {\centernot \implies} \argmin_{\mathbf{e}[1:N]} w_H (\mathbf{e}[1:N],\mathbf{L}[1:N])
\end{equation}
In \cite{battail_decodage_1986,battail_we_1993,fang_decodage_1987,battail_decodage_1983}, the author{s} suggested a the optimized version of Algorithm \ref{alg:pattern-gen} to be used to generate error patterns on POSD. Similarly,
\cite{ma_guessing_2024} suggests using it on GCD.
Regardless of the proven performance gains with GRAND, no hardware architecture has been proposed of GRAND with this \ac{ML} error pattern generation technique due to its hardware incompatibility resulting from its sequential nature and the dynamic memory allocation.

\subsubsection{Empirical Error Patterns { (LUT)}}
Following the methodology established in \cite{condo_fixed_2022} {for GRAND}, \ac{TEP}s can be generated according to their empirical likelihood of occurrence. After sorting the received channel signal by increasing reliability, Monte-Carlo simulations can be conducted to record the most frequently occurring \ac{TEP}s. This empirical set, stored in memory, can be utilized with both GRAND \cite{condo_fixed_2022}, \ac{OSD} and POSD{\cite{jalaleddine_partial_2023}. We refer to these error patterns in this work as lookup table (LUT) error patterns.}

\section{Proposed Theoretical Framework, The \ac{EW} Order}

{
 The probability of occurrence of an error pattern with a certain Hamming weight given the probability of error $p$ is given by the binomial term:
\begin{equation}
    P(hw'(\mathbf{e})=x) = \binom{n}{x} p^{\,x}\,(1 - p)^{n-x},
\end{equation}
Considering the example of an AWGN channel with BPSK modulation and with $\frac{E_s}{N_0}=8dB$, the probability of error can be directly calculated as:
\begin{equation}
    p=Q\!\left(\sqrt{2\,\frac{E_s}{N_0}}\right) = 1.91 \times 10^{-4}.
\end{equation}}
{
Hence, the chance of encountering an error pattern with $hw'(\mathbf{e})=3$ in a codeword of length $128$ bits can be calculated as:}
\begin{equation}
    P(hw'(\mathbf{e})=3) = \binom{128}{3} (1.91 \times 10^{-4})^{\,3}\,(1 - 1.91 \times 10^{-4})^{125}=2.3\times 10^{-6}.
\end{equation}

{
In other words, among $10^6$ error patterns, only approximately 
$2.3$ are expected to have a Hamming weight of $3$. Although such error patterns rarely occur, their impact on frame error rate (FER) performance is significant. If error patterns of Hamming weight 3 are not queried, the FER performance of the decoder is bounded from below by $2.3 \times 10 ^{-6}$ irrespective of the error correcting capability of the underlying code. This FER floor might be unsuitable for some URLCC applications \cite{feng_ultra-reliable_2021,hampel_5g_2019} where extremely low target FERs are required. Moreover, characterizing these rare error patterns through Monte Carlo simulation is inherently challenging precisely because of their scarcity.} To overcome the time complexity of generating the error patterns using Monte-Carlo simulations and to have a more robust ordering of the error patterns, we develop a systematic approach that aims to reduce the expected value of the weighted Hamming distance. We discuss two variations, one where the distribution of the LLRs exists in closed form and another where the distribution of the LLRs is difficult to define in closed form:
\subsection{Using the LLR distribution} \label{sec:tep_gen_llr}
This analysis assumes the use of BPSK modulation on a continuous-value memoryless channel. 
{Let $f_{L}$ denote} the probability density function of the LLR at the receiver.The operations that GRAND, OSD, {POSD}, rely on is first determining the reliability of the signal through calculating the absolute value of the LLRs followed by sorting the LLRs in terms of their reliabilities and finally generating error patterns based the order of bit reliabilities (or their values in the ML error patterns). In the following algorithm we will follow step-by-step what happens to the \ac{PDF} of the individual bits throughout the steps.
\subsubsection{Finding the reliabilities}
Calculating the reliability of a bit is equivalent to calculating the absolute value of the LLRs. This will provide us a measure of how far away we are from the $0$ threshold which determines whether the bit is $0$ or $1$. After applying the absolute value on the LLRs, the \ac{PDF} defined by the $f_{L}(l)$ function and \ac{CDF} defined by $F_{L}(l)$ function of the distribution of the LLR is {``}folded" providing the following distribution:
\begin{equation}
    f_{|\mathbf{L}|}(l)= f_{\mathbf{L}} (l) + f_{\mathbf{L}} (-l), \hspace{1cm} l \geq 0 \label{eq:abs_LLRs_pdf}
\end{equation}
\begin{equation}
    F_{|\mathbf{L}|}(l)= F_{\mathbf{L}} (l) + F_{\mathbf{L}} (-l), \hspace{1cm} y \geq 0
\end{equation}
\subsubsection{Sorting the reliabilities}
After the reliabilities are determined, the reliabilities are sorted. Assuming sorting happens in ascending order of reliabilities, the \ac{CDF} of the ordered statistics of the resulting distribution follows a binomial distribution \cite{george_casella_statistical_2024}:
\begin{equation}
    F_{\phi(|\mathbf{L}|)[i]}(y) = \sum_{i=k}^N \binom{N}{i} \left[F_{|\mathbf{L}|}(y)\right]^i \left[1 - F_{|\mathbf{L}|}(y)\right]^{N - i}.
\end{equation}
The corresponding \ac{PDF} can be obtained through differentiating the \ac{CDF} to obtain:
\begin{equation}
    f_{\phi(|\mathbf{L}|)[i]}(y)= \frac{N!}{(i-1)!(N-i)!} \left[ F_{|\mathbf{L}|} (y) \right]^{i-1}   \times \left[ 1- F_{|\mathbf{L}|} (y) \right]^{N-i}  f_{|\mathbf{L}|}(y),
\label{eq:binomial_dist}
\end{equation}

where $i\in [1,N]$, and the full derivation can be seen in \cite{george_casella_statistical_2024}.
\subsubsection{Calculating the expected values}
{Given the \ac{PDF} of the sorted reliabilities at each bit location, we can compute the expected values of the bit positions in the ordered reliability statistics}:
\begin{equation}
    \mathbb{E}(\phi(|\mathbf{L}|)[i]) = \int_0^\infty  f_{\phi(|\mathbf{L}|)[i]} (l)\times l \hspace{0.2cm} dl  \label{eq:expected_value_calculation}
\end{equation}
This gives us an expected value of the LLR in each of the position in the received signal.

From this expected value, we can calculate the expected weighted Hamming distance of any error pattern GRAND, POSD and \ac{OSD} applies. Owing to the linearity of expectation and given that the error patterns are deterministic constants, we can calculate the expected value of $w_H$ of an error pattern as:
\begin{equation}
\mathbb{E} (w_H) =  \mathbb{E}\left(\sum_{i=1} ^N    \mathbf{e}[i] \times |\mathbf{L}[i]|\right)=\sum_{i=1} ^N    \mathbf{e}[i] \times  \mathbb{E}(|\mathbf{L}[i]|).
\end{equation}
\subsubsection{Generating Error Patterns}
Rather than trying to produce error patterns that minimize $w_H$ which can only be done with the knowledge of the current set of received LLRs, we choose to generate \ac{TEP}s offline which minimize the expected value of $w_H$,
$
   \mathbb{E} [w'_H(\mathbf{e},\mathbf{L})].
$
Computing lists of \ac{TEP}s that minimize $\mathbb{E} (w_H)$ is done using the Algorithm \ref{alg:pattern-gen} since the weighted Hamming distance, a non-negative number, satisfies the monotonicity property. The first condition specified in (\ref{eq:minweight_1}) is satisfied by noting that the expected value of the weighted Hamming distance is always a non-negative number :

\begin{align}
        \mathbb{E}(w_H(\mathbf{e} \cup \mathbf{1}_j)) &= \left(\sum_{i=1} ^N    e[i] \times  \mathbb{E}(|\mathbf{L}[i]|)\right) +  \mathbb{E}(|\mathbf{L}[i]|), \\
       &= \mathbb{E}(w_H(\mathbf{e}))+  \mathbb{E}(w_H(\mathbf{1}_j))\geq  \mathbb{E}(w_H(\mathbf{e}))\quad . 
\end{align}
For the second condition specified in (\ref{eq:minweight_2}), : 
\begin{align}
     \mathbb{E}(w_H(\mathbf{e})) \leq  \mathbb{E}(w_H(\mathbf{e}')) &\implies  \mathbb{E}(w_H(\mathbf{e})) + \mathbb{E}(w_H(1_j)) \leq  \mathbb{E}(w_H(\mathbf{e}')) + \mathbb{E}(w_H(1_j)), \\ &\implies  \mathbb{E}(w_H(\mathbf{e} \cup 1_j))  \leq  \mathbb{E}(w_H(\mathbf{e}' \cup 1_j)) .
\end{align}

Using Algorithm \ref{alg:pattern-gen} allows us to generate a list of TEPs to be used with any memoryless channel given the knowledge of the LLR distribution of the channel. We refer to the TEPs produced using this method as EW TEPs.

\subsection{Using the distribution of the received signal}\label{sec:tep_gen_no_llr}
 Sometimes, the formulation of the LLR is difficult to calculate especially when $\mathcal{L}(y)$ is a non-linear, non-invertible function of $y$. In these cases, finding the distribution of the LLR analytically requires solving a difficult transformation-of-random-variables problem. In this case, we suggest an alternative method which depends on $f_y(y)$, the \ac{PDF} of the received signal. 

\subsubsection{Sort the received channel signal} The, algorithm starts by sorting $y$ in ascending order. This step gives us the distribution of each of the received bits in ascending value of received signal: 
\begin{equation}
    f_{\phi(\mathbf{y})[i]}(y)= \frac{N!}{(i-1)!(N-i)!} \left[ F_{\mathbf{y}} (y) \right]^{i-1}  \times \left[ 1- F_{y} (\mathbf{y}) \right]^{N-i}  f_{\mathbf{y}}(y)
\end{equation}
\subsubsection{Calculate the expected value of the received signal} Then the expected value of each received signal on each bit can be calculated as: 
\begin{equation}
    \mathbb{E}(\phi(\mathbf{y})[i]) = \int_0^\infty  f_{\phi(\mathbf{y})[i]}(y) \times y \hspace{0.2cm}dy. 
\end{equation}
\subsubsection{Calculate the LLR of the expected values of the received signal} The LLR of each $\mathbb{E}(\phi(|\mathbf{y}|)[i])$ is calculated: 

\begin{equation}
    \mathcal{L}(\mathbb{E}(\phi(\mathbf{y})[i])) = \ln \frac{P(\mathbb{E}(\phi(\mathbf{y})[i])|c=0)}{P(\mathbb{E}(\phi(\mathbf{y})[i])|c=1)} .
\end{equation}
Then the reliabilities of the bits  can be determined calculating the absolute values of the LLRs $ \left| \mathcal{L}(\mathbb{E}(\phi(\mathbf{y})[i]))\right|$.

\subsubsection{Generating Error Patterns}

After sorting the LLRs in increasing $ \left| \mathcal{L}(\mathbb{E}(\phi(\mathbf{y})[i]))\right|$ to produce $ \phi(\left| \mathcal{L}(\mathbb{E}(\phi(\mathbf{y})[i])\right|)$, Algorithm \ref{alg:pattern-gen} can be used to generate lists of binary error vectors by increasing:
$$ \sum_{i=1}^{N}\mathbf{e}[i]\times \left| \mathcal{L}(\mathbb{E}(\phi(\mathbf{y}))\right|[i].$$


\subsection{Applicability on GRAND,OSD and POSD}
It is important to know that GRAND applies \ac{TEP}s on all the recieved bits. The LLRs associated with the bits follow the distribution of :
\begin{equation}
    f_{\phi(|\mathbf{L}|)[i]}(y)= \frac{n!}{(i-1)!(n-i)!} \left[ F_{|\mathbf{L}|} (y) \right]^{i-1}   \times \left[ 1- F_{|\mathbf{L}|} (y) \right]^{n-i}  f_{|\mathbf{L}|}(y).
\end{equation}
POSD queries against the first $k$ message bits which follow the distribution :
\begin{equation}
    f_{\phi(|\mathbf{L}|)[i]}(y)= \frac{k!}{(i-1)!(k-i)!} \left[ F_{|\mathbf{L}|} (y) \right]^{i-1}  
    \times \left[ 1- F_{|\mathbf{L}|} (y) \right]^{k-i}  f_{|\mathbf{L}|}(y)
\label{eq:binomial_dist_posd}
\end{equation}
where $i\in [1,k]$.
For OSD, the distribution we observe is the one shown in (\ref{eq:binomial_dist}); however, since \ac{OSD} operates on the set of most reliable bits, the least reliable bit in the sequence is the $(n-k) ^{th}$ bit from the total distribution:

\begin{equation}
    f_{\phi(|\mathbf{L}|)[i]}(y)= \frac{n!}{(n-k+i-1)!(k-i)!} \left[ F_{|\mathbf{L|}} (y) \right]^{n-k+i-1}  
    \times \left[ 1- F_{|\mathbf{L}|} (y) \right]^{k-i}  f_{|\mathbf{L}|}(y)
\end{equation}
where $i\in [1,k]$.

Hence, the distributions of the LLRs that GRAND and POSD operate on are the same distribution of the entire channel (with different sampling sizes). Therefore, it follows that the error patterns used with GRAND would be similar to the ones used with POSD. However, OSD only uses the most reliable basis that exhibits a different \ac{PDF}. We will show in later sections that POSD and GRAND usually require similar error patterns, with those error patterns usually differing from the error patterns used with OSD.

\textbf{Note}: The error pattern $\mathbf{e}$ of GRAND is entirely generated without any input of the generator matrix. This is different from that of POSD or \ac{OSD} which is partially pre-generated (for $k$ bits) and the rest is computed up by using the generator matrix. Hence, the expected value of the weighted Hamming distance of a GRAND error pattern can be known in advance without any relation to the structure of the code; while that of a POSD / \ac{OSD} error pattern depends on the structure of the code. For OSD, it is difficult to approximate the effect of the structure of the code on the weighted Hamming distance of the generated error patterns as $\mathbf{G''}$ changes with each received signal, while for POSD, it is possible to approximate the effect of the structure of the code as one generator matrix is used irregardless of the received signal.

\subsection{Complexity Analysis}
{All the steps required to generate these error patterns can be performed prior to deployment in the field. Our proposed method requires complex calculations which are often difficult to do by hand. To simplify the necessary calculations needed by our methods, it is generally sufficient to use numerical integration techniques to approximate the required integrals. As such, the set of most frequent TEPs could be generated rapidly, which is a significant advantage over Monte Carlo simulations which demand long runtimes to gather a representative set of frequent TEPs. \\
Fig. \ref{fig:timing_TEP_generation} shows the time needed to generate \acp{TEP} with Monte Carlo simulations and our proposed method on an AWGN channel. We run the simulations using MATLAB on an AMD Ryzen 5 7600X 6-Core Processor (4.70 GHz) with 32 GB of RAM. To ensure reliable frequency-based ranking, Monte Carlo simulations require a minimum of 10 occurrences for each \ac{TEP}. Fig. \ref{fig:timing_TEP_generation} shows that our proposed method is consistently able to generate $10000$ \acp{TEP} in less than $4$ seconds across all simulated $\frac{E_b}{N_0}$. On the other hand, the time needed to generate $10000$ \acp{TEP} with Monte Carlo simulations exceeds 2 days at $\frac{E_b}{N_0}=8$ dB, $45471 \times$ that of the EW \acp{TEP} generation time.\\
In terms of implementation, the main overhead lies in memory, as our pre-generated \acp{TEP} must be stored. As a result, a hardware architecture similar to the fixed-latency LUT-based GRAND \cite{condo_fixed_2022} is well suited. For example, storing $MQ$ \acp{TEP} would require $n\times Q$ bits with GRAND \cite{condo_fixed_2022} (or $(n-k) \times MQ$ if only the syndromes are stored \cite{abbas_high-throughput_2022}) and $k\times MQ$ bits with OSD/POSD, where $MQ$ refers to a maximum number of queries. For $MQ=1024$ TEPs and $n=128, k=64$, the number of stored bits is $131072$ bits (16.384 kB) with GRAND using a hardware similar to \cite{condo_fixed_2022} and $65536$ bits (8.192 kB) with GRAND using a hardware similar to \cite{abbas_high-throughput_2022}. For OSD and POSD we need $65536$ bits (8.192 kB) to store the TEPs. Finally, the proposed method can be combined with the piecewise linear approach to refine estimation, particularly in regions that are difficult to linearize.
}
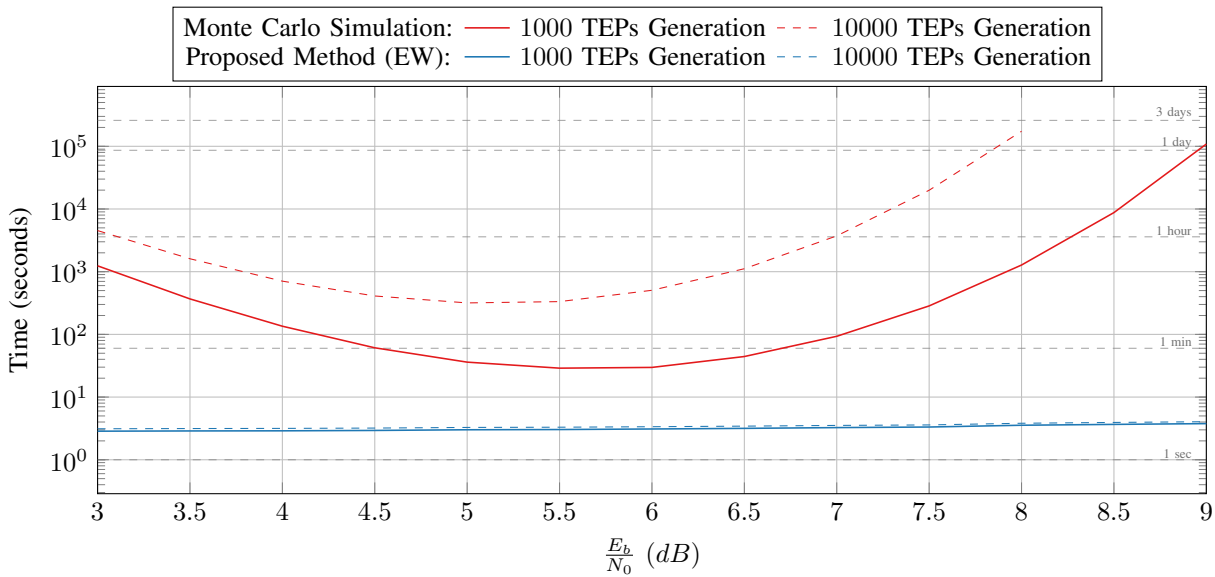
\begin{figure}
    \centering
\begin{tikzpicture}
    \begin{axis}[height=7cm,  width=0.9\columnwidth,    xlabel={$\frac{E_b}{N_0} $ $(dB)$},
        ylabel={Time (seconds) },
        grid=major,ymode=log,
        xmin=3, xmax=9,xtick={3,3.5,...,9},legend style={at={(0.03,0.97)}, anchor=north west, legend cell align=left,     align=left, draw=black,font=\footnotesize}
    ]
\tikzset{refstylee/.style={gray, dashed, opacity=0.7}}
\tikzset{labelstylee/.style={pos=1, anchor=south east, font=\tiny, text=black!80, xshift=-2pt,yshift=-1mm}}
\addplot[refstylee, forget plot] coordinates {(3, 1) (9, 1)} node[labelstylee] {1 sec};
\addplot[refstylee, forget plot] coordinates {(3, 60) (9, 60)} node[labelstylee] {1 min};
\addplot[refstylee, forget plot] coordinates {(3, 3600) (9, 3600)} node[labelstylee] {1 hour};
\addplot[refstylee, forget plot] coordinates {(3, 86400) (9, 86400)} node[labelstylee] {1 day};
\addplot[refstylee, forget plot] coordinates {(3, 259200) (9, 259200)} node[labelstylee] {3 days};
    \addplot[mark=none, Paired-1,semithick] table[
        x index=0,  
        y index=1,  
        header=true
    ] {Figures/AWGN/timing/mine.tex};\label{plot:timing-mine-1k}
   \addplot[mark=none, Paired-1,dashed] table[
        x index=0,  
        y index=1,  
        header=true
    ] {Figures/AWGN/timing/mine10k.tex};\label{plot:timing-mine-10k}
      \addplot[mark=none, Paired-5,semithick] table[
        x index=0,  
        y index=1,  
        header=true
    ] {Figures/AWGN/timing/monte.tex};\label{plot:timing-monte-1k}
   \addplot[mark=none, Paired-5,dashed] table[
        x index=0,  
        y index=1,  
        header=true
    ] {Figures/AWGN/timing/monte10k.tex};\label{plot:timing-monte-10k}
    
    \end{axis}
     \matrix[
            matrix of nodes,
            anchor=north,
            draw,
            inner sep=0.2em
        ] at ([yshift=30pt,xshift=10pt]current bounding box.north)
        {Monte Carlo Simulation:& \ref{plot:timing-monte-1k}   $1000$ TEPs Generation  &
            \ref{plot:timing-monte-10k} $10000$ TEPs Generation  \\ Proposed Method (EW):&  \ref{plot:timing-mine-1k}  $1000$ TEPs Generation  &
            \ref{plot:timing-mine-10k} $10000$ TEPs Generation  \\
        };
    \end{tikzpicture}
        \caption{Time needed to generate TEPs with Monte-Carlo Simulations and our proposed method as a function of $\frac{E_b}{N_0}$.}
\label{fig:timing_TEP_generation}
\end{figure}
\section{Additive White Gaussian Noise }

For the additive white Gaussian channel, the received channel signal can be represented as:
\begin{equation}
    \mathbf{y}= \mathbf{x}+ \mathbf{n} ,
\end{equation} where $\mathbf{x}$ is the BPSK modulated signal and $\mathbf{n}$ is a Gaussian distribution with mean 0 and variance $\sigma^2$.
After receiving the signal, the LLR of $\mathbf{y}$ can be calculated as :

\begin{equation}
L=\mathcal{L}(y) = \ln \frac{P(y|c=0)}{P(y|c=1)} = \ln \frac{\frac{1}{\sqrt{2\pi\sigma^2}} e^{-\frac{(y-1)^2}{2\sigma^2}}}{ \frac{1}{\sqrt{2\pi\sigma^2}} e^{-\frac{(y+1)^2}{2\sigma^2}}} =\frac{(y+1)^2-(y-1)^2}{2\sigma^2}=\frac{2y}{\sigma^2}\label{eq:LLR_formula_gen}
\end{equation}

\subsection{Generating the list of TEPs}

\subsubsection{Finding the \ac{PDF} of the reliabilities}
The \ac{PDF} of the LLR $L$ of each of the received symbols can be represented as \cite{hou_performance_2001}:
\begin{align}
    f_{\mathbf{L}}(l|c=0) &= \frac{1}{ \sqrt{8\pi /\sigma^2} } e^{- \frac{\left(l -\frac{2}{\sigma^2}\right)^2}{\frac{8}{\sigma^2}}}  \hspace{1cm }, \\
        f_{\mathbf{L}}(l|c=1) &= \frac{1}{ \sqrt{8\pi /\sigma^2} } e^{- \frac{\left(l +\frac{2}{\sigma^2}\right)^2}{\frac{8}{\sigma^2}}}  \hspace{1cm }. 
\end{align}
Fig. \ref{fig:pdf_manipulations_awgn} (a) shows a sample distribution of $L$. The highlighted area in blue shows the area  where $c=1$ is interpreted as $0$ and the highlighted area in red shows the area where a $x=0$ is interpreted as $1$.

Owing to the law of total probability, we can represent the \ac{PDF} of the LLR as:
\begin{align}
   f_{\mathbf{L}}(l)= &P(x=+1)f_{\mathbf{L}}(l|x=+1)+ 
        P(x=-1)f_{\mathbf{L}}(l|x=-1) ,\\
        = & \frac{1}{2} \times \frac{1}{ \sqrt{8\pi /\sigma^2} } e^{- \frac{\left(l -\frac{2}{\sigma^2}\right)^2}{\frac{8}{\sigma^2}}} 
        + \frac{1}{2} \times \frac{1}{ \sqrt{8\pi /\sigma^2} } e^{- \frac{\left(l +\frac{2}{\sigma^2}\right)^2}{\frac{8}{\sigma^2}}} . 
\end{align}

\pgfmathdeclarefunction{sortedgauss}{2}{%
\pgfmathparse{#2/(sqrt(8*pi))*exp(-(((x-#1*(2/#2^2))^2)/(8/(#2^2)) + #2/(sqrt(8*pi))*exp(-(((x+#1*(2/#2^2))^2)/(8/(#2^2))}}
\pgfmathdeclarefunction{gauss}{2}{%
  \pgfmathparse{#2/(sqrt(8*pi))*exp(-(((x-#1*(2/#2^2))^2)/(8/(#2^2))}}
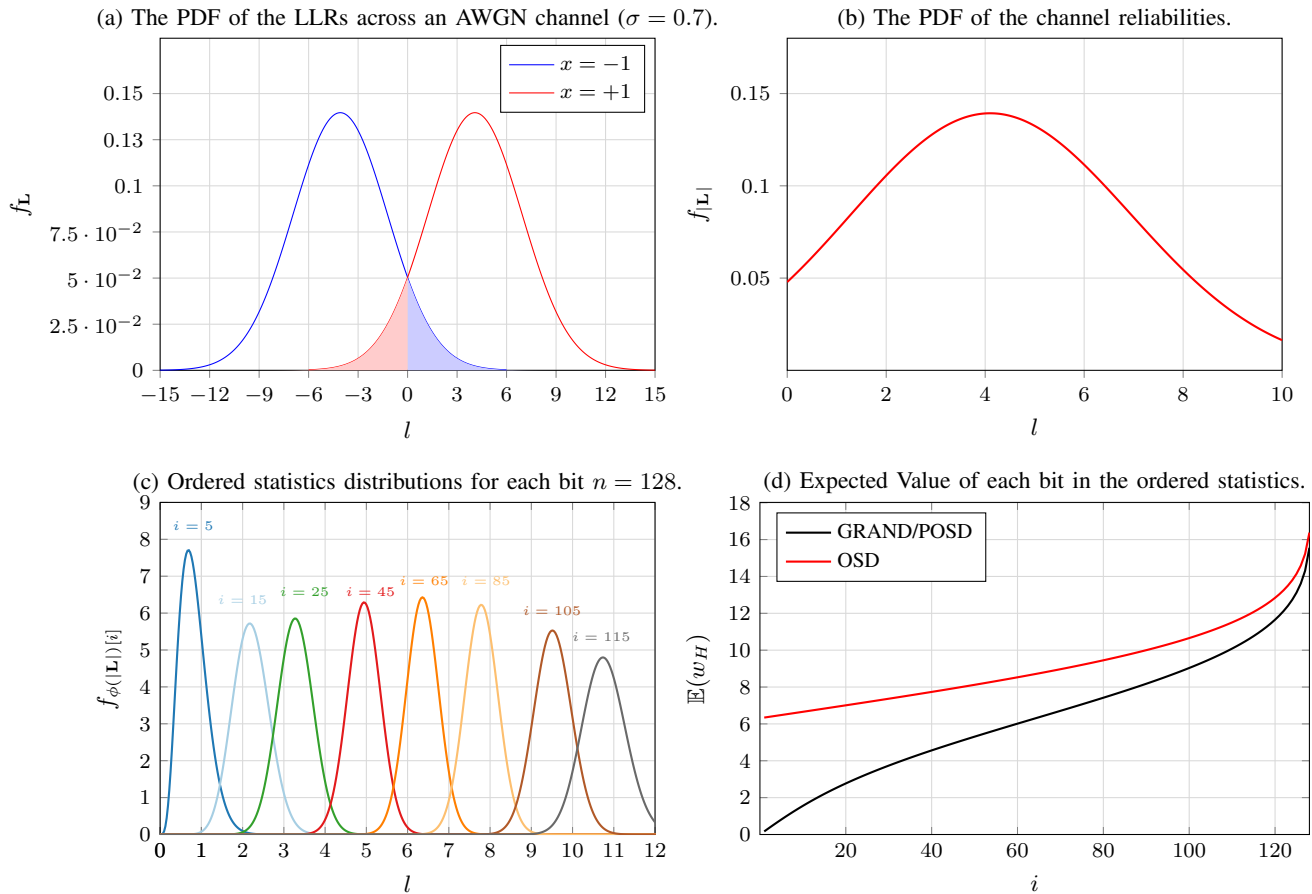
\begin{figure}
\centering
 \begin{tikzpicture}
    \begin{groupplot}[group style={group name=pdfmixture, group size= 2 by 2, horizontal sep=50pt, vertical sep=50pt},
          footnotesize,
          height=7cm,  width=\columnwidth,
          xlabel=$y$,
          xmin=-5, xmax=5, xtick={-5,...,1,...,20},
          tick align=inside,
          grid=both, grid style={gray!30},
          /pgfplots/table/ignore chars={|},
          ] 
          \nextgroupplot[title= (a) The \ac{PDF} of the LLRs across an AWGN channel {($\sigma=0.7$).},footnotesize, width= 0.45\linewidth, height=6cm,    
            xtick={-15,-12,-9,...,15},ylabel=$f_{\mathbf{L}}$,xlabel=$l$,
            tick align=outside, ytick={0,0.025,...,0.15}, xmax=15,xmin=-15,ymin=0,ymax=0.18,tickpos=left,
            every axis plot post/.append style={
  mark=none,samples=100,smooth}, 
  grid=both, grid style={gray!30},
            tick align=outside, tickpos=left] 
\addplot [blue, domain=-15:6]{gauss(-1,0.7)}; \addlegendentry{$x=-1$}  \label{plot:awgn_neg_pdf}
 
\addplot [red, domain=-6:15]{gauss(1,0.7)};\addlegendentry{$x=+1$} \label{plot:awgn_pos_pdf}
 \addplot [fill=blue!20, draw=none, domain=0:15] {gauss(-1,0.7)}\closedcycle;
  \addplot [fill=red!20, draw=none, domain=-15:0] {gauss(1,0.7)}\closedcycle;
\nextgroupplot[title= (b) {The \ac{PDF} of the channel reliabilities.},
height=6cm,width=0.45\columnwidth,ytick={0.05,0.1,0.15},scaled ticks=false,y tick label style={/pgf/number format/fixed},   xtick={-10,-8,...,10},ylabel=$f_{|\mathbf{L}|}$,xlabel=$l$,xmax=10,xmin=0,ymin=0,ymax=0.18, tick align=outside, tickpos=left,  every axis plot post/.append style={ mark=none,samples=100,smooth}, 
  grid=both, grid style={gray!30} ] 
\addplot [red,thick, domain=0:10]{sortedgauss(1,0.7)}; \label{plot:abs_pdf_awgn}


\nextgroupplot[title=(c) {Ordered statistics distributions for each bit  $n=128$.},
      footnotesize,
      height=6cm,  width=0.45\columnwidth,  
      xlabel=$l$,
      xmin=0, xmax=12, xtick={-5,...,1,...,12},
      tick align=inside,
      grid=both, grid style={gray!30},
      /pgfplots/table/ignore chars={|},ylabel= $f_{\phi(|\mathbf{L}|)[i]}$,ytick pos=left,ymin=0, ymax = 9
      ] 
\addplot [thick, Paired-1] file {Figures/AWGN/dist5_sampled_20.txt};\label{dist_5} 
\node [above] at (axis cs:  0.8,  8) {\color{Paired-1}\tiny$i=5$};
\addplot[thick,Paired-2] file {Figures/AWGN/dist15_sampled_20.txt};
\label{dist_15};\node [above] at (axis cs:  2,  6) {\color{Paired-2}\tiny$i=15$};
\addplot[thick,Paired-3] file {Figures/AWGN/dist25_sampled_20.txt};
\label{dist_25};\node [above] at (axis cs:  3.5,  6.2) {\color{Paired-3}\tiny$i=25$};
\addplot[thick,Paired-5] file {Figures/AWGN/dist45_sampled_20.txt};
\label{dist_45};\node [above] at (axis cs:  5.1,  6.2) {\color{Paired-5}\tiny$i=45$};
\addplot[thick,Paired-7] file {Figures/AWGN/dist65_sampled_20.txt};
\label{dist_65};\node [above] at (axis cs:  6.4,  6.5) {\color{Paired-7}\tiny$i=65$};
\addplot[thick,Paired-8] file {Figures/AWGN/dist85_sampled_20.txt};
\label{dist_85};\node [above] at (axis cs:  7.9,  6.5) {\color{Paired-8}\tiny$i=85$};
\addplot[thick,Paired-11] file {Figures/AWGN/dist105_sampled_20.txt};
\label{dist_105};\node [above] at (axis cs:  9.5,  5.7) {\color{Paired-11}\tiny$i=105$};
\addplot[thick,Paired-12] file {Figures/AWGN/dist115_sampled_20.txt};
\label{dist_115};\node [above] at (axis cs:  10.7,  5) {\color{Paired-12}\tiny$i=115$};
\label{dist_125};
   \coordinate (top) at (rel axis cs:0,1);
      \coordinate (spypoint1) at (axis cs:7.45,2e-7);
      \coordinate (magnifyglass1) at (axis cs:2.6,1.1e-5);
      \coordinate (bot) at (rel axis cs:1,0);
 \path (top|-current bounding box.north) -- coordinate(legendpos) (bot|-current bounding box.north);
\nextgroupplot[title=(d) Expected Value of each bit in the ordered statistics., height=6cm,width=0.49\columnwidth,
    xlabel={$i$},
    ylabel={$\mathbb{E} (w_H)$ },
    grid=major,
    xmin=0, xmax=128,xtick={20,40,...,120},ymin=0,legend style={at={(0.03,0.97)}, anchor=north west, legend cell align=left,     align=left, draw=black,font=\footnotesize}
]
\addplot[mark=none,black,thick] table[
    x index=0,  
    y index=1,  
    header=false
] {Figures/AWGN/expected_values_128_latex.txt}; \label{plot:expected_values_grand_awgn}\addlegendentry{GRAND/POSD}
\addplot[mark=none,red,thick] table[
    x index=0,  
    y index=1,  
    header=false
] {Figures/AWGN/expected_values_osd_128.txt};\label{plot:expected_values_osd_awgn} \addlegendentry{OSD}

    \end{groupplot}
    \end{tikzpicture}
    \caption{The steps followed to generate the \ac{EW} error patterns in an AWGN channel.}\label{fig:pdf_manipulations_awgn}
\end{figure}

\input{Figures/AWGN/pdf_abs_awgn}
After generating the absolute values of the LLRs, the distribution changes based on (\ref{eq:abs_LLRs_pdf}) and we obtain :
\begin{align}
    f_{|\mathbf{L}|}(l) &= \frac{1}{ \sqrt{8\pi /\sigma^2} } \left(e^{- \frac{\left(l -\frac{2}{\sigma^2}\right)^2}{\frac{8}{\sigma^2}}}  + e^{- \frac{\left(l +\frac{2}{\sigma^2}\right)^2}{\frac{8}{\sigma^2}}} \right) \hspace{0.2cm } &, l \in [0,+\infty),\label{eq:abs_dist_awgn} \\
    F_{|\mathbf{L}|}(l) &= \int_0^lf_{|\mathbf{L}|}(x) dx = \frac{1}{2}\left( erfc\left(\frac{\sqrt{2/\sigma^2}+l}{2\sqrt{2/\sigma^2}}\right)- erfc\left(\frac{\sqrt{2/\sigma^2}-l}{2\sqrt{2/\sigma^2}}\right)\right) &, l \in [0,+\infty).\label{eq:abs_cdf_awgn}
\end{align}

The resulting distribution of the absolute value of the LLR is plotted in Fig. \ref{fig:pdf_manipulations_awgn} (b).

\subsubsection{Sorting the reliabilities}

\input{Figures/AWGN/distribution_plot}
Plugging  $F_{|\mathbf{L}|}$ (\ref{eq:abs_cdf_awgn}) and $f_{|\mathbf{L}|}$ (\ref{eq:abs_dist_awgn})  into (\ref{eq:binomial_dist}), we obtain the distributions of the individual sorted LLRs. We use MATLAB to calculate and plot the individual distributions of a subset of the bits in Fig. \ref{fig:pdf_manipulations_awgn} (c). 
\input{Figures/AWGN/combined_performance_awgn}
\subsubsection{Expected Value Calculation}
From these distributions we can use (\ref{eq:expected_value_calculation}) to calculate the expected values of the individual bits which is plotted in Fig. \ref{fig:pdf_manipulations_awgn} (d).

\subsubsection{Generating Error Patterns}
The algorithm in section \ref{sec:optimizing_binary} is applied on the expected values of the LLRs calculated in the previous step to generate the list of EW \ac{TEP}s.
\input{Figures/AWGN/expected_value_plot}

\subsection{Decoding Performance Comparison}

 Fig. \ref{fig:performance_grand_awgn} shows the performance of \ac{GRAND}, \ac{POSD} and \ac{OSD} with the proposed \ac{EW}, and the existing \ac{LW}, \ac{HW}, \ac{ILW} and \ac{ML} \ac{TEP}s. {High-rate codes are decoded using GRAND, whereas medium-rate to low-rate codes are decoded using POSD and OSD. This is because GRAND needs to guess the entire error pattern affecting the codeword, and it struggles to match the decoding performance of POSD and OSD when there are many bits in error.} We discuss the performance of each of these decoders with different \ac{TEP}s and varying codes.
\subsubsection{\ac{GRAND}}
We simulate GRAND on \ac{RLC} [127,85] and 5G {\ac{CA}-Polar code [128,105+11] (with 11 Cyclic Redundancy Check (CRC) bits)}. 
For a maximum number of queries of $10^6$, we can observe that the \ac{EW} error patterns result in a $0.2 \rightarrow 0.3$ dB gain compared to LW error patterns and $0.1$ dB gain compared to ILW error patterns at target FER $10^{-5}$. Using the \ac{ML} \ac{TEP}s result in $0.3$ dB performance gain compared to the \ac{EW} error patterns with RLC [127,85] code and a $0.1$ dB performance gain compared to the \ac{EW} error patterns with CA-Polar [128,105+11] code. \ac{HW} \ac{TEP}s result in a performance loss of $\approx2\rightarrow2.5 $ dB compared to the EW error patterns target FER $10^{-5}$.

{For a maximum number of queries of $10^3$, the ML, EW, ILW, LW, and HW TEPs exhibit the same trend of decoding performance as what we observed with $MQ=10^6$. However, for RLC [127,85] and a target FER $10^{-5}$, we observe a slight $0.1$ dB gain of the LUT error patterns compared to the EW error patterns. This performance gain is not observed with CA-Polar [128,105+11] code as both EW and LUT error patterns perform similarly.}
 \subsubsection{\ac{POSD}}
We simulate POSD on BCH [127,64] and RLC code [128,32]. For a maximum number of queries of $10^5$, we can observe that the \ac{EW} error patterns and the \ac{LW} error patterns result in the same performance. Additionally, we can observe a $0.2 \rightarrow 0.5$ dB gain compared to \ac{ILW} error patterns at target FER $10^{-4}$. Using the \ac{ML} \ac{TEP}s results in $0.1$ dB performance gain compared to the \ac{EW} error patterns with BCH [127,64] and RLC code [128,32] at target FER $10^{-4}$. \ac{HW}
\ac{TEP}s also achieve poor performance as they result in a performance loss of $\approx2.3$ dB compared to the EW error patterns at target FER $10^{-4}$.

{For a maximum number of queries of $10^3$ we observe that the performance of the error patterns changes between BCH [127,64] and RLC [128,32]. At target FER $10^{-4}$ with BCH [127,64], LUT TEPs outperform EW and ILW TEPs by $0.1$ dB. Additionally, ILW and EW error pattern outperform LW TEPs by $0.1$ dB. However, for RLC [128,32] and a target FER $10^{-4}$, we observe a slight $0.1$ dB gain of the EW error patterns compared to the to the LUT error patterns. For this code, the LW and LUT error patterns produce almost the same decoding performance, but obtain a $0.3$ dB gain compared to the LW TEPs at target FER $10^{-4}$.}

 \subsubsection{\ac{OSD}}
We simulate OSD on BCH [127,64] and RLC code [128,32]. For a maximum number of queries of $10^4$, we can observe that the \ac{EW}, \ac{HW}, \ac{ML} \ac{TEP}s result in the same decoding performance. Additionally, we can observe a $0.1 \rightarrow 0.3$ dB gain compared to \ac{ILW} \ac{TEP}s at target FER $10^{-5}$. \ac{LW}
\ac{TEP}s achieve poor performance as they result in a performance loss of $\approx1.5 \rightarrow 2.5$ dB compared to the EW error patterns at target FER $10^{-4}$. 
{For a maximum number of queries of $10^3$ we observe that the LUT, HW, EW and ML TEP performance is the same. At target FER $10^{-5}$ we observe that the ILW TEPs attain a $0.4$ dB to $0.5$ dB loss compared to the LUT, HW, EW and ML. This degradation in performance is more pronounced than that observed with $MQ=10^4$.}

\subsection{\ac{TEP} analysis}
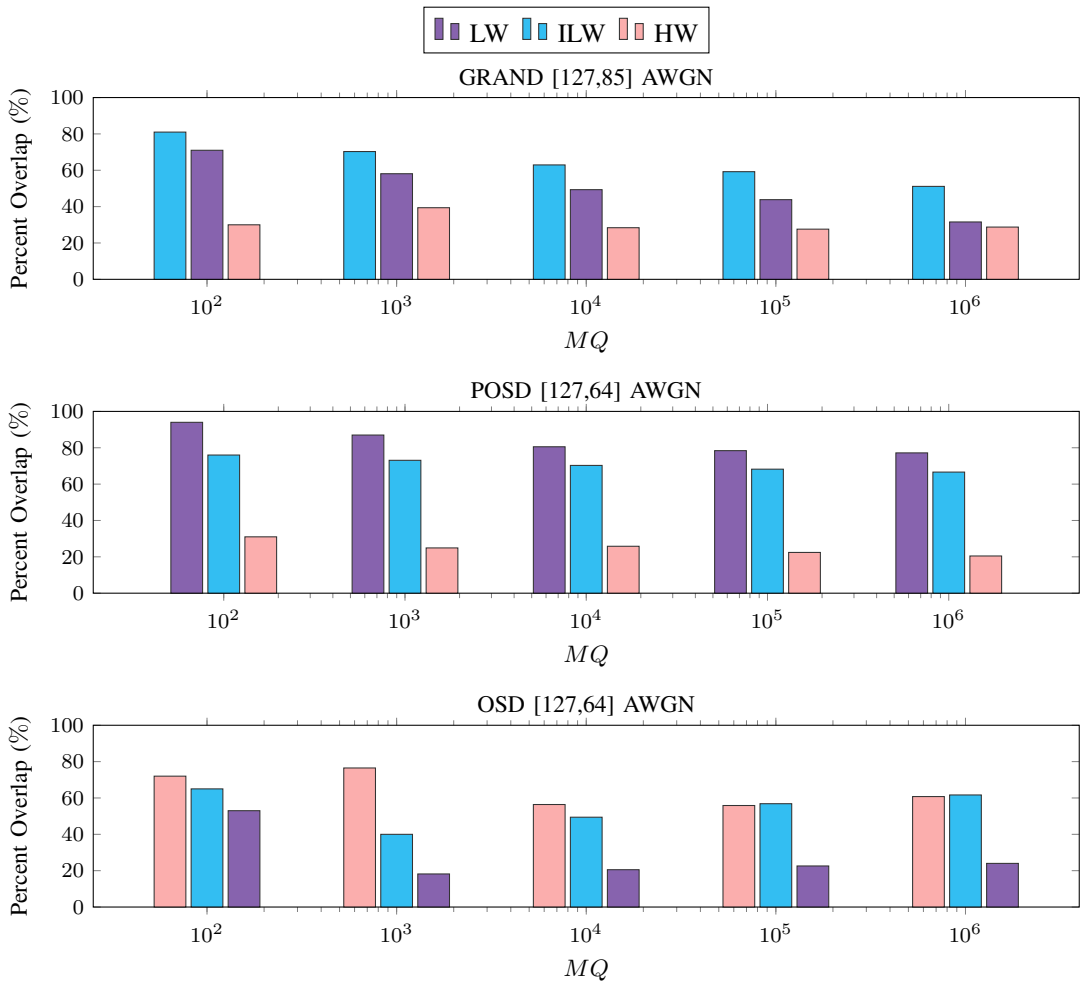
\begin{figure}[t]
\centering
\begin{tikzpicture}
 \begin{groupplot}[group style={group name=pdfmixture, group size= 1 by 3, horizontal sep=25pt, vertical sep=50pt},footnotesize,
      height=4cm,  width=0.81\columnwidth,
      xlabel={$MQ$}, /pgfplots/table/ignore chars={|}]
\nextgroupplot[title= GRAND {[127,85] }AWGN,ylabel= Percent Overlap ($\%$), ytick pos=left, ,ymin=0, ymax = 100, 
    ybar,
    xmode=log,
    bar width=12pt, 
    xtick=data, 
    enlarge x limits=0.15, 
    legend entries={},
    legend style={draw=none, fill=none}]
    \addplot[fill=cyan, opacity=0.8] table[
    x index=0,  
    y index=4,  
    header=false] {Figures/AWGN/grand_awgn_7.txt};\label{plot:ilwo_grand_awgn_perc}
\addplot[fill=Paired-9, opacity=0.8] table[
    x index=0,  
    y index=2,  
    header=false] {Figures/AWGN/grand_awgn_7.txt};\label{plot:lw_grand_awgn_perc}

\addplot[fill=Paired-6, opacity=0.8] table[
    x index=0,  
    y index=3,  
    header=false] {Figures/AWGN/grand_awgn_7.txt};\label{plot:hw_grand_awgn_perc}
\nextgroupplot[title= POSD {[127,64]} AWGN,ylabel= Percent Overlap ($\%$), ymin=0, ymax = 100,
    ymin=0, 
    ybar,
    xmode=log,
    bar width=12pt, 
    xtick=data, 
    enlarge x limits=0.18, 
    legend entries={},
    legend style={draw=none, fill=none}]

        \addplot[fill=Paired-9, opacity=0.8] table[
    x index=0,  
    y index=2,  
    header=false] {Figures/AWGN/posd_awgn.txt};\label{plot:lw_posd_awgn_perc}
            \addplot[fill=cyan, opacity=0.8] table[
    x index=0,  
    y index=4,  
    header=false] {Figures/AWGN/posd_awgn.txt};\label{plot:ilwo_posd_awgn_perc}

\addplot[fill=Paired-6, opacity=0.8] table[
    x index=0,  
    y index=3,  
    header=false] {Figures/AWGN/posd_awgn.txt};\label{plot:hw_posd_awgn_perc}
        \coordinate (top) at (rel axis cs:0,1);
      \coordinate (spypoint1) at (axis cs:7.45,2e-7);
      \coordinate (magnifyglass1) at (axis cs:2.6,1.1e-5);
      \coordinate (bot) at (rel axis cs:1,0);
    \nextgroupplot[title=OSD {[127,64]} AWGN, ymin=0, ymax = 100,
    ymin=0, ylabel= Percent Overlap ($\%$),
    ybar,
    xmode=log,
    bar width=12pt, 
    xtick=data, 
    enlarge x limits=0.15, 
    legend entries={},
    legend style={draw=none, fill=none}]
    \addplot[fill=Paired-6, opacity=0.8] table[
    x index=0,  
    y index=3,  
    header=false] {Figures/AWGN/osd_awgn.txt};\label{plot:hw_osd_awgn_perc}
    \addplot[fill=cyan, opacity=0.8] table[
    x index=0,  
    y index=4,  
    header=false] {Figures/AWGN/osd_awgn.txt};\label{plot:ilwo_osd_awgn_perc}
    \addplot[fill=Paired-9, opacity=0.8] table[
    x index=0,  
    y index=2,  
    header=false] {Figures/AWGN/osd_awgn.txt};\label{plot:lw_osd_awgn_perc}
\end{groupplot}
    \matrix[
        matrix of nodes,
        anchor=north,
        draw,
        inner sep=0.2em
    ] at ([yshift=20pt,xshift=10pt]current bounding box.north)
    {
        \ref{plot:lw_grand_awgn_perc}&  LW &[1pt] \ref{plot:ilwo_grand_awgn_perc}&  ILW &[1pt]  \ref{plot:hw_grand_awgn_perc}&  HW  \\
    };
\end{tikzpicture}
\caption{{Percentage of overlap of the pre-generated test error patterns (TEPs) with EW TEPs for GRAND, POSD, and OSD over an AWGN channel.}}\label{fig:percent_intersection_teps_combined_awgn}
\end{figure}
Fig. \ref{fig:percent_intersection_teps_combined_awgn} shows the percentage of overlap between the EW TEPs and the LW, ILW and HW TEPs with GRAND, POSD and \ac{OSD}. We discuss the correlation between the percentage overlap of TEPs and the decoding performance below.

\subsubsection{GRAND}
ILW error patterns have the largest percentage of overlap with the EW TEPs of $81 \% $ at $MQ=10^2$  and  $51 \%$ at $MQ=10^6$. This explains why the performance of the ILW error patterns supersedes that of the LW error patterns with $71\%$ common TEPs with EW TEPs at $MQ=10^2$ and $31 \%$ at $MQ=10^6$ . The HW TEPs consistently have low number of overlapping \ac{TEP}s with EW with $30\%$ common \ac{TEP}s at $MQ=10^2$ and $29\%$ at $MQ=10^6$. This results in $\approx1.5 \rightarrow 2.5$ dB performance loss compared to EW TEPs.

\subsubsection{POSD}

 The LW error patterns have the largest overlap with the EW TEPs of $94 \% \to 78 \%$ at $MQ=10^2 \to 10^5$ compared to ILW which has $76 \% \to 68\% $ overlapping \ac{TEP}s with EW error patterns. This lower number of common \ac{TEP}s explains why ILW error patterns are less performing for POSD than LW error patterns. HW TEPs share $31\% \to 20\%$ overlapping TEPs with EW TEPs, which results in $\approx1.5 \rightarrow 2.5$ dB performance loss compared to EW TEPs.

\subsubsection{OSD}
HW TEPs share the highest percentage of common TEPs with EW, with an overlap of $72\%$ at $MQ=10^2$ and $56\%$ at $MQ=10^4$. This is followed by ILW TEPs with $65\%$ at $MQ=10^2$ and $49\%$ at $MQ=10^4$. In contrast,  LW TEPs show a larger drop in common TEPs with EW, falling from  $53\%$ at $MQ=10^2$ to approximately $21\%$ at $MQ=10^4$.
\subsection{Conclusion}
We can consistently see that the performance of the predetermined \ac{TEP}s correlates with the percentage of common TEPs with the EW \ac{TEP}s. { In general, EW TEPs either obtain a similar performance or outperform other TEPs except for the ML TEPs which are hardware incompatible. EW TEPs generally obtain the same performance as the LUT error patterns, but suffer a small performance degradation of $0.1$ dB with GRAND on RLC [127,85] and POSD [127,64] due to slight variations in the error patterns produced by the LUT method and the EW method with $MQ=10^3$ TEPs. This degradation in performance is generally acceptable in light of our ability to generate more EW error patterns on demand, without requiring long simulation times.}


\section{Mixture of White Gaussian Channel}

\begin{figure}[t]
  \centering
  \begin{minipage}[t]{0.45\textwidth}
  \vspace{-6.0cm}
    \centering
    { \refstepcounter{table}  
    \text{Table \thetable:\label{tab:gmm_params} } Parameters of the Gaussian distributions in the mixed Gaussian channel.\newline.}
    
    \vspace{1em}
    \begin{tabular}{c|ccc|ccc}
      \toprule
      \multirow{2}{*}{$i$} & \multicolumn{3}{c|}{\textbf{Channel 1}} & \multicolumn{3}{c}{\textbf{Channel 2}} \\
      \cmidrule(lr){2-4} \cmidrule(lr){5-7}
        & $\mathbf{\mu}[i]$ & $\mathbf{\omega}[i]$ & $\mathbf{\sigma}[i]$ & $\mathbf{\mu}[i]$ & $\mathbf{\omega}[i]$ & $\mathbf{\sigma}[i]$ \\
      \midrule
      1 & -3.0 & 0.29 & 0.3555 & -2.7 & 0.29 & 0.3555 \\
      2 & -0.1 & 0.01 & 0.13   & -0.1 & 0.01 & 0.13   \\
      3 &  0.0 & 0.40 & 0.10   &  0.0 & 0.40 & 0.10   \\
      4 &  0.1 & 0.01 & 0.13   &  0.1 & 0.01 & 0.13   \\
      5 &  3.0 & 0.29 & 0.3555 &  2.7 & 0.29 & 0.3555 \\
      \bottomrule
    \end{tabular}
  \end{minipage}
  \hfill
  \begin{minipage}[t]{0.5\textwidth}
  \refstepcounter{figure}
    \centering
    \begin{tikzpicture}
    \begin{groupplot}[group style={group name=pdfmixture, group size= 1 by 1, horizontal sep=5pt, vertical sep=50pt},
          footnotesize,
          height=7cm,  width=\columnwidth,  
          xlabel=$y$,
          xmin=-5, xmax=5, xtick={-5,...,1,...,20},
          tick align=inside,
          grid=both, grid style={gray!30},
          /pgfplots/table/ignore chars={|},
          ] 
          \nextgroupplot[ylabel= pdf, ytick pos=left, 
          ymin=0, ymax = 2]
         \addplot[ Paired-9 , thick]  table[x=x, y=pdf] {Figures/Mixture_of_Gaussians/HOSD/pdf_hosd_p.tex}; \label{mixed:positive_pdf_hosd}
              \addplot[ Paired-3 , thick]  table[x=x, y=pdf] {Figures/Mixture_of_Gaussians/HOSD/pdf_hosd_n.tex}; \label{mixed:negative_pdf_hosd}
       \coordinate (top) at (rel axis cs:0,1);
          \coordinate (spypoint1) at (axis cs:7.45,2e-7);
          \coordinate (magnifyglass1) at (axis cs:2.6,1.1e-5);
          \coordinate (bot) at (rel axis cs:1,0);
    \end{groupplot}
      
     \path (top|-current bounding box.north) -- coordinate(legendpos) (bot|-current bounding box.north);
     \matrix[
      matrix of nodes,
      anchor=south,
      draw,
      inner sep=0.2em
    ] at(legendpos) 
        {
        \ref{mixed:positive_pdf_hosd}&  $c=0$ &[1pt] 
        \ref{mixed:negative_pdf_hosd}&   $c=1$ &[1pt] \\ 
          };
    \end{tikzpicture}
        \text{Fig. \thefigure:} The distribution of the \ac{PDF} of the transmitted codeword bits along the mixed Gaussian channel 2. \label{fig:pdf_mxgs}
  \end{minipage}
  \end{figure}
  \begin{figure}
 \begin{minipage}[t]{0.49\textwidth}
  \refstepcounter{figure}
    \begin{tikzpicture}
    \begin{axis}[height=7cm,  width=\columnwidth,    xlabel={$y$},
        ylabel={$L$ },
        xmin=-5, xmax=5,xtick={-5,-4,...,5},ymin=-40,
        ymax=40,grid=both,grid style={line width=.1pt, draw=gray!10}
    ]
    \addplot[red,thick] table[
        x index=0,  
        y index=1,  
        header=false
    ] {Figures/Mixture_of_Gaussians/GRAND/grand_llr.txt}; \label{line:mapping_llr_grand}
    \addplot[blue,thick] table[
        x index=0,  
        y index=1,  
        header=false
    ] {Figures/Mixture_of_Gaussians/HOSD/HOSD_llr.txt};\label{line:mapping_llr_osd}
     \draw[] (-5, 0) -- (5, 0)  ;
      \draw[] (0, -40) -- (0, 40) ;
    \end{axis}
       \matrix[
            matrix of nodes,
            anchor=north,
            draw,
            inner sep=0.2em
        ] at ([yshift=10pt,xshift=15pt]current bounding box.north)
        { \ref{line:mapping_llr_grand} &  Channel 1 &[1pt] \ref{line:mapping_llr_osd}&  Channel 2 \\
        };
    \end{tikzpicture}
    \text{Fig. \thefigure:} The LLR value as a function of the  received signal. \label{tab:llr_mapping_mxgs}
    
    \end{minipage}
    \hfill
    \begin{minipage}[t]{0.49\textwidth}
  \refstepcounter{figure}
    \begin{tikzpicture}
    \begin{axis}[height=7cm,  width=0.9\columnwidth,    xlabel={$i$},
        ylabel={$\mathbb{E} (w_H)$ },
        grid=major,
        xmin=0, xmax=128,xtick={0,20,...,120},ymin=0,legend style={at={(0.03,0.97)}, anchor=north west, legend cell align=left,     align=left, draw=black,font=\footnotesize}
    ]
    \addplot[mark=none, Paired-1,semithick] table[
        x index=0,  
        y index=1,  
        header=true
    ] {Figures/Mixture_of_Gaussians/GRAND/Expected_GRAND_llr.txt};\label{plot:expected_values_grand_custom_3}
   \addplot[mark=none, Paired-5,semithick] table[
        x index=0,  
        y index=1,  
        header=true
    ] {Figures/Mixture_of_Gaussians/HOSD/expected_128_custom_3.txt};\label{plot:expected_values_HOSD_custom_3}
    \addplot [mark=none, Paired-3,semithick] table[
        x index=0,  
        y index=1,  
        header=true
    ] {Figures/Mixture_of_Gaussians/OSD/Expected_OSD_llr.txt};
    \label{plot:expected_values_osd_custom_3}
    \end{axis}
     \matrix[
            matrix of nodes,
            anchor=north,
            draw,
            inner sep=0.2em
        ] at ([yshift=30pt,xshift=10pt]current bounding box.north)
        {
            \ref{plot:expected_values_grand_custom_3}  POSD/GRAND Channel 1  \\  \ref{plot:expected_values_HOSD_custom_3} POSD Channel 2  \ref{plot:expected_values_osd_custom_3}  OSD Channel 2  \\
        };
    \end{tikzpicture}

       \text{Fig. \thefigure:} The expected value of the ordered statistics of the absolute values of the LLRs for 128 bits used in each algorithm for channels 1 and 2 as a function of ordered bit position. \label{fig:expected_vals_mxgs}
    \end{minipage}
\end{figure}

In this section we discuss the case of a mixture of white Gaussian channel. We choose this type of channel since it can approximate any smooth density function with any specific nonzero amount of error with enough components \cite{goodfellow_deep_2016}. We consider an example of this channel and produce \ac{TEP}s catered for this channel using the method discussed in section \ref{sec:tep_gen_no_llr}. We then compare against the SOTA \ac{TEP}s suggested in Section \ref{section:TEPs}.

The density function of a mixture of white Gaussian channel noise with $\gamma$ Gaussian elements can be represented as \cite{le_approximation_2016}: 
\begin{equation}
    f_{\mathbf{y}}(y)= \sum_{i=1}^{\gamma} \frac{\mathbf{\omega}[i]}{\sqrt{2\pi\sigma[i]^2}} e^{-\frac{y^2}{2\sigma[i]^2}}
\end{equation}
where:
\begin{equation}
    \sum_{i=1}^{\gamma} \mathbf{\omega}[i] =1 .
\end{equation}

The approximation of the LLR in this case is not always trivial as it is not always linear with respect to the received signal. Hence, we use the method discussed in Section \ref{sec:tep_gen_no_llr}. Rather than finding the $f_{\mathbf{L}}(l)$, we focus on finding $f_\mathbf{y}(y)$, the \ac{PDF} of the received signal. Due to our assumption about BPSK modulation, this is the shifted \ac{PDF} of the noise signal. For a modulated signal of $+1$, the distribution of the received signal at the receiver end is:
\begin{equation}
    f_{\mathbf{y}}(y|c=0)=  \sum_{i=1}^{\gamma} \frac{\mathbf{\omega}[i]}{\sqrt{2\pi\sigma[i]^2}} e^{-\frac{(y-1)^2}{2\sigma[i]^2}}.
\end{equation}
For a modulated signal of $-1$ , the distribution of the received signal at the receiver end is:
\begin{equation}
    f_{\mathbf{y}}(y|c=1)=  \sum_{i=1}^{\gamma} \frac{\mathbf{\omega}[i]}{\sqrt{2\pi\sigma[i]^2}} e^{-\frac{(y+1)^2}{2\sigma[i]^2}}.
\end{equation}

Following the law of total probabilities, the distribution of the received signal is hence: 
\begin{align}
   f_\mathbf{y}(y) &= P(x=+1) \times f_{\mathbf{y}}(y|x=+1) + f_{\mathbf{y}}(y|x=-1) \times P(x=-1) \\
   &= \frac{1}{2} \sum_{i=1}^{\gamma} \frac{\mathbf{\omega}[i]}{\sqrt{2\pi\sigma[i]^2}} e^{-\frac{(y-1)^2}{2\sigma[i]^2}} + \frac{1}{2}  \sum_{i=1}^{\gamma} \frac{\mathbf{\omega}[i]}{\sqrt{2\pi\sigma[i]^2}} e^{-\frac{(y+1)^2}{2\sigma[i]^2}}.
\end{align}

Additionally the LLR of each received signal 
$y$ is:

\begin{equation}
    \mathcal{L}(y) = \ln \frac{P(y|x=1)}{P(y|x=-1)} = \ln \frac{\sum_{i=1}^{\gamma} \frac{\mathbf{\omega}[i]}{\sqrt{2\pi\sigma[i]^2}} e^{-\frac{y^2}{2\sigma[i]^2}}}{ \sum_{i=1}^{\gamma} \frac{\mathbf{\omega}[i]}{\sqrt{2\pi\sigma[i]^2}} e^{-\frac{y^2}{2\sigma[i]^2}}} \label{eq:LLR_mxgs_formula_gen}
\end{equation}

\subsection{Channel Models}
\begin{table}[t]
\centering
\caption{Approximation of the LLR function for a mixture of Gaussian distributions.}
\begin{tabular}{c|ccc|ccc}
\toprule
\multirow{2}{*}{Segment} & \multicolumn{3}{c|}{\textbf{Channel 1}} & \multicolumn{3}{c}{\textbf{Channel 2}} \\
\cmidrule(lr){2-4} \cmidrule(lr){5-7}
 & Slope & Intercept & Domain & Slope & Intercept & Domain \\
\midrule
1 & 15.8 & 47.4  & $[-\infty,\,-1.7)$     & 15.8 & 42.7  & $[-\infty,\,-1.65)$ \\
2 & -35.3 & -42.9 & $[-1.7,\,-0.94)$  & -31.1 & -36.5 & $[-1.65,\,-0.94)$ \\
3 & 11.5 & 3.8   & $[-0.94,\,-0.17)$  & 13.8 & 8.1   & $[-0.94,\,-0.24)$ \\
4 & -30.5 & 0 & $[-0.17,\,0.3)$   & -26.5 & 0  & $[-0.24,\,0.38)$ \\
5 & 23.8 & -14.12 & $[0.3,\,0.96)$    & 24.7 & -17.26 & $[0.38,\,0.94)$ \\
6 & -27 & 34.19 & $[0.96,\,1.83)$     & -22.9 & 27.53  & $[0.94,\,1.78)$ \\
7 & 15.8 & -47.43 & $[1.83,\,+\infty)$      & 15.8 & -42.7 & $[1.78,\,+\infty)$ \\
\bottomrule
\end{tabular}
\end{table}

Assume two channels defined by the parameters in TABLE \ref{tab:gmm_params}. We have a mixture of 5 Gaussian distributions for each channel. For custom channel 1, the main components are situated at $-3$, $0$ and $3$. For custom channel 2, the main components are situated at $-2.7$, $0$ and $2.7$. We plot the distribution of channel 2 in Fig. \ref{fig:pdf_mxgs}. The LLR corresponding to the received signal distributions in channel 1 and 2 can be seen in Fig. \ref{tab:llr_mapping_mxgs}. We note that this LLR is non-linear with multiple x-axis intersections. Any received signal that maps to a positive LLR is considered the bit $0$ and every received signal that maps to a negative LLR is considered a $1$ bit.

After completing the method described in Section \ref{sec:tep_gen_no_llr}, we obtain the expected values of the received signal. Then the LLR of the received signal is calculated as: The LLR of each 
$\mathbb{E}(\phi(|y|)[i])$ is:

\begin{equation}
    \mathcal{L}(\mathbb{E}(\phi(|y|)[i])) = \ln \frac{P(\mathbb{E}(\phi(|y|)[i])|x=1)}{P(\mathbb{E}(\phi(|y|)[i])|x=-1)}= \ln \frac{\sum_{i=1}^{\gamma} \frac{\mathbf{\omega}[i]}{\sqrt{2\pi\sigma[i]^2}} e^{-\frac{(\mathbb{E}(\phi(|y|)[i])-1)^2}{2\sigma[i]^2}}}{ \sum_{i=1}^{\gamma} \frac{\mathbf{\omega}[i]}{\sqrt{2\pi\sigma[i]^2}} e^{-\frac{(\mathbb{E}(\phi(|y|)[i])+1)^2}{2\sigma[i]^2}}} ,\label{eq:LLR_mxgs_formula_gen_e}
\end{equation} and it is shown in Fig. \ref{fig:expected_vals_mxgs}. We note that channel 1 is more reliable than channel 2 hence, we will analyze channel 1 with POSD and GRAND that have limited error correcting capabilities and channel 2 with OSD.

\subsection{Performance Analysis}

Since this channel has a fixed SNR, we discuss the FER performance of the decoders as a function of maximum number of queries shown in Fig \ref{fig:FER_MXGS}. { Due to the computational complexity of simulating these channels and calculating the LLRs, obtaining the LUT error patterns proved to be infeasible within a reasonable time duration of 7 days with 50 CPU cores. Hence, we will present the performance of the remaining TEPs with this mixture of white Gaussian channel.}
\subsubsection{GRAND}
With channel 1, we can observe that for \ac{RLC} (127,85) and a target FER $10^{-2}$, GRAND with ILW and LW requires $3.8 \times$ and  $13 \times$ the number of queries required with EW error patterns respectively. Compared to the ML TEPs, $40\%$ more EW TEPs are needed to achieve target FER  $10^{-2}$. 
Similarly with CA-Polar (128,105) code, and a target FER $10^{-2}$, GRAND with ILW and LW require $1.8 \times$ and  $14.7 \times$ the number of queries required with EW error patterns respectively. Compared to the ML TEPs, $30\%$ more EW TEPs are needed to achieve target FER  $10^{-2}$.
The HW TEPs do not reach target FER  $10^{-2}$ in both simulations.

\subsubsection{POSD} 
For POSD, we can observe with channel 1 for \ac{BCH} (127,64) and a target FER $10^{-3}$, GRAND LW require $3.3 \times$ the number of queries required with EW error patterns. Additionally, $50 \%$ more EW TEPs are needed to achieve target FER  $10^{-3}$ compared to the ML TEPs. The HW and ILW TEPs do not reach target FER  $10^{-3}$ in this simulation.
Alternatively with channel 2, we can observe  for \ac{RLC} (128,32) and a target FER $10^{-3}$, GRAND with LW error patterns requires the almost the same number of queries as required with EW error patterns. ILW error patterns require $35\%$ more TEPs than EW.  Additionally, $30\% $ more EW TEPs are needed to achieve target FER  $10^{-3}$ compared to the ML TEPs. The HW TEPs do not reach target FER  $10^{-3}$ in this simulation.

\subsubsection{OSD}
For OSD, we can observe with channel 2 for \ac{BCH} (127,64) and a target FER $10^{-4}$, GRAND with HW and LW TEPs require $3 \times$ and $3.7 \times$ the number of queries required with EW error patterns respectively. In this case, the EW, ML and ILW TEPs provide the same performance.
Alternatively with \ac{RLC} (127,85) and a target FER $10^{-2}$, GRAND with LW error patterns requires $11.5 \times$ the number of queries required with EW TEPs to obtain the same performance. ILW error patterns require $37\%$ more TEPs than EW, and the HW TEPs perform similarly to the EW error patterns.  Additionally, $10\% $ more EW TEPs are needed to achieve target FER  $10^{-2}$ compared to the ML TEPs. In this graph, we also observe that between $MQ=10^5$ and $MQ=10^6$ increasing the $MQ$ of ILW, EW and ML TEPs results in little improvement in FER performance.

\subsection{Similarity between \ac{TEP}s}
Fig. \ref{fig:percent_intersection_teps_combined_mxgs} shows the percentage of overlap between LW, ILW and HW TEPs with EW \ac{TEP}s with GRAND, POSD and \ac{OSD} along mixed Gaussian channels 1 and 2. For each maximum number of queries, we compare the percentage of overlap and relate it to the decoding performance.

\subsubsection{GRAND}LW error patterns have the largest overlap with the EW TEPs of $94 \% $ at $MQ=10^2$ and goes down to $46.5 \%$ at $MQ=10^6$. ILW TEPs start with $69\%$ common TEPs with EW TEP at $MQ=10^2$ and reach at $51.6 \%$ at $MQ=10^6$. Notably, the ILW TEPs' overlap surpasses LW at high query counts and correlates with how ILW TEPs outperform LW TEPs in our simulations which are conducted with $MQ=10^6$. The HW TEPs consistently share a low percentage of overlap with EW of $14\%$ overlap \ac{TEP}s at $MQ=10^2$ and $11.8\%$ at $MQ=10^6$, which results in a large decoding performance loss compared to EW TEPs.

\subsubsection{POSD} For POSD with channel 1, the LW error patterns have the largest overlap with the EW TEPs of $83 \%$ at $MQ=10^2$  and $ 62 \%$ at $MQ=10^5$. This is also the case with channel 2 with an overlap of $ 88 \%$ at $MQ=10^2$ and  $ 62.1 \%$ at $MQ=10^5$. The ILW TEPs have a lower percentage overlap of $68 \%~/ ~72\% $ common \ac{TEP}s with EW error patterns at $MQ=10^2$ decreasing to $59 \% ~/~ 60\% $ at $MQ=10^5$ on channels 1~/~2 respectively. This lower number of common \ac{TEP}s correlates with why ILW error patterns are less performing for POSD than LW error patterns. HW TEPs share $30\% \to 35\%$ common TEPs with EW at $MQ=10^2$ dropping to $\approx 20\%$ at $MQ=10^5$ on channel 1 and 2 respectively.
\subsubsection{OSD}
ILW TEPs share the highest percentage of overlapping TEPs with EW, with an overlap of $82\%$ at $MQ=10^2$ increasing to $86\%$ at $MQ=10^6$. LW TEPs follow with an overlap of overlap, moving from $56\%$ at $MQ=10^2$ to approximately $55\%$ at $MQ=10^6$. Meanwhile, HW TEPs show a significant drop in common TEPs with EW, falling from $49\%$ at $MQ=10^2$ (close to that obtained with LW) to approximately $32\%$ at $MQ=10^6$. 
\subsection{Conclusion}We can see that the performance of the predetermined \ac{TEP}s generally correlates with the percentage of common TEPs with the EW \ac{TEP}s along channels 1 and 2. Additionally, we have shown  that the EW error patterns perform most closely to the ML error patterns even without the need to derive the \ac{PDF} of the LLRs. Consistently, EW TEPs are still able to map to newly encountered additive noise profiles.

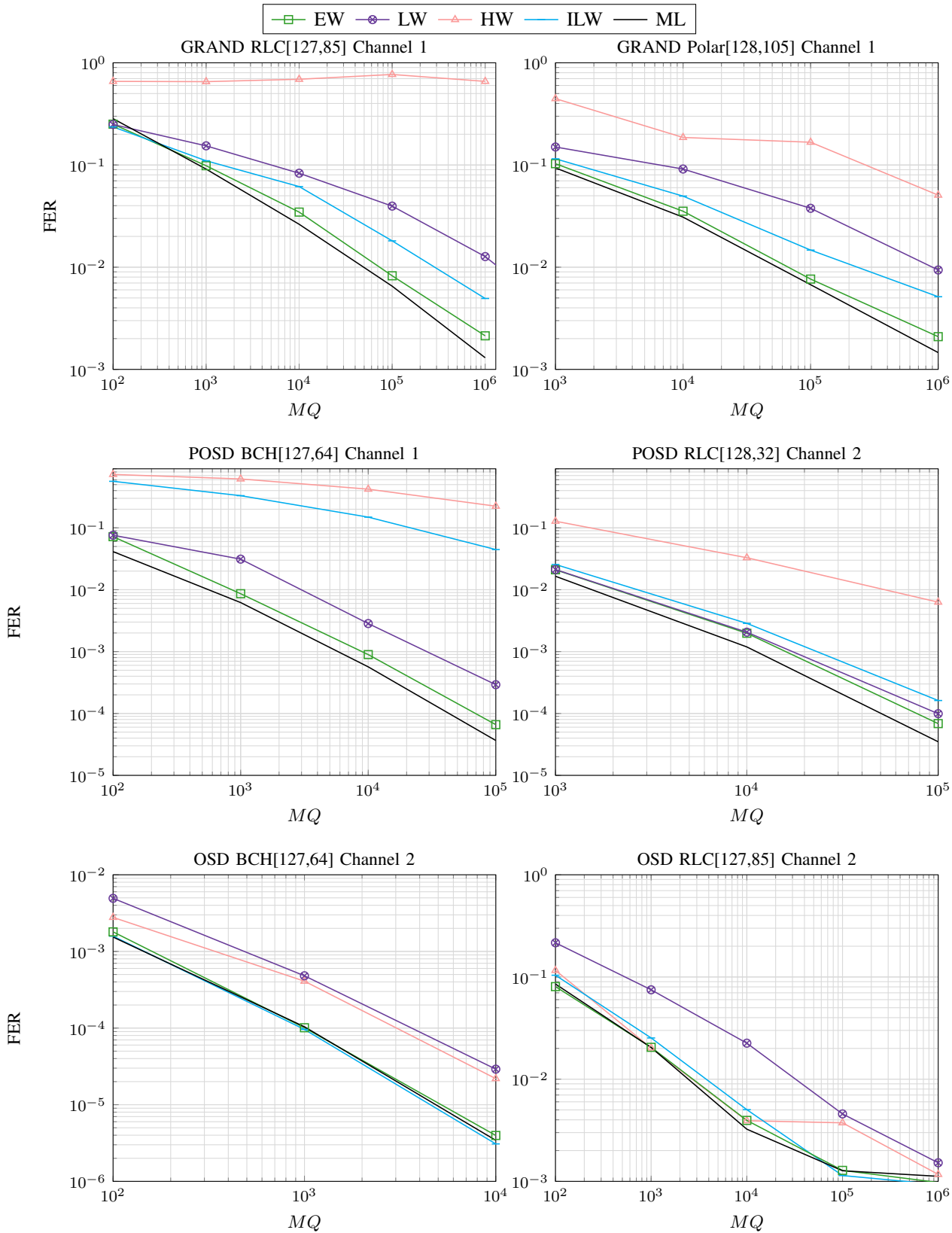
\begin{figure}
\centering
  \begin{tikzpicture}[]
    \begin{groupplot}[group style={group name=fer_queries, group size= 2 by 3, horizontal sep=30pt, vertical sep=50pt},
      footnotesize,
      height=7cm,  width=0.46\columnwidth,  
      xlabel={$MQ$},xmode=log,
      ymode=log,
      tick align=inside,
      grid=both, grid style={gray!30},
      /pgfplots/table/ignore chars={|},
      ] 
      \nextgroupplot[title=GRAND RLC{[127,85]} Channel 1 ,ylabel= FER, ytick pos=left, 
      ymin=1e-3, ymax = 1, xmin=1e2, xmax=1.3e6]
    \addplot[mark=square, Paired-3, semithick]  table[x=MaxQ, y=FER] {Figures/Mixture_of_Gaussians/GRAND/EW.tex}; \label{gp:MXGS_EW_perf_grand}
       \addplot[mark=otimes  , Paired-9, semithick]  table[x=MaxQ, y=FER] {Figures/Mixture_of_Gaussians/GRAND/BCH_127_85_GRAND_LW_0_5_5_MQ_100_SIM_mHW_127_CUSTOM_2.res} ;\label{gp:MXGS_LW_perf_grand}
           \addplot[mark=-, cyan, semithick]  table[x=MaxQ, y=FER] {Figures/Mixture_of_Gaussians/GRAND/BCH_127_85_GRAND_ILW_0_5_5_MQ_100_SIM_mHW_127_CUSTOM_2.res} ;\label{gp:MXGS_ILW_perf_grand}
        \addplot[mark=triangle  , Paired-6 , semithick]  table[x=MaxQ, y=FER] {Figures/Mixture_of_Gaussians/GRAND/BCH_127_85_GRAND_HW_0_5_5_MQ_100_SIM_mHW_127_CUSTOM_2.res} ;\label{gp:MXGS_HW_perf_grand}
        \addplot[mark=none, black , semithick]  table[x=MaxQ, y=FER] {Figures/Mixture_of_Gaussians/GRAND/RLC_127_85_GRAND_ML_0_5_5_MQ_1000_SIM_mHW_85_CUSTOM_2.tex} ;\label{gp:MXGS_ML_perf_grand}
        \nextgroupplot[title=GRAND Polar{[128,105]} Channel 1, ytick pos=left, 
      ymin=1e-3, ymax = 1, xmin=1e3, xmax=1e6]
          \addplot[mark=square, Paired-3 , semithick]  table[x=MaxQ, y=FER] {Figures/Mixture_of_Gaussians/GRAND/Polar_128_105_GRAND_EW_0_5_5_MQ_1000_SIM_mHW_105_CUSTOM_2.res}; \label{gp:MXGS_EW_perf_2_grand}
       \addplot[mark=otimes  , Paired-9, semithick]  table[x=MaxQ, y=FER] {Figures/Mixture_of_Gaussians/GRAND/Polar_128_105_GRAND_LW_0_5_5_MQ_1000_SIM_mHW_105_CUSTOM_2.res} ;\label{gp:MXGS_LW_perf_2_grand}
           \addplot[mark=-, cyan, semithick]  table[x=MaxQ, y=FER] {Figures/Mixture_of_Gaussians/GRAND/Polar_128_105_GRAND_ILW_0_5_5_MQ_1000_SIM_mHW_105_CUSTOM_2.res} ;\label{gp:MXGS_ILW_perf_2_grand}
        \addplot[mark=triangle  , Paired-6 , semithick]  table[x=MaxQ, y=FER] {Figures/Mixture_of_Gaussians/GRAND/Polar_128_105_GRAND_HW_0_5_5_MQ_1000_SIM_mHW_105_CUSTOM_2.res} ;\label{gp:MXGS_HW_perf_2_grand}
        \addplot[mark=none, black , semithick]  table[x=MaxQ, y=FER] {Figures/Mixture_of_Gaussians/GRAND/Polar_128_105_GRAND_ML_0_5_5_MQ_1000_SIM_mHW_105_CUSTOM_2.res} ;\label{gp:MXGS_ML_perf_2_grand}
        \nextgroupplot[title=POSD BCH{[127,64]} Channel 1 ,ylabel=FER, ytick pos=left, y label style={at={(axis description cs:-0.225,.5)},anchor=south},ymin=1e-5, ymax = 9e-1, xmin=1e2, xmax=1e5]
    \addplot[mark=triangle  , Paired-6 , semithick]  table[x=MaxQ, y=FER] {Figures/Mixture_of_Gaussians/HOSD/HW_1C2.tex}; \label{gp:MXGS_HW_perf_HOSD}
       \addplot[mark=-, cyan, semithick]  table[x=MaxQ, y=FER] {Figures/Mixture_of_Gaussians/HOSD/ILW_1C2.tex} ;\label{gp:MXGS_ILW_perf_HOSD}
           \addplot[mark=square, Paired-3 , semithick]  table[x=MaxQ, y=FER] {Figures/Mixture_of_Gaussians/HOSD/EW_1C2.tex}; \label{gp:MXGS_EW_perf_HOSD}
       \addplot[mark=otimes  , Paired-9, semithick]  table[x=MaxQ, y=FER] {Figures/Mixture_of_Gaussians/HOSD/LW_1C2.tex} ;\label{gp:MXGS_LW_perf_HOSD}
              \addplot[mark=none  , black, semithick]  table[x=MaxQ, y=FER] {Figures/Mixture_of_Gaussians/HOSD/BCH_127_64_HOSD_ML_0_5_5_MQ_1000_SIM_mHW_64_CUSTOM_2.res} ;\label{gp:MXGS_ML_perf_HOSD}
           \coordinate (top) at (rel axis cs:0,1);
      \coordinate (spypoint1) at (axis cs:7.45,2e-7);
      \coordinate (magnifyglass1) at (axis cs:2.6,1.1e-5);
      \coordinate (bot) at (rel axis cs:1,0);
       \nextgroupplot[title=POSD RLC{[128,32]} Channel 2, ytick pos=left, y label style={at={(axis description cs:-0.225,.5)},anchor=south},ymin=1e-5, ymax = 9e-1, xmin=1e3, xmax=1e5]
       \addplot[mark=triangle  , Paired-6 , semithick]  table[x=MaxQ, y=FER] {Figures/Mixture_of_Gaussians/HOSD/RLC_128_32_HOSD_HW_0_5_5_MQ_1000_SIM_mHW_32_CUSTOM_3.res}; \label{gp:MXGS_HW_perf_2_HOSD}
       \addplot[mark=-, cyan, semithick]  table[x=MaxQ, y=FER] {Figures/Mixture_of_Gaussians/HOSD/RLC_128_32_HOSD_ILW_0_5_5_MQ_1000_SIM_mHW_32_CUSTOM_3.res} ;\label{gp:MXGS_ILW_perf_2_HOSD}
           \addplot[mark=square, Paired-3 , semithick]  table[x=MaxQ, y=FER] {Figures/Mixture_of_Gaussians/HOSD/RLC_128_32_HOSD_EW_0_5_5_MQ_1000_SIM_mHW_32_CUSTOM_3.res}; \label{gp:MXGS_EW_perf_2_HOSD}
       \addplot[mark=otimes  , Paired-9, semithick]  table[x=MaxQ, y=FER] {Figures/Mixture_of_Gaussians/HOSD/RLC_128_32_HOSD_LW_0_5_5_MQ_1000_SIM_mHW_32_CUSTOM_3.res} ;\label{gp:MXGS_LW_perf_2_HOSD}
              \addplot[mark=none  , black, semithick]  table[x=MaxQ, y=FER] {Figures/Mixture_of_Gaussians/HOSD/RLC_128_32_HOSD_ML_0_5_5_MQ_1000_SIM_mHW_32_CUSTOM_3.res} ;\label{gp:MXGS_ML_perf_2_HOSD}
    \nextgroupplot[ title=OSD BCH{[127,64]} Channel 2 , ytick pos=left,ylabel= FER, y label style={at={(axis description cs:-0.225,.5)},anchor=south},ymin=1e-6, ymax = 1e-2, xmin=1e2, xmax=1e4]
    \addplot[mark=triangle  , Paired-6 , semithick]  table[x=MaxQ, y=FER] {Figures/Mixture_of_Gaussians/OSD/BCH_127_64_OSD_HW_0_5_5_MQ_100_SIM_mHW_127_CUSTOM_3.res}; \label{gp:MXGS_HW_perf_osd}
       \addplot[mark=-, cyan, semithick]  table[x=MaxQ, y=FER] {Figures/Mixture_of_Gaussians/OSD/BCH_127_64_OSD_ILW_0_5_5_MQ_100_SIM_mHW_127_CUSTOM_3.res} ;\label{gp:MXGS_ILW_perf_osd}
           \addplot[mark=square, Paired-3 , semithick]  table[x=MaxQ, y=FER] {Figures/Mixture_of_Gaussians/OSD/BCH_127_64_OSD_EW_0_5_5_MQ_100_SIM_mHW_127_CUSTOM_3.res}; \label{gp:MXGS_EW_perf_osd}
       \addplot[mark=otimes  , Paired-9, semithick]  table[x=MaxQ, y=FER] {Figures/Mixture_of_Gaussians/OSD/BCH_127_64_OSD_LW_0_5_5_MQ_100_SIM_mHW_127_CUSTOM_3.res} ;\label{gp:MXGS_LW_perf_osd}
           \addplot[mark=none  , black, semithick]  table[x=MaxQ, y=FER] {Figures/Mixture_of_Gaussians/OSD/BCH_127_64_OSD_ML_0_5_5_MQ_1000_SIM_mHW_64_CUSTOM_3.res} ;\label{gp:MXGS_ML_perf_osd}
           \nextgroupplot[ title=OSD RLC{[127,85]} Channel 2, ytick pos=left, y label style={at={(axis description cs:-0.225,.5)},anchor=south},ymin=1e-3, ymax = 1e-0, xmin=1e2, xmax=1e6]
    \addplot[mark=triangle  , Paired-6 , semithick]  table[x=MaxQ, y=FER] {Figures/Mixture_of_Gaussians/OSD/RLC_127_85_OSD_HW_0_5_5_MQ_100_SIM_mHW_85_CUSTOM_3.res}; \label{gp:MXGS_HW_perf_osd_2}
       \addplot[mark=-, cyan, semithick]  table[x=MaxQ, y=FER] {Figures/Mixture_of_Gaussians/OSD/RLC_127_85_OSD_ILW_0_5_5_MQ_100_SIM_mHW_85_CUSTOM_3.res} ;\label{gp:MXGS_ILW_perf_osd_2}
           \addplot[mark=square, Paired-3 , semithick]  table[x=MaxQ, y=FER] {Figures/Mixture_of_Gaussians/OSD/RLC_127_85_OSD_EW_0_5_5_MQ_100_SIM_mHW_85_CUSTOM_3.res}; \label{gp:MXGS_EW_perf_osd_2}
       \addplot[mark=otimes  , Paired-9, semithick]  table[x=MaxQ, y=FER] {Figures/Mixture_of_Gaussians/OSD/RLC_127_85_OSD_LW_0_5_5_MQ_100_SIM_mHW_85_CUSTOM_3.res} ;\label{gp:MXGS_LW_perf_osd_2}
           \addplot[mark=none  , black, semithick]  table[x=MaxQ, y=FER] {Figures/Mixture_of_Gaussians/OSD/RLC_127_85_OSD_ML_0_5_5_MQ_100_SIM_mHW_85_CUSTOM_3.res} ;\label{gp:MXGS_ML_perf_osd_2}
    \end{groupplot}
    \path (current bounding box.north west) -- coordinate (legendpos)
      (current bounding box.north east);
    \matrix[
    matrix of nodes,
    anchor=south,
    draw,
    inner sep=0.2em,
    draw
    ]at(legendpos) 
    {
    \ref{gp:MXGS_EW_perf_grand}&   EW &[1pt] 
    \ref{gp:MXGS_LW_perf_grand}&   LW &[1pt] \ref{gp:MXGS_HW_perf_grand}&   HW &[1pt] \ref{gp:MXGS_ILW_perf_grand}&   ILW &[1pt]  \ref{gp:MXGS_ML_perf_grand}&   ML &[1pt]\\ \\
      };
  \end{tikzpicture}
  \caption{\label{fig:FER_MXGS} Decoding performance of the \ac{TEP}s on custom mixture of Gaussian channels.}
\end{figure}
\begin{figure}[t]
\centering
\begin{tikzpicture}
 \begin{groupplot}[group style={group name=pdfmixture, group size= 1 by 4, horizontal sep=25pt, vertical sep=50pt},footnotesize,
      height=4cm,  width=0.81\columnwidth,
      xlabel={$MQ$}, /pgfplots/table/ignore chars={|}]
      
\nextgroupplot[title= GRAND {[127,85] }Channel 1,ylabel= Percent Overlap ($\%$), ytick pos=left,
ymin=0, ymax = 100, 
    ybar,
    xmode=log,
    bar width=12pt, 
    xtick=data, 
    enlarge x limits=0.15, 
    legend entries={},
    legend style={draw=none, fill=none}]
\addplot[fill=Paired-9, opacity=0.8] table[
    x index=0,  
    y index=2,  
    header=false] {Figures/Mixture_of_Gaussians/GRAND/grand_custom_2.txt};\label{plot:lw_grand_mxgs_perc}
    \addplot[fill=cyan, opacity=0.8] table[
    x index=0,  
    y index=4,  
    header=false] {Figures/Mixture_of_Gaussians/GRAND/grand_custom_2.txt};\label{plot:ilwo_grand_mxgs_perc}
\addplot[fill=Paired-6, opacity=0.8] table[
    x index=0,  
    y index=3,  
    header=false] {Figures/Mixture_of_Gaussians/GRAND/grand_custom_2.txt};\label{plot:hw_grand_mxgs_perc}
\nextgroupplot[title= POSD {[127,64] }Channel 1,ylabel= Percent Overlap ($\%$),  ymin=0, ymax = 100,
    ymin=0, 
    ybar,
    xmode=log,
    bar width=12pt, 
    xtick=data, 
    enlarge x limits=0.15, 
    legend entries={},
    legend style={draw=none, fill=none}]
        \addplot[fill=Paired-9, opacity=0.8] table[
    x index=0,  
    y index=2,  
    header=false] {Figures/Mixture_of_Gaussians/HOSD/posd_custom_2.txt};\label{plot:lw_hosd_mxgs_perc}
    \addplot[fill=cyan, opacity=0.8] table[
    x index=0,  
    y index=4,  
    header=false] {Figures/Mixture_of_Gaussians/HOSD/posd_custom_2.txt};\label{plot:ilwo_hosd_mxgs_perc}

\addplot[fill=Paired-6, opacity=0.8] table[
    x index=0,  
    y index=3,  
    header=false] {Figures/Mixture_of_Gaussians/HOSD/posd_custom_2.txt};\label{plot:hw_hosd_mxgs_perc}
\nextgroupplot[title= POSD  {[128,32] }Channel 2,ylabel= Percent Overlap ($\%$), ymin=0, ymax = 100,
    ymin=0, 
    ybar,
    xmode=log,
    bar width=12pt, 
    xtick=data, 
    enlarge x limits=0.15, 
    legend entries={},
    legend style={draw=none, fill=none}]
\addplot[fill=Paired-9, opacity=0.8] table[
    x index=0,  
    y index=2,  
    header=false] {Figures/Mixture_of_Gaussians/HOSD/posd_custom_3.txt};\label{plot:lw_hosd_mxgs_perc_2}
    \addplot[fill=cyan, opacity=0.8] table[
    x index=0,  
    y index=4,  
    header=false] {Figures/Mixture_of_Gaussians/HOSD/posd_custom_3.txt};\label{plot:ilwo_hosd_mxgs_perc_2}
\addplot[fill=Paired-6, opacity=0.8] table[
    x index=0,  
    y index=3,  
    header=false] {Figures/Mixture_of_Gaussians/HOSD/posd_custom_3.txt};\label{plot:hw_hosd_mxgs_perc_2}
        \coordinate (top) at (rel axis cs:0,1);
      \coordinate (spypoint1) at (axis cs:7.45,2e-7);
      \coordinate (magnifyglass1) at (axis cs:2.6,1.1e-5);
      \coordinate (bot) at (rel axis cs:1,0);
    \nextgroupplot[title=OSD {[127,64] }Channel 2,ylabel= Percent Overlap ($\%$), ymin=0, ymax = 100,
    ymin=0, 
    ybar,
    xmode=log,
    bar width=12pt, 
    xtick=data, 
    enlarge x limits=0.15, 
    legend entries={},
    legend style={draw=none, fill=none}]
    \addplot[fill=cyan, opacity=0.8] table[
    x index=0,  
    y index=4,  
    header=false] {Figures/Mixture_of_Gaussians/OSD/osd_custom_3.tex};\label{plot:ilwo_osd_mxgs_perc}
    \addplot[fill=Paired-9, opacity=0.8] table[
    x index=0,  
    y index=2,  
    header=false] {Figures/Mixture_of_Gaussians/OSD/osd_custom_3.tex};\label{plot:lw_osd_mxgs_perc}
\addplot[fill=Paired-6, opacity=0.8] table[
    x index=0,  
    y index=3,  
    header=false] {Figures/Mixture_of_Gaussians/OSD/osd_custom_3.tex};\label{plot:hw_osd_mxgs_perc}
\end{groupplot}
    \matrix[
        matrix of nodes,
        anchor=north,
        draw,
        inner sep=0.2em
    ] at ([yshift=20pt,xshift=10pt]current bounding box.north)
    {
        \ref{plot:lw_grand_mxgs_perc}&  LW &[1pt] \ref{plot:ilwo_grand_mxgs_perc}&  ILW &[1pt]  \ref{plot:hw_grand_mxgs_perc}&  HW  &[1pt] \\
    };
\end{tikzpicture}
\caption{Percentage of the overlap of the TEPs generated by \ac{SOTA} methods with the EW error patterns with a mixture of Gaussian distribution channels.}\label{fig:percent_intersection_teps_combined_mxgs}
\end{figure}
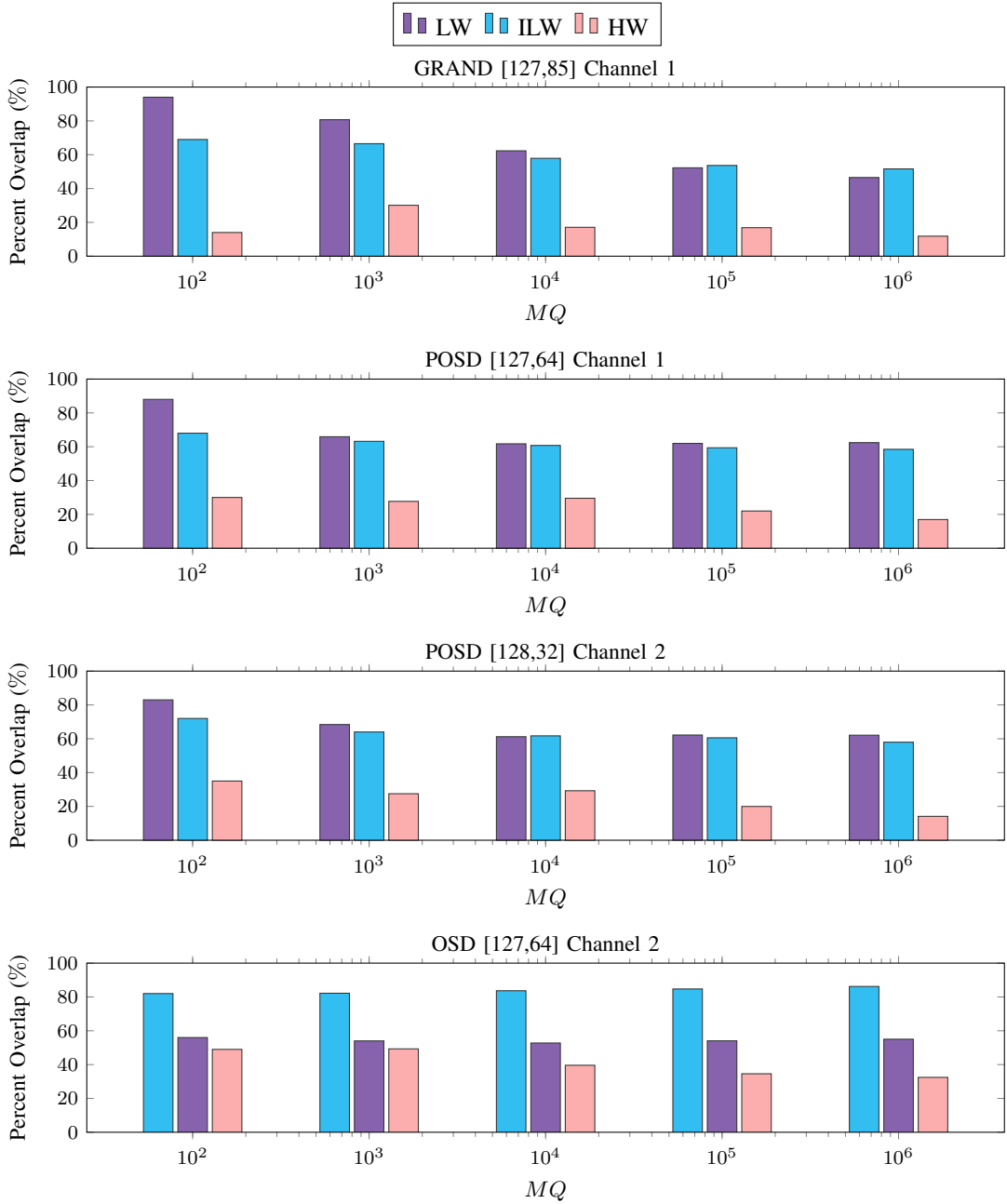

\section{Uncorrelated Fast Fading}

The Rayleigh fading channel models the case where there is no direct line of sight between transmitter and receiver. The transmitted signal is scattered off surfaces and the same transmitted signal interferes either destructuctively or constructively with itself. This process can be modeled by a fading gain $h$ multiplied to the modulated signal. As such, in a Rayleigh fading channel, the receiver receives a channel signal with value:
\begin{equation}
    y= h \times x + n
\end{equation}
where $n\sim \mathcal{N}(0,\sigma^2)$ and $h$ is Rayleigh distributed with a average power of 1 ($E[h^2] =1$).  

The distribution of the channel signal conditioned on fading coefficient and the sent code bit is\cite{hou_performance_2001}:
\begin{equation}
    f_{\mathbf{y}}(y|h,x)= \frac{1}{\sqrt{2\pi\sigma^2}} e^{-\left( \frac{(y-hx)^2}{2\sigma^2}\right)}
\end{equation}
We can easily observe that the channel is output-symmetric defined by the notion of $f(-y|h,-x)=f(y|h,x)$ \cite{richardson_capacity_2001}:
\begin{equation}
f_{\mathbf{y}}(-y|h,-x) = \frac{1}{\sqrt{2\pi\sigma^2}} e^{-\left( \frac{(-y+hx)^2}{2\sigma^2}\right)}= \frac{1}{\sqrt{2\pi\sigma^2}} e^{-\left( \frac{(y-hx)^2} {2\sigma^2}\right)}=f_{\mathbf{y}}(y|h,x).
\end{equation}

The notion of output-channel symmetry helps in simplifying the analysis. First remark that:

\begin{equation}
       f_{\mathbf{L}}(l)= P(x=+1)f_{L}(l|x=+1)+ 
        P(x=-1)f_{L}(l|x=-1). 
\end{equation}

Hence, to calculate the reliability of the individual bits, we can expand (\ref{eq:abs_LLRs_pdf}) with $l\geq0$:

\begin{align}
    f_{|\mathbf{L}|}(l)=& f_{L} (l) + f_{\mathbf{L}} (-l),  \\
    =&{\color{blue}  P(x=+1)}{\color{Paired-7}f_{\mathbf{L}}(l|x=+1)}+ 
        {\color{blue} P(x=-1)}{\color{purple}f_{\mathbf{L}}(l|x=-1)}, \\
        &+ {\color{blue} P(x=+1)}{\color{purple}f_{\mathbf{L}}(-l|x=+1)}+ 
        {\color{blue} P(x=-1)}{\color{Paired-7}f_{\mathbf{L}}(-l|x=-1)},\\
        =& {\color{Paired-7}f_{\mathbf{L}}(l|x=+1)} +{\color{purple}f_{\mathbf{L}}(-l|x=+1)}. 
\end{align}
Where the last equation follows from the fact that the similary colored terms are equal. Specifically, we can see that:
\begin{align}
  P(x=+1)=& P(x=-1) = 0.5, \quad &(\text{Equiprobable input symbols}),\\
  f_{\mathbf{L}}(l|x=+1)=& f_{\mathbf{L}}(-l|x=-1), &(\text{Symmetry}),\\
  f_{\mathbf{L}}(l|x=-1)=&
       f_{\mathbf{L}}(-l|x=+1), & (\text{Symmetry}).  
\end{align}

Hence, to simplify the analysis, we can use the distribution of the LLR conditioned on the $x=1 $ or $c=0$ without the need to calculate the conditional probability given that $x=-1$ being transmitted. In this section we will first discuss how EW TEPs can be generated for a Rayleigh fading channel with perfect knowledge of \ac{CSI} and then discuss how they can be generated for a Rayleigh fading channel without perfect knowledge of channel state information \ac{NCSI}.
\subsection{Ideal \acf{CSI}}
In the uncorrelated fast fading channel with ideal channel state information (CSI), the value of $h$ is known for the receiver. Hence, after the signal is received, the fading gain is removed from the received signal producing:
\begin{equation}
   \frac{ {y}}{{h}}= {x}+{\frac{n}{h}}.
\end{equation}
The LLR in this case can be calculated as \cite{hou_performance_2001}:
\begin{equation}
    L=\frac{2}{\sigma^2}y  \times h
\end{equation}

 The distribution of the LLRs conditioned on $c=0$ being sent is \cite{hou_performance_2001}:
\begin{figure}
  \centering
  \begin{tikzpicture}[]
    \begin{groupplot}[group style={group name=fer_queries, group size= 2 by 3, horizontal sep=30pt, vertical sep=50pt},
      footnotesize,
      height=7cm,  width=0.46\columnwidth,  
      xlabel=$\frac{E_b}{N_0}$ (dB),
      ymode=log,
      tick align=inside,
      grid=both, grid style={gray!30},
      /pgfplots/table/ignore chars={|},
      ] 
      \nextgroupplot[title=GRAND RLC {[127,85]},ylabel= FER, ytick pos=left,ymin=1e-6, ymax = 1, xmin=1, xmax=20]
    \addplot[mark=square, Paired-3 , semithick]  table[x=Eb/N0, y=FER] {Figures/RAYLEIGH/RLC/GRAND/RLC_127_85_GRAND_EW_0_1_20_MQ_1000000_SIM_mHW_127_RAYLEIGH.res}; \label{gp:RAYLEIGH_GRAND_EW_perf}
       \addplot[mark=triangle  , Paired-6, semithick]  table[x=Eb/N0, y=FER] {Figures/RAYLEIGH/RLC/GRAND/RLC_127_85_GRAND_HW_0_1_20_MQ_1000000_SIM_mHW_127_RAYLEIGH.res} ;\label{gp:RAYLEIGH_GRAND_HW_perf}
      \addplot[mark=none  , black, semithick]  table[x=Eb/N0, y=FER] {Figures/RAYLEIGH/RLC/GRAND/RLC_127_85_GRAND_ML_0_1_20_MQ_1000000_SIM_mHW_85_RAYLEIGH.res} ;\label{gp:RAYLEIGH_GRAND_ML_perf}
         \addplot[mark=otimes  , Paired-9, semithick]  table[x=Eb/N0, y=FER] {Figures/RAYLEIGH/RLC/GRAND/RLC_127_85_GRAND_LW_0_1_20_MQ_1000000_SIM_mHW_127_RAYLEIGH.res} ;\label{gp:RAYLEIGH_GRAND_LW_perf}
     \addplot[mark=-, cyan , semithick]  table[x=Eb/N0, y=FER] {Figures/RAYLEIGH/RLC/GRAND/RLC_127_85_GRAND_ILW_0_1_20_MQ_1000000_SIM_mHW_127_RAYLEIGH.res};\label{gp:RAYLEIGH_GRAND_ILW_perf}
     \addplot[mark=square, Paired-3 , dashed]  table[x=Eb/N0, y=FER] {Figures/RAYLEIGH/RLC/GRAND/RLC_127_85_GRAND_EW_0_0_20_MQ_1024_SIM_mHW_85_RAYLEIGH.res}; \label{gp:RAYLEIGH_GRAND_EW_perf_3}
       \addplot[mark=triangle  , Paired-6, dashed]  table[x=Eb/N0, y=FER] {Figures/RAYLEIGH/RLC/GRAND/RLC_127_85_GRAND_HW_0_0_20_MQ_1024_SIM_mHW_85_RAYLEIGH.res} ;\label{gp:RAYLEIGH_GRAND_HW_perf_3}
      \addplot[mark=none  , black, dashed]  table[x=Eb/N0, y=FER] {Figures/RAYLEIGH/RLC/GRAND/RLC_127_85_GRAND_ML_0_0_20_MQ_1024_SIM_mHW_85_RAYLEIGH.res} ;\label{gp:RAYLEIGH_GRAND_ML_perf_3}
         \addplot[mark=otimes  , Paired-9, dashed]  table[x=Eb/N0, y=FER] {Figures/RAYLEIGH/RLC/GRAND/RLC_127_85_GRAND_LW_0_0_20_MQ_1024_SIM_mHW_85_RAYLEIGH.res} ;\label{gp:RAYLEIGH_GRAND_LW_perf_3}
     \addplot[mark=-, cyan , dashed]  table[x=Eb/N0, y=FER] {Figures/RAYLEIGH/RLC/GRAND/RLC_127_85_GRAND_ILW_0_0_20_MQ_1024_SIM_mHW_85_RAYLEIGH.res};\label{gp:RAYLEIGH_GRAND_ILW_perf_3}
       \addplot[mark=none, red , dashed]  table[x=Eb/N0, y=FER] {Figures/RAYLEIGH/RLC/GRAND/RLC_127_85_GRAND_LUT_0_0_20_MQ_1024_SIM_mHW_85_RAYLEIGH.res};\label{gp:RAYLEIGH_GRAND_LUT_perf_3}
           \nextgroupplot[title=GRAND Polar {[128,105+11]}, ytick pos=left,ymin=1e-6, ymax = 1, xmin=1, xmax=20]
 \addplot[mark=square, Paired-3 , semithick]  table[x=Eb/N0, y=FER] {Figures/RAYLEIGH/Polar/GRAND/Polar_128_105_GRAND_EW_0_0_20_MQ_1000000_SIM_mHW_105_RAYLEIGH.res}; \label{gp:RAYLEIGH_GRAND_EW_perf_2}
       \addplot[mark=triangle  , Paired-6, semithick]  table[x=Eb/N0, y=FER] {Figures/RAYLEIGH/Polar/GRAND/Polar_128_105_GRAND_HW_0_0_20_MQ_1000000_SIM_mHW_105_RAYLEIGH.res} ;\label{gp:RAYLEIGH_GRAND_HW_perf_2}
      \addplot[mark=none  , black, semithick]  table[x=Eb/N0, y=FER] {Figures/RAYLEIGH/Polar/GRAND/Polar_128_105_GRAND_ML_0_0_20_MQ_1000000_SIM_mHW_105_RAYLEIGH.res} ;\label{gp:RAYLEIGH_GRAND_ML_perf_2}
         \addplot[mark=otimes  , Paired-9, semithick]  table[x=Eb/N0, y=FER] {Figures/RAYLEIGH/Polar/GRAND/Polar_128_105_GRAND_LW_0_0_20_MQ_1000000_SIM_mHW_105_RAYLEIGH.res} ;\label{gp:RAYLEIGH_GRAND_LW_perf_2}
     \addplot[mark=-, cyan , semithick]  table[x=Eb/N0, y=FER] {Figures/RAYLEIGH/Polar/GRAND/Polar_128_105_GRAND_ILW_0_0_20_MQ_1000000_SIM_mHW_105_RAYLEIGH.res};\label{gp:RAYLEIGH_GRAND_ILW_perf_2}
     \addplot[mark=square, Paired-3 , dashed]  table[x=Eb/N0, y=FER] {Figures/RAYLEIGH/Polar/GRAND/Polar_128_105_GRAND_EW_0_0_20_MQ_1024_SIM_mHW_105_RAYLEIGH.res}; \label{gp:RAYLEIGH_GRAND_EW_perf_5}
       \addplot[mark=triangle  , Paired-6, dashed]  table[x=Eb/N0, y=FER] {Figures/RAYLEIGH/Polar/GRAND/Polar_128_105_GRAND_HW_0_0_20_MQ_1024_SIM_mHW_105_RAYLEIGH.res} ;\label{gp:RAYLEIGH_GRAND_HW_perf_5}
      \addplot[mark=none  , black, dashed]  table[x=Eb/N0, y=FER] {Figures/RAYLEIGH/Polar/GRAND/Polar_128_105_GRAND_ML_0_0_20_MQ_1024_SIM_mHW_105_RAYLEIGH.res} ;\label{gp:RAYLEIGH_GRAND_ML_perf_5}
         \addplot[mark=otimes  , Paired-9, dashed]  table[x=Eb/N0, y=FER] {Figures/RAYLEIGH/Polar/GRAND/Polar_128_105_GRAND_LW_0_0_20_MQ_1024_SIM_mHW_105_RAYLEIGH.res} ;\label{gp:RAYLEIGH_GRAND_LW_perf_5}
     \addplot[mark=-, cyan , dashed]  table[x=Eb/N0, y=FER] {Figures/RAYLEIGH/Polar/GRAND/Polar_128_105_GRAND_ILW_0_0_20_MQ_1024_SIM_mHW_105_RAYLEIGH.res};\label{gp:RAYLEIGH_GRAND_ILW_perf_5}
          \addplot[mark=none, red , dashed]  table[x=Eb/N0, y=FER] {Figures/RAYLEIGH/Polar/GRAND/Polar_128_105_GRAND_LUT_0_0_20_MQ_1024_SIM_mHW_105_RAYLEIGH.res};\label{gp:RAYLEIGH_GRAND_LUT_perf_5}
            \nextgroupplot[ title=POSD BCH {[127,64]},ylabel= FER, ytick pos=left, y label style={at={(axis description cs:-0.225,.5)},anchor=south},ymin=1e-5, ymax = 1, xmin=1, xmax=20]
    \addplot[mark=square, Paired-3 , semithick]  table[x=Eb/N0, y=FER] {Figures/RAYLEIGH/BCH/POSD/BCH_127_64_HOSD_EW_0_0_20_MQ_100000_SIM_mHW_127_RAYLEIGH.res}; \label{gp:RAYLEIGH_HOSD_EW_perf}
        \addplot[mark=none, black , semithick]  table[x=Eb/N0, y=FER] {Figures/RAYLEIGH/BCH/POSD/BCH_127_64_HOSD_ML_0_0_20_MQ_100000_SIM_mHW_127_RAYLEIGH.res}; \label{gp:RAYLEIGH_HOSD_ML_perf}
       \addplot[mark=triangle  , Paired-6, semithick]  table[x=Eb/N0, y=FER] {Figures/RAYLEIGH/BCH/POSD/BCH_127_64_HOSD_HW_0_0_20_MQ_100000_SIM_mHW_127_RAYLEIGH.res} ;\label{gp:RAYLEIGH_HOSD_HW_perf}
         \addplot[mark=otimes  , Paired-9, semithick]  table[x=Eb/N0, y=FER] {Figures/RAYLEIGH/BCH/POSD/BCH_127_64_HOSD_LW_0_0_20_MQ_100000_SIM_mHW_127_RAYLEIGH.res} ;\label{gp:RAYLEIGH_HOSD_LW_perf}
     \addplot[mark=-, cyan , semithick]  table[x=Eb/N0, y=FER] {Figures/RAYLEIGH/BCH/POSD/BCH_127_64_HOSD_ILW_0_0_20_MQ_100000_SIM_mHW_127_RAYLEIGH.res};\label{gp:RAYLEIGH_HOSD_ILW_perf}
      \addplot[mark=square, Paired-3 , dashed]  table[x=Eb/N0, y=FER] {Figures/RAYLEIGH/BCH/POSD/BCH_127_64_HOSD_EW_0_0_20_MQ_1024_SIM_mHW_64_RAYLEIGH.res}; \label{gp:RAYLEIGH_HOSD_EW_perf_4}
        \addplot[mark=none, black , dashed]  table[x=Eb/N0, y=FER] {Figures/RAYLEIGH/BCH/POSD/BCH_127_64_HOSD_ML_0_0_20_MQ_1024_SIM_mHW_64_RAYLEIGH.res}; \label{gp:RAYLEIGH_HOSD_ML_perf_4}
       \addplot[mark=triangle  , Paired-6, dashed]  table[x=Eb/N0, y=FER] {Figures/RAYLEIGH/BCH/POSD/BCH_127_64_HOSD_HW_0_0_20_MQ_1024_SIM_mHW_64_RAYLEIGH.res} ;\label{gp:RAYLEIGH_HOSD_HW_perf_4}
         \addplot[mark=otimes  , Paired-9, dashed]  table[x=Eb/N0, y=FER] {Figures/RAYLEIGH/BCH/POSD/BCH_127_64_HOSD_LW_0_0_20_MQ_1024_SIM_mHW_64_RAYLEIGH.res} ;\label{gp:RAYLEIGH_HOSD_LW_perf_4}
     \addplot[mark=-, cyan , dashed]  table[x=Eb/N0, y=FER] {Figures/RAYLEIGH/BCH/POSD/BCH_127_64_HOSD_ILW_0_0_20_MQ_1024_SIM_mHW_64_RAYLEIGH.res};\label{gp:RAYLEIGH_HOSD_ILW_perf_4}
          \addplot[mark=none, red , dashed]  table[x=Eb/N0, y=FER] {Figures/RAYLEIGH/BCH/POSD/BCH_127_64_HOSD_LUT_0_0_20_MQ_1024_SIM_mHW_64_RAYLEIGH.res};\label{gp:RAYLEIGH_HOSD_LUT_perf_4}
       \nextgroupplot[ title=POSD RLC{[128,32]}, ytick pos=left, y label style={at={(axis description cs:-0.225,.5)},anchor=south},ymin=1e-5, ymax = 1, xmin=1, xmax=20]
    \addplot[mark=square, Paired-3 , semithick]  table[x=Eb/N0, y=FER] {Figures/RAYLEIGH/RLC/POSD/RLC_128_32_HOSD_EW_0_0_20_MQ_100000_SIM_mHW_32_RAYLEIGH.res}; \label{gp:RAYLEIGH_HOSD_EW_perf_2}
    \addplot    [mark=none, black , semithick]    table[x=Eb/N0, y=FER] {Figures/RAYLEIGH/RLC/POSD/RLC_128_32_HOSD_ML_0_0_20_MQ_100000_SIM_mHW_32_RAYLEIGH.res}; \label{gp:RAYLEIGH_HOSD_ML_perf_2}
       \addplot[mark=triangle  , Paired-6, semithick]  table[x=Eb/N0, y=FER] {Figures/RAYLEIGH/RLC/POSD/RLC_128_32_HOSD_HW_0_0_20_MQ_100000_SIM_mHW_32_RAYLEIGH.res} ;\label{gp:RAYLEIGH_HOSD_HW_perf_2}
         \addplot[mark=otimes  , Paired-9, semithick]  table[x=Eb/N0, y=FER] {Figures/RAYLEIGH/RLC/POSD/RLC_128_32_HOSD_LW_0_0_20_MQ_100000_SIM_mHW_32_RAYLEIGH.res} ;\label{gp:RAYLEIGH_HOSD_LW_perf_2}
     \addplot[mark=-, cyan , semithick]  table[x=Eb/N0, y=FER] {Figures/RAYLEIGH/RLC/POSD/RLC_128_32_HOSD_ILW_0_0_20_MQ_100000_SIM_mHW_32_RAYLEIGH.res};\label{gp:RAYLEIGH_HOSD_ILW_perf_2}
        \addplot[mark=square, Paired-3 , dashed]  table[x=Eb/N0, y=FER] {Figures/RAYLEIGH/RLC/POSD/RLC_128_32_HOSD_EW_0_0_20_MQ_1024_SIM_mHW_32_RAYLEIGH.res}; \label{gp:RAYLEIGH_HOSD_EW_perf_5}
    \addplot    [mark=none, black , dashed]    table[x=Eb/N0, y=FER] {Figures/RAYLEIGH/RLC/POSD/RLC_128_32_HOSD_ML_0_0_20_MQ_1024_SIM_mHW_32_RAYLEIGH.res}; \label{gp:RAYLEIGH_HOSD_ML_perf_5}
       \addplot[mark=triangle  , Paired-6, dashed]  table[x=Eb/N0, y=FER] {Figures/RAYLEIGH/RLC/POSD/RLC_128_32_HOSD_HW_0_0_20_MQ_1024_SIM_mHW_32_RAYLEIGH.res} ;\label{gp:RAYLEIGH_HOSD_HW_perf_5}
         \addplot[mark=otimes  , Paired-9, dashed]  table[x=Eb/N0, y=FER] {Figures/RAYLEIGH/RLC/POSD/RLC_128_32_HOSD_LW_0_0_20_MQ_1024_SIM_mHW_32_RAYLEIGH.res} ;\label{gp:RAYLEIGH_HOSD_LW_perf_5}
     \addplot[mark=-, cyan , dashed]  table[x=Eb/N0, y=FER] {Figures/RAYLEIGH/RLC/POSD/RLC_128_32_HOSD_ILW_0_0_20_MQ_1024_SIM_mHW_32_RAYLEIGH.res};\label{gp:RAYLEIGH_HOSD_ILW_perf_5}
          \addplot[mark=none, red , dashed]  table[x=Eb/N0, y=FER] {Figures/RAYLEIGH/RLC/POSD/RLC_128_32_HOSD_LUT_0_0_20_MQ_1024_SIM_mHW_32_RAYLEIGH.res};\label{gp:RAYLEIGH_HOSD_LUT_perf_5}
      \coordinate (top) at (rel axis cs:0,1);
      \coordinate (spypoint1) at (axis cs:7.45,2e-7);
      \coordinate (magnifyglass1) at (axis cs:2.6,1.1e-5);
      \coordinate (bot) at (rel axis cs:1,0);
    \nextgroupplot[title=OSD BCH {[127,64]},ylabel= FER, ytick pos=left, y label style={at={(axis description cs:-0.225,.5)},anchor=south},ymin=1e-6, ymax = 1, xmin=1, xmax=7]
    \addplot[mark=square, Paired-3 , semithick]  table[x=Eb/N0, y=FER] {Figures/RAYLEIGH/BCH/OSD/BCH_127_64_OSD_EW_0_0_20_MQ_10000_SIM_mHW_127_RAYLEIGH.res}; \label{gp:RAYLEIGH_OSD_EW_perf}
       \addplot[mark=triangle, Paired-6, semithick]  table[x=Eb/N0, y=FER] {Figures/RAYLEIGH/BCH/OSD/BCH_127_64_OSD_HW_0_0_20_MQ_10000_SIM_mHW_127_RAYLEIGH.res} ;\label{gp:RAYLEIGH_OSD_HW_perf}
         \addplot[mark=otimes, Paired-9, semithick]  table[x=Eb/N0, y=FER] {Figures/RAYLEIGH/BCH/OSD/BCH_127_64_OSD_LW_0_0_20_MQ_10000_SIM_mHW_127_RAYLEIGH.res} ;\label{gp:RAYLEIGH_OSD_LW_perf}
     \addplot[mark=-, cyan , semithick]  table[x=Eb/N0, y=FER] {Figures/RAYLEIGH/BCH/OSD/BCH_127_64_OSD_ILW_0_0_20_MQ_10000_SIM_mHW_127_RAYLEIGH.res};\label{gp:RAYLEIGH_OSD_ILW_perf}
     \addplot[mark=none, black , semithick]  table[x=Eb/N0, y=FER] {Figures/RAYLEIGH/BCH/OSD/BCH_127_64_OSD_ML_0_0_20_MQ_10000_SIM_mHW_64_RAYLEIGH.res};\label{gp:RAYLEIGH_OSD_ML_perf}
         \addplot[mark=square, Paired-3 , dashed]  table[x=Eb/N0, y=FER] {Figures/RAYLEIGH/BCH/OSD/BCH_127_64_OSD_EW_0_0_20_MQ_1024_SIM_mHW_64_RAYLEIGH.res}; \label{gp:RAYLEIGH_OSD_EW_perf_4}
       \addplot[mark=triangle, Paired-6, dashed]  table[x=Eb/N0, y=FER] {Figures/RAYLEIGH/BCH/OSD/BCH_127_64_OSD_HW_0_0_20_MQ_1024_SIM_mHW_64_RAYLEIGH.res} ;\label{gp:RAYLEIGH_OSD_HW_perf_4}
         \addplot[mark=otimes, Paired-9, dashed]  table[x=Eb/N0, y=FER] {Figures/RAYLEIGH/BCH/OSD/BCH_127_64_OSD_LW_0_0_20_MQ_1024_SIM_mHW_64_RAYLEIGH.res} ;\label{gp:RAYLEIGH_OSD_LW_perf_4}
     \addplot[mark=-, cyan , dashed]  table[x=Eb/N0, y=FER] {Figures/RAYLEIGH/BCH/OSD/BCH_127_64_OSD_ILW_0_0_20_MQ_1024_SIM_mHW_64_RAYLEIGH.res};\label{gp:RAYLEIGH_OSD_ILW_perf_4}
     \addplot[mark=none, black , dashed]  table[x=Eb/N0, y=FER] {Figures/RAYLEIGH/BCH/OSD/BCH_127_64_OSD_ML_0_0_20_MQ_1024_SIM_mHW_64_RAYLEIGH.res};\label{gp:RAYLEIGH_OSD_ML_perf_4}
          \addplot[mark=none, red , dashed]  table[x=Eb/N0, y=FER] {Figures/RAYLEIGH/BCH/OSD/BCH_127_64_OSD_LUT_0_0_20_MQ_1024_SIM_mHW_64_RAYLEIGH.res};\label{gp:RAYLEIGH_OSD_LUT_perf_4}
         \nextgroupplot[title=OSD RLC {[128,32]}, ytick pos=left, y label style={at={(axis description cs:-0.225,.5)},anchor=south},ymin=1e-6, ymax = 1, xmin=1, xmax=7]
    \addplot[mark=square, Paired-3 , semithick]  table[x=Eb/N0, y=FER] {Figures/RAYLEIGH/RLC/OSD/RLC_128_32_OSD_EW_0_0_20_MQ_10000_SIM_mHW_32_RAYLEIGH.res}; \label{gp:RAYLEIGH_OSD_EW_perf_2}
       \addplot[mark=triangle, Paired-6, semithick]  table[x=Eb/N0, y=FER] {Figures/RAYLEIGH/RLC/OSD/RLC_128_32_OSD_HW_0_0_20_MQ_10000_SIM_mHW_32_RAYLEIGH.res} ;\label{gp:RAYLEIGH_OSD_HW_perf_2}
         \addplot[mark=otimes, Paired-9, semithick]  table[x=Eb/N0, y=FER] {Figures/RAYLEIGH/RLC/OSD/RLC_128_32_OSD_LW_0_0_20_MQ_10000_SIM_mHW_32_RAYLEIGH.res} ;\label{gp:RAYLEIGH_OSD_LW_perf_2}
     \addplot[mark=-, cyan , semithick]  table[x=Eb/N0, y=FER] {Figures/RAYLEIGH/RLC/OSD/RLC_128_32_OSD_ILW_0_0_20_MQ_10000_SIM_mHW_32_RAYLEIGH.res};\label{gp:RAYLEIGH_OSD_ILW_perf_2}
     \addplot[mark=none, black , semithick]  table[x=Eb/N0, y=FER] {Figures/RAYLEIGH/RLC/OSD/RLC_128_32_OSD_ML_0_0_20_MQ_10000_SIM_mHW_32_RAYLEIGH.res};\label{gp:RAYLEIGH_OSD_ML_perf_2}
      \addplot[mark=square, Paired-3 , dashed]  table[x=Eb/N0, y=FER] {Figures/RAYLEIGH/RLC/OSD/RLC_128_32_OSD_EW_0_0_20_MQ_1024_SIM_mHW_32_RAYLEIGH.res}; \label{gp:RAYLEIGH_OSD_EW_perf_5}
       \addplot[mark=triangle, Paired-6, dashed]  table[x=Eb/N0, y=FER] {Figures/RAYLEIGH/RLC/OSD/RLC_128_32_OSD_HW_0_0_20_MQ_1024_SIM_mHW_32_RAYLEIGH.res} ;\label{gp:RAYLEIGH_OSD_HW_perf_5}
         \addplot[mark=otimes, Paired-9, dashed]  table[x=Eb/N0, y=FER] {Figures/RAYLEIGH/RLC/OSD/RLC_128_32_OSD_LW_0_0_20_MQ_1024_SIM_mHW_32_RAYLEIGH.res} ;\label{gp:RAYLEIGH_OSD_LW_perf_5}
     \addplot[mark=-, cyan , dashed]  table[x=Eb/N0, y=FER] {Figures/RAYLEIGH/RLC/OSD/RLC_128_32_OSD_ILW_0_0_20_MQ_1024_SIM_mHW_32_RAYLEIGH.res};\label{gp:RAYLEIGH_OSD_ILW_perf_5}
     \addplot[mark=none, black , dashed]  table[x=Eb/N0, y=FER] {Figures/RAYLEIGH/RLC/OSD/RLC_128_32_OSD_ML_0_0_20_MQ_1024_SIM_mHW_32_RAYLEIGH.res};\label{gp:RAYLEIGH_OSD_ML_perf_5}
          \addplot[mark=none, red , dashed]  table[x=Eb/N0, y=FER] {Figures/RAYLEIGH/RLC/OSD/RLC_128_32_OSD_LUT_0_0_20_MQ_1024_SIM_mHW_32_RAYLEIGH.res};\label{gp:RAYLEIGH_OSD_LUT_perf_5}
    \end{groupplot}
     \path (current bounding box.north west) -- coordinate (legendpos)
      (current bounding box.north east);
    \matrix[
    matrix of nodes,
    anchor=south,
    draw,
    inner sep=0.2em,
    draw
    ]at(legendpos) 
    {
    \ref{gp:RAYLEIGH_GRAND_EW_perf}&  EW &[1pt] 
    \ref{gp:RAYLEIGH_GRAND_LW_perf}&   LW &[1pt] \ref{gp:RAYLEIGH_GRAND_HW_perf}&   HW &[1pt] \ref{gp:RAYLEIGH_GRAND_ILW_perf}&   ILW &[1pt] \ref{gp:RAYLEIGH_GRAND_ML_perf}&  EW &[1pt] \\
      };
  \end{tikzpicture}
  \caption{Comparison of decoding performance of different error patterns with GRAND, POSD and OSD on a Rayleigh fading channel with perfect knowledge of \ac{CSI} .{ Straight lines represent simulations run with $10^6$, $10^5$ and $10^4$ maximum number of queries with GRAND, POSD and OSD respectively. Dashed lines represent simulations run with a maximum number of queries of $10^3$.}}\label{fig:performance_RAYLEIGH}
\end{figure}
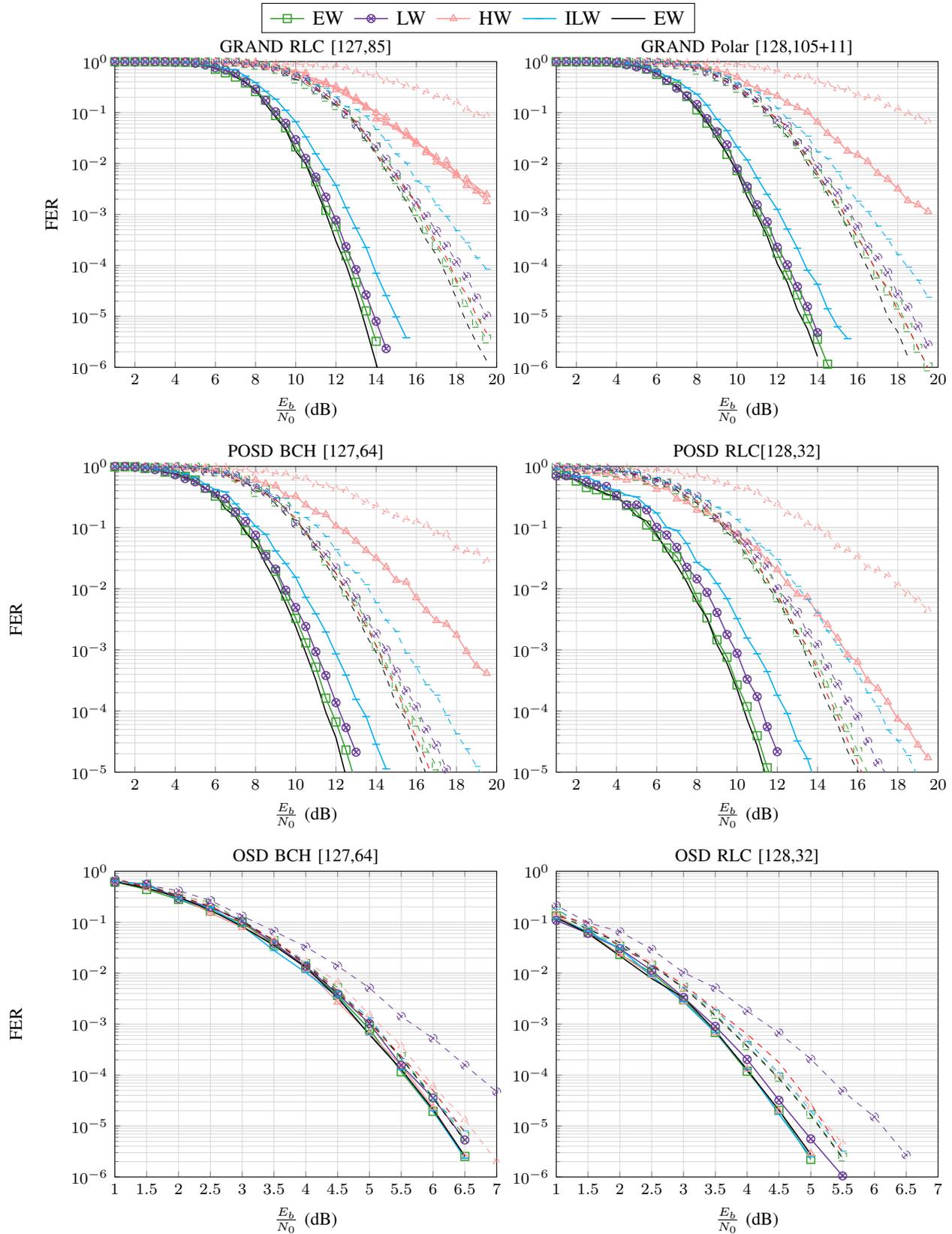
    \begin{equation}
        f_{\mathbf{L}}(l|c=0)=\frac{\sigma}{\sqrt{2\pi}}   e^{\frac{-l(\sqrt{2\sigma^2+1}-1)}{2}} \int_0^\infty  e^{-\frac{\left(\frac{\sigma^2}{2h}l-h\sqrt{2\sigma^2+1}\right)^2}{2\sigma^2}} dh.
    \end{equation}
Hence, the \ac{PDF} of the reliabilities of the bits can be calculated for $l\geq 0$ as:
\begin{align}
    f_{|\mathbf{L}|}(l)=&\frac{\sigma}{\sqrt{2\pi}}   e^{\frac{-l(\sqrt{2\sigma^2+1}-1)}{2}} \int_0^\infty  e^{-\frac{\left(\frac{\sigma^2}{2h}l-h\sqrt{2\sigma^2+1}\right)^2}{2\sigma^2}} dh +\frac{\sigma}{\sqrt{2\pi}}   e^{\frac{l(\sqrt{2\sigma^2+1}-1)}{2}} \int_0^\infty  e^{-\frac{\left(\frac{\sigma^2}{2h}l+h\sqrt{2\sigma^2+1}\right)^2}{2\sigma^2}} dh.
\end{align}
We use the same remaining steps as described in Section \ref{sec:tep_gen_llr} to generate the \ac{TEP}s for this channel.

\subsubsection{Decoding Performance Comparison}

 Fig. \ref{fig:performance_RAYLEIGH} shows the performance of \ac{GRAND}, \ac{POSD} and \ac{OSD} with the proposed \ac{EW}, and the existing \ac{LW}, \ac{HW}, \ac{ILW} and \ac{ML} \ac{TEP}s. We discuss the performance of each of these decoders with different \ac{TEP}s and varying codes.

\textit{\ac{GRAND}:}
We simulate GRAND on RLC [127,85] and 5G \ac{CA}-Polar code [128,105+11]. For a maximum number of queries of $10^6$, we can observe that the \ac{EW} error patterns result in a $0.1 \rightarrow 0.3$ dB gain compared to LW error patterns and $1.3$ dB gain compared to ILW error patterns at target FER $10^{-5}$. Using the \ac{ML} \ac{TEP}s result in $0.3$ dB performance gain compared to the \ac{EW} error patterns with RLC [127,85] code and a $0.1$ dB performance gain compared to the \ac{EW} error patterns with CA-Polar [128,105+11] code. \ac{HW}
\ac{TEP}s do not reach the target FER $10^{-5}$ within the simulated range of \ac{SNR}s.
{For a maximum number of queries of $10^3$ we observe the same trend of decoding performance of the EW, LW, ILW and the HW error patterns as what we observed with $MQ=10^6$. LUT TEPs perform similarly to the EW TEPs with $MQ=10^3$ and RLC [127,85] and Polar [128,105+11].}

 \textit{\ac{POSD}:}
We simulate POSD on BCH [127,64] and RLC code [128,32]. For a maximum number of queries of $10^5$, we can observe that the \ac{EW} error patterns result in $0.4 \rightarrow 0.6$ dB gain in performance compared to LW TEPs at target FER $10^{-4}$. Additionally, we can observe a $1.5 \rightarrow 2$ dB gain compared to \ac{ILW} error patterns at target FER $10^{-4}$. Using the \ac{ML} \ac{TEP}s results in $0.1\rightarrow 0.3$ dB performance gain compared to the \ac{EW} error patterns . \ac{HW}
\ac{TEP}s also achieve poor performance and they do not reach target FER $10^{-4}$ in the simulated SNR range.
{For a maximum number of queries of $10^3$ we observe the same pattern in decoding performance of the EW, LW, ILW and HW error patterns compared to the $MQ=10^5$ case. A wider performance loss is achieved using ILW TEPs with $MQ=10^3$ of $\approx 2.5$ dB at target FER $10^{-4}$ compared to the $1.5 \rightarrow 2$ dB loss in performance at $MQ=10^5$. LUT TEPs outperform EW TEPs slightly (by less than $0.1$ dB) at target FER $10^{-4}$ with $MQ=10^3$.}

 \textit{\ac{OSD}:}
We simulate OSD on BCH [127,64] and RLC code [128,32]. For a maximum number of queries of $10^4$, we can observe that the \ac{EW}, \ac{HW}, \ac{ML}, \ac{ILW} \ac{TEP}s result in the same performance. \ac{LW}
\ac{TEP}s result in a performance loss of $\approx0.1$ dB compared to the EW error patterns at target FER $10^{-5}$. {For a maximum number of queries of $10^3$ we observe that the ILW, ML and EW TEP performance is approximately the same. At target FER $10^{-5}$ we observe that the HW TEPs attain a $0.1$ dB to $0.2$ dB loss compared to the ILW, EW and ML. We also observe that the LUT TEPs achieve almost the same performance as the EW and ML TEPs except with RLC [128,32] at target FER $10^{-5}$ where they $\approx 0.1$ dB loss compared to EW and ML TEPs.}

\subsection{No Channel Side Information (NCSI)}
In the absence of CSI, the LLR of this channel can be approximated as \cite{hou_performance_2001}:
\begin{equation}
    L=\frac{2}{\sigma^2}y  \times E[h]
\end{equation} with the \ac{PDF} of the LLR \cite{hou_performance_2001}:

    \begin{equation}
          f_{\mathbf{L}}(l|c=0)=  \frac{\sigma \Delta^2}{2E[h]}   e^{\frac{-\Delta^2 \sigma^2 l^2}{4E[h]^2}} \times \left[ \sqrt{\frac{2}{\pi}}e^{-\frac{\Delta^2 l^2}{8E[h]^2}}  + \frac{\Delta l}{E[h]}Q\left(\frac{-\Delta l}{2E[h]}\right)\right],
    \end{equation}
where $\Delta =\sqrt{\frac{\sigma^2}{2\sigma^2 +1}} ,~E[h]=0.8862,~ Q(x)=\frac{1}{2} erfc(\frac{x}{\sqrt{2}})$.
The same error generation procedure is followed as was done in the case of ideal \ac{CSI}.

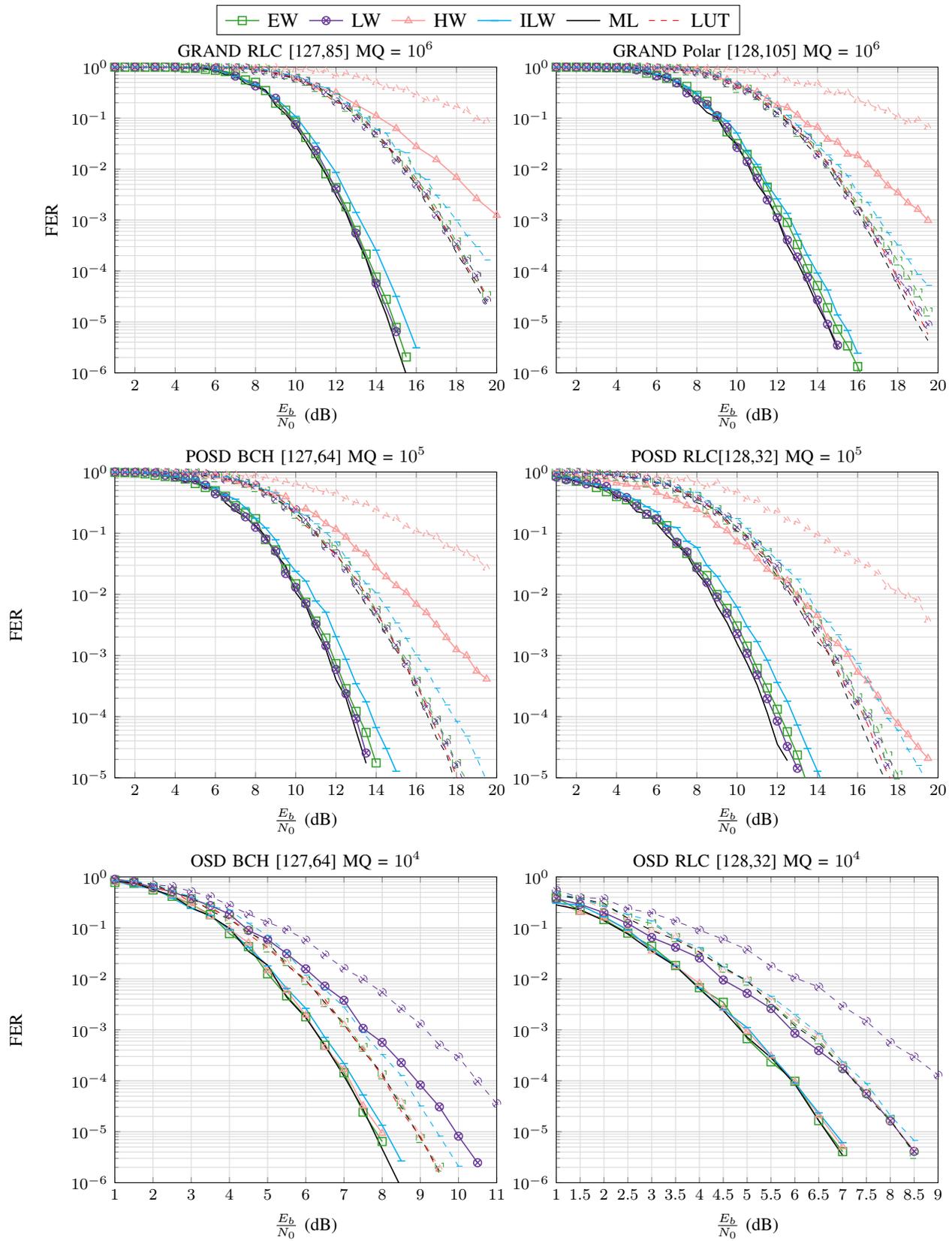
\begin{figure}
  \centering
  \begin{tikzpicture}[]
    \begin{groupplot}[group style={group name=fer_queries, group size= 2 by 3, horizontal sep=30pt, vertical sep=50pt},
      footnotesize,
      height=7cm,  width=0.46\columnwidth,  
      xlabel=$\frac{E_b}{N_0}$ (dB),
      ymode=log,
      tick align=inside,
      grid=both, grid style={gray!30},
      /pgfplots/table/ignore chars={|},
      ] 
      \nextgroupplot[title=GRAND RLC {[127,85]} MQ {= $10^6$},ylabel= FER, ytick pos=left,ymin=1e-6, ymax = 1, xmin=1, xmax=20]
    \addplot[mark=square, Paired-3 , semithick]  table[x=Eb/N0, y=FER] {Figures/Rayleigh_no_CSI/RLC/GRAND/RLC_127_85_GRAND_EW_0_1_20_MQ_1000000_SIM_mHW_127_RAYLEIGH_NO_CSI.res}; \label{gp:RAYLEIGH_NO_CSI_GRAND_EW_perf}
       \addplot[mark=triangle  , Paired-6, semithick]  table[x=Eb/N0, y=FER] {Figures/Rayleigh_no_CSI/RLC/GRAND/RLC_127_85_GRAND_HW_0_1_20_MQ_1000000_SIM_mHW_127_RAYLEIGH_NO_CSI.res} ;\label{gp:RAYLEIGH_NO_CSI_GRAND_HW_perf}
         \addplot[mark=otimes  , Paired-9, semithick]  table[x=Eb/N0, y=FER] {Figures/Rayleigh_no_CSI/RLC/GRAND/RLC_127_85_GRAND_LW_0_1_20_MQ_1000000_SIM_mHW_127_RAYLEIGH_NO_CSI.res} ;\label{gp:RAYLEIGH_NO_CSI_GRAND_LW_perf}
     \addplot[mark=-, cyan , semithick]  table[x=Eb/N0, y=FER] {Figures/Rayleigh_no_CSI/RLC/GRAND/RLC_127_85_GRAND_ILW_0_1_20_MQ_1000000_SIM_mHW_127_RAYLEIGH_NO_CSI.res};\label{gp:RAYLEIGH_NO_CSI_GRAND_ILW_perf}
         \addplot[mark=none, black , semithick]  table[x=Eb/N0, y=FER] {Figures/Rayleigh_no_CSI/RLC/GRAND/RLC_127_85_GRAND_ML_0_0_20_MQ_1000000_SIM_mHW_127_RAYLEIGH_NO_CSI.res};\label{gp:RAYLEIGH_NO_CSI_GRAND_ML_perf}
             \addplot[mark=square, Paired-3 , dashed]  table[x=Eb/N0, y=FER] {Figures/Rayleigh_no_CSI/RLC/GRAND/RLC_127_85_GRAND_EW_0_0_20_MQ_1024_SIM_mHW_85_RAYLEIGH_NO_CSI.res}; \label{gp:RAYLEIGH_NO_CSI_GRAND_EW_perf_4}
       \addplot[mark=triangle  , Paired-6, dashed]  table[x=Eb/N0, y=FER] {Figures/Rayleigh_no_CSI/RLC/GRAND/RLC_127_85_GRAND_HW_0_0_20_MQ_1024_SIM_mHW_85_RAYLEIGH_NO_CSI.res} ;\label{gp:RAYLEIGH_NO_CSI_GRAND_HW_perf_4}
         \addplot[mark=otimes  , Paired-9, dashed]  table[x=Eb/N0, y=FER] {Figures/Rayleigh_no_CSI/RLC/GRAND/RLC_127_85_GRAND_LW_0_0_20_MQ_1024_SIM_mHW_85_RAYLEIGH_NO_CSI.res} ;\label{gp:RAYLEIGH_NO_CSI_GRAND_LW_perf_4}
     \addplot[mark=-, cyan , dashed]  table[x=Eb/N0, y=FER] {Figures/Rayleigh_no_CSI/RLC/GRAND/RLC_127_85_GRAND_ILW_0_0_20_MQ_1024_SIM_mHW_85_RAYLEIGH_NO_CSI.res};\label{gp:RAYLEIGH_NO_CSI_GRAND_ILW_perf_4}
         \addplot[mark=none, black , dashed]  table[x=Eb/N0, y=FER] {Figures/Rayleigh_no_CSI/RLC/GRAND/RLC_127_85_GRAND_ML_0_0_20_MQ_1024_SIM_mHW_85_RAYLEIGH_NO_CSI.res};\label{gp:RAYLEIGH_NO_CSI_GRAND_ML_perf_4}
            \addplot[mark=none, red , dashed]  table[x=Eb/N0, y=FER] {Figures/Rayleigh_no_CSI/RLC/GRAND/RLC_127_85_GRAND_LUT_0_0_20_MQ_1024_SIM_mHW_85_RAYLEIGH_NO_CSI.res};\label{gp:RAYLEIGH_NO_CSI_GRAND_LUT_perf_4}
     \nextgroupplot[title=GRAND Polar {[128,105]} MQ {= $10^6$}, ytick pos=left,ymin=1e-6, ymax = 1, xmin=1, xmax=20]
    \addplot[mark=square, Paired-3 , semithick]  table[x=Eb/N0, y=FER] {Figures/Rayleigh_no_CSI/Polar/GRAND/Polar_128_105_GRAND_EW_0_0_20_MQ_1000000_SIM_mHW_105_RAYLEIGH_NO_CSI.res}; \label{gp:RAYLEIGH_NO_CSI_GRAND_EW_perf_2}
       \addplot[mark=triangle  , Paired-6, semithick]  table[x=Eb/N0, y=FER] {Figures/Rayleigh_no_CSI/Polar/GRAND/Polar_128_105_GRAND_HW_0_0_20_MQ_1000000_SIM_mHW_105_RAYLEIGH_NO_CSI.res} ;\label{gp:RAYLEIGH_NO_CSI_GRAND_HW_perf_2}
         \addplot[mark=otimes  , Paired-9, semithick]  table[x=Eb/N0, y=FER] {Figures/Rayleigh_no_CSI/Polar/GRAND/Polar_128_105_GRAND_LW_0_0_20_MQ_1000000_SIM_mHW_105_RAYLEIGH_NO_CSI.res} ;\label{gp:RAYLEIGH_NO_CSI_GRAND_LW_perf_2}
     \addplot[mark=-, cyan , semithick]  table[x=Eb/N0, y=FER] {Figures/Rayleigh_no_CSI/Polar/GRAND/Polar_128_105_GRAND_ILW_0_0_20_MQ_1000000_SIM_mHW_105_RAYLEIGH_NO_CSI.res};\label{gp:RAYLEIGH_NO_CSI_GRAND_ILW_perf_2}
          \addplot[mark=none, black , semithick]  table[x=Eb/N0, y=FER] {Figures/Rayleigh_no_CSI/Polar/GRAND/Polar_128_105_GRAND_ML_0_0_20_MQ_1000000_SIM_mHW_105_RAYLEIGH_NO_CSI.res};\label{gp:RAYLEIGH_NO_CSI_GRAND_ML_perf_2}
             \addplot[mark=square, Paired-3 , dashed]  table[x=Eb/N0, y=FER] {Figures/Rayleigh_no_CSI/Polar/GRAND/Polar_128_105_GRAND_EW_0_0_20_MQ_1024_SIM_mHW_105_RAYLEIGH_NO_CSI.res}; \label{gp:RAYLEIGH_NO_CSI_GRAND_EW_perf_5}
       \addplot[mark=triangle  , Paired-6, dashed]  table[x=Eb/N0, y=FER] {Figures/Rayleigh_no_CSI/Polar/GRAND/Polar_128_105_GRAND_HW_0_0_20_MQ_1024_SIM_mHW_105_RAYLEIGH_NO_CSI.res} ;\label{gp:RAYLEIGH_NO_CSI_GRAND_HW_perf_5}
         \addplot[mark=otimes  , Paired-9, dashed]  table[x=Eb/N0, y=FER] {Figures/Rayleigh_no_CSI/Polar/GRAND/Polar_128_105_GRAND_LW_0_0_20_MQ_1024_SIM_mHW_105_RAYLEIGH_NO_CSI.res} ;\label{gp:RAYLEIGH_NO_CSI_GRAND_LW_perf_5}
     \addplot[mark=-, cyan , dashed]  table[x=Eb/N0, y=FER] {Figures/Rayleigh_no_CSI/Polar/GRAND/Polar_128_105_GRAND_ILW_0_0_20_MQ_1024_SIM_mHW_105_RAYLEIGH_NO_CSI.res};\label{gp:RAYLEIGH_NO_CSI_GRAND_ILW_perf_5}
          \addplot[mark=none, black , dashed]  table[x=Eb/N0, y=FER] {Figures/Rayleigh_no_CSI/Polar/GRAND/Polar_128_105_GRAND_ML_0_0_20_MQ_1024_SIM_mHW_105_RAYLEIGH_NO_CSI.res};\label{gp:RAYLEIGH_NO_CSI_GRAND_ML_perf_5}
            \addplot[mark=none, red , dashed]  table[x=Eb/N0, y=FER] {Figures/Rayleigh_no_CSI/Polar/GRAND/Polar_128_105_GRAND_LUT_0_0_20_MQ_1024_SIM_mHW_105_RAYLEIGH_NO_CSI.res};\label{gp:RAYLEIGH_NO_CSI_GRAND_LUT_perf_5}
            \nextgroupplot[ title=POSD BCH {[127,64]} MQ {= $10^5$},ylabel=FER, ytick pos=left, y label style={at={(axis description cs:-0.225,.5)},anchor=south},ymin=1e-5, ymax = 1, xmin=1, xmax=20]
    \addplot[mark=square, Paired-3 , semithick]  table[x=Eb/N0, y=FER] {Figures/Rayleigh_no_CSI/BCH/POSD/BCH_127_64_HOSD_EW_0_0_20_MQ_100000_SIM_mHW_127_RAYLEIGH_NO_CSI.res}; \label{gp:RAYLEIGH_NO_CSI_HOSD_EW_perf}
        \addplot[mark=none, black , semithick]  table[x=Eb/N0, y=FER] {Figures/Rayleigh_no_CSI/BCH/POSD/BCH_127_64_HOSD_ML_0_0_20_MQ_100000_SIM_mHW_64_RAYLEIGH_NO_CSI.res}; \label{gp:RAYLEIGH_NO_CSI_HOSD_ML_perf}
       \addplot[mark=triangle  , Paired-6, semithick]  table[x=Eb/N0, y=FER] {Figures/Rayleigh_no_CSI/BCH/POSD/BCH_127_64_HOSD_HW_0_0_20_MQ_100000_SIM_mHW_127_RAYLEIGH_NO_CSI.res} ;\label{gp:RAYLEIGH_NO_CSI_HOSD_HW_perf}
         \addplot[mark=otimes  , Paired-9, semithick]  table[x=Eb/N0, y=FER] {Figures/Rayleigh_no_CSI/BCH/POSD/BCH_127_64_HOSD_LW_0_0_20_MQ_100000_SIM_mHW_127_RAYLEIGH_NO_CSI.res} ;\label{gp:RAYLEIGH_NO_CSI_HOSD_LW_perf}
     \addplot[mark=-, cyan , semithick]  table[x=Eb/N0, y=FER] {Figures/Rayleigh_no_CSI/BCH/POSD/BCH_127_64_HOSD_ILW_0_0_20_MQ_100000_SIM_mHW_127_RAYLEIGH_NO_CSI.res};\label{gp:RAYLEIGH_NO_CSI_HOSD_ILW_perf}
      \addplot[mark=square, Paired-3 , dashed]  table[x=Eb/N0, y=FER] {Figures/Rayleigh_no_CSI/BCH/POSD/BCH_127_64_HOSD_EW_0_0_20_MQ_1024_SIM_mHW_64_RAYLEIGH_NO_CSI.res}; \label{gp:RAYLEIGH_NO_CSI_HOSD_EW_perf_4}
        \addplot[mark=none, black , dashed]  table[x=Eb/N0, y=FER] {Figures/Rayleigh_no_CSI/BCH/POSD/BCH_127_64_HOSD_ML_0_0_20_MQ_1024_SIM_mHW_64_RAYLEIGH_NO_CSI.res}; \label{gp:RAYLEIGH_NO_CSI_HOSD_ML_perf_4}
       \addplot[mark=triangle  , Paired-6, dashed]  table[x=Eb/N0, y=FER] {Figures/Rayleigh_no_CSI/BCH/POSD/BCH_127_64_HOSD_HW_0_0_20_MQ_1024_SIM_mHW_64_RAYLEIGH_NO_CSI.res} ;\label{gp:RAYLEIGH_NO_CSI_HOSD_HW_perf_4}
         \addplot[mark=otimes  , Paired-9, dashed]  table[x=Eb/N0, y=FER] {Figures/Rayleigh_no_CSI/BCH/POSD/BCH_127_64_HOSD_LW_0_0_20_MQ_1024_SIM_mHW_64_RAYLEIGH_NO_CSI.res} ;\label{gp:RAYLEIGH_NO_CSI_HOSD_LW_perf_4}
     \addplot[mark=-, cyan , dashed]  table[x=Eb/N0, y=FER] {Figures/Rayleigh_no_CSI/BCH/POSD/BCH_127_64_HOSD_ILW_0_0_20_MQ_1024_SIM_mHW_64_RAYLEIGH_NO_CSI.res};\label{gp:RAYLEIGH_NO_CSI_HOSD_ILW_perf_4}
         \addplot[mark=none, red , dashed]  table[x=Eb/N0, y=FER] {Figures/Rayleigh_no_CSI/BCH/POSD/BCH_127_64_HOSD_LUT_0_0_20_MQ_1024_SIM_mHW_64_RAYLEIGH_NO_CSI.res};\label{gp:RAYLEIGH_NO_CSI_HOSD_LUT_perf_4}
       \nextgroupplot[ title=POSD RLC{[128,32]} MQ {= $10^5$}, ytick pos=left, y label style={at={(axis description cs:-0.225,.5)},anchor=south},ymin=1e-5, ymax = 1, xmin=1, xmax=20]
    \addplot[mark=square, Paired-3 , semithick]  table[x=Eb/N0, y=FER] {Figures/Rayleigh_no_CSI/RLC/POSD/RLC_128_32_HOSD_EW_0_0_20_MQ_100000_SIM_mHW_32_RAYLEIGH_NO_CSI.res}; \label{gp:RAYLEIGH_NO_CSI_HOSD_EW_perf_2}
       \addplot[mark=triangle  , Paired-6, semithick]  table[x=Eb/N0, y=FER] {Figures/Rayleigh_no_CSI/RLC/POSD/RLC_128_32_HOSD_HW_0_0_20_MQ_100000_SIM_mHW_32_RAYLEIGH_NO_CSI.res} ;\label{gp:RAYLEIGH_NO_CSI_HOSD_HW_perf_2}
         \addplot[mark=otimes  , Paired-9, semithick]  table[x=Eb/N0, y=FER] {Figures/Rayleigh_no_CSI/RLC/POSD/RLC_128_32_HOSD_LW_0_0_20_MQ_100000_SIM_mHW_32_RAYLEIGH_NO_CSI.res} ;\label{gp:RAYLEIGH_NO_CSI_HOSD_LW_perf_2}
     \addplot[mark=-, cyan , semithick]  table[x=Eb/N0, y=FER] {Figures/Rayleigh_no_CSI/RLC/POSD/RLC_128_32_HOSD_ILW_0_0_20_MQ_100000_SIM_mHW_32_RAYLEIGH_NO_CSI.res};\label{gp:RAYLEIGH_NO_CSI_HOSD_ILW_perf_2}     
          \addplot[mark=none, black , semithick]  table[x=Eb/N0, y=FER] {Figures/Rayleigh_no_CSI/RLC/POSD/RLC_128_32_HOSD_ML_0_0_20_MQ_100000_SIM_mHW_32_RAYLEIGH_NO_CSI.res};\label{gp:RAYLEIGH_NO_CSI_HOSD_ML_perf_2}  
    \addplot[mark=square, Paired-3 , dashed]  table[x=Eb/N0, y=FER] {Figures/Rayleigh_no_CSI/RLC/POSD/RLC_128_32_HOSD_EW_0_0_20_MQ_1024_SIM_mHW_32_RAYLEIGH_NO_CSI.res}; \label{gp:RAYLEIGH_NO_CSI_HOSD_EW_perf_5}
       \addplot[mark=triangle  , Paired-6, dashed]  table[x=Eb/N0, y=FER] {Figures/Rayleigh_no_CSI/RLC/POSD/RLC_128_32_HOSD_HW_0_0_20_MQ_1024_SIM_mHW_32_RAYLEIGH_NO_CSI.res} ;\label{gp:RAYLEIGH_NO_CSI_HOSD_HW_perf_5}
         \addplot[mark=otimes  , Paired-9, dashed]  table[x=Eb/N0, y=FER] {Figures/Rayleigh_no_CSI/RLC/POSD/RLC_128_32_HOSD_LW_0_0_20_MQ_1024_SIM_mHW_32_RAYLEIGH_NO_CSI.res} ;\label{gp:RAYLEIGH_NO_CSI_HOSD_LW_perf_5}
     \addplot[mark=-, cyan , dashed]  table[x=Eb/N0, y=FER] {Figures/Rayleigh_no_CSI/RLC/POSD/RLC_128_32_HOSD_ILW_0_0_20_MQ_1024_SIM_mHW_32_RAYLEIGH_NO_CSI.res};\label{gp:RAYLEIGH_NO_CSI_HOSD_ILW_perf_5}     
          \addplot[mark=none, black , dashed]  table[x=Eb/N0, y=FER] {Figures/Rayleigh_no_CSI/RLC/POSD/RLC_128_32_HOSD_ML_0_0_20_MQ_1024_SIM_mHW_32_RAYLEIGH_NO_CSI.res};\label{gp:RAYLEIGH_NO_CSI_HOSD_ML_perf_5}  
                \addplot[mark=none, red , dashed]  table[x=Eb/N0, y=FER] {Figures/Rayleigh_no_CSI/RLC/POSD/RLC_128_32_HOSD_LUT_0_0_20_MQ_1024_SIM_mHW_32_RAYLEIGH_NO_CSI.res};\label{gp:RAYLEIGH_NO_CSI_HOSD_LUT_perf_5}  
          \coordinate (top) at (rel axis cs:0,1);
      \coordinate (spypoint1) at (axis cs:7.45,2e-7);
      \coordinate (magnifyglass1) at (axis cs:2.6,1.1e-5);
      \coordinate (bot) at (rel axis cs:1,0);
    \nextgroupplot[title=OSD BCH {[127,64]} MQ {= $10^4$},ylabel=FER, ytick pos=left, y label style={at={(axis description cs:-0.225,.5)},anchor=south},ymin=1e-6, ymax = 1, xmin=1, xmax=11]
    \addplot[mark=square, Paired-3 , semithick]  table[x=Eb/N0, y=FER] {Figures/Rayleigh_no_CSI/BCH/OSD/BCH_127_64_OSD_EW_0_0_20_MQ_10000_SIM_mHW_127_RAYLEIGH_NO_CSI.res}; \label{gp:RAYLEIGH_NO_CSI_OSD_EW_perf}
       \addplot[mark=triangle, Paired-6, semithick]  table[x=Eb/N0, y=FER] {Figures/Rayleigh_no_CSI/BCH/OSD/BCH_127_64_OSD_HW_0_1_10_MQ_10000_SIM_mHW_64_RAYLEIGH_NO_CSI.res} ;\label{gp:RAYLEIGH_NO_CSI_OSD_HW_perf}
         \addplot[mark=otimes, Paired-9, semithick]  table[x=Eb/N0, y=FER] {Figures/Rayleigh_no_CSI/BCH/OSD/BCH_127_64_OSD_LW_0_0_20_MQ_10000_SIM_mHW_127_RAYLEIGH_NO_CSI.res} ;\label{gp:RAYLEIGH_NO_CSI_OSD_LW_perf}
     \addplot[mark=-, cyan , semithick]  table[x=Eb/N0, y=FER] {Figures/Rayleigh_no_CSI/BCH/OSD/BCH_127_64_OSD_ILW_0_0_20_MQ_10000_SIM_mHW_127_RAYLEIGH_NO_CSI.res};\label{gp:RAYLEIGH_NO_CSI_OSD_ILW_perf}
     \addplot[mark=none, black , semithick]  table[x=Eb/N0, y=FER] {Figures/Rayleigh_no_CSI/BCH/OSD/RLC_127_64_OSD_ML_0_0_20_MQ_10000_SIM_mHW_64_RAYLEIGH_NO_CSI.res};\label{gp:RAYLEIGH_NO_CSI_OSD_ML_perf}  
        \addplot[mark=square, Paired-3 , dashed]  table[x=Eb/N0, y=FER] {Figures/Rayleigh_no_CSI/BCH/OSD/BCH_127_64_OSD_EW_0_0_20_MQ_1024_SIM_mHW_64_RAYLEIGH_NO_CSI.res}; \label{gp:RAYLEIGH_NO_CSI_OSD_EW_perf_4}
       \addplot[mark=triangle, Paired-6, dashed]  table[x=Eb/N0, y=FER] {Figures/Rayleigh_no_CSI/BCH/OSD/BCH_127_64_OSD_HW_0_0_20_MQ_1024_SIM_mHW_64_RAYLEIGH_NO_CSI.res} ;\label{gp:RAYLEIGH_NO_CSI_OSD_HW_perf_4}
         \addplot[mark=otimes, Paired-9, dashed]  table[x=Eb/N0, y=FER] {Figures/Rayleigh_no_CSI/BCH/OSD/BCH_127_64_OSD_LW_0_0_20_MQ_1024_SIM_mHW_64_RAYLEIGH_NO_CSI.res} ;\label{gp:RAYLEIGH_NO_CSI_OSD_LW_perf_4}
     \addplot[mark=-, cyan , dashed]  table[x=Eb/N0, y=FER] {Figures/Rayleigh_no_CSI/BCH/OSD/BCH_127_64_OSD_ILW_0_0_20_MQ_1024_SIM_mHW_64_RAYLEIGH_NO_CSI.res};\label{gp:RAYLEIGH_NO_CSI_OSD_ILW_perf_4}
     \addplot[mark=none, black , dashed]  table[x=Eb/N0, y=FER] {Figures/Rayleigh_no_CSI/BCH/OSD/BCH_127_64_OSD_ML_0_0_20_MQ_1024_SIM_mHW_64_RAYLEIGH_NO_CSI.res};\label{gp:RAYLEIGH_NO_CSI_OSD_ML_perf_4}  
          \addplot[mark=none, red , dashed]  table[x=Eb/N0, y=FER] {Figures/Rayleigh_no_CSI/BCH/OSD/BCH_127_64_OSD_LUT_0_0_20_MQ_1024_SIM_mHW_64_RAYLEIGH_NO_CSI.res};\label{gp:RAYLEIGH_NO_CSI_OSD_LUT_perf_4}  
     \nextgroupplot[title=OSD RLC {[128,32]} MQ {= $10^4$}, ytick pos=left, y label style={at={(axis description cs:-0.225,.5)},anchor=south},ymin=1e-6, ymax = 1, xmin=1, xmax=9]
    \addplot[mark=square, Paired-3 , semithick]  table[x=Eb/N0, y=FER] {Figures/Rayleigh_no_CSI/RLC/OSD/RLC_128_32_OSD_EW_0_0_20_MQ_10000_SIM_mHW_32_RAYLEIGH_NO_CSI.res}; \label{gp:RAYLEIGH_NO_CSI_OSD_EW_perf_2}
       \addplot[mark=triangle, Paired-6, semithick]  table[x=Eb/N0, y=FER] {Figures/Rayleigh_no_CSI/RLC/OSD/RLC_128_32_OSD_HW_0_0_20_MQ_10000_SIM_mHW_32_RAYLEIGH_NO_CSI.res} ;\label{gp:RAYLEIGH_NO_CSI_OSD_HW_perf_2}
         \addplot[mark=otimes, Paired-9, semithick]  table[x=Eb/N0, y=FER] {Figures/Rayleigh_no_CSI/RLC/OSD/RLC_128_32_OSD_LW_0_0_20_MQ_10000_SIM_mHW_32_RAYLEIGH_NO_CSI.res} ;\label{gp:RAYLEIGH_NO_CSI_OSD_LW_perf_2}
     \addplot[mark=-, cyan , semithick]  table[x=Eb/N0, y=FER] {Figures/Rayleigh_no_CSI/RLC/OSD/RLC_128_32_OSD_ILW_0_0_20_MQ_10000_SIM_mHW_32_RAYLEIGH_NO_CSI.res};\label{gp:RAYLEIGH_NO_CSI_OSD_ILW_perf_2}
     \addplot[mark=none, black , semithick]  table[x=Eb/N0, y=FER] {Figures/Rayleigh_no_CSI/RLC/OSD/RLC_128_32_OSD_ML_0_0_20_MQ_10000_SIM_mHW_32_RAYLEIGH_NO_CSI.res};\label{gp:RAYLEIGH_NO_CSI_OSD_ML_perf_2}    
        \addplot[mark=square, Paired-3 , dashed]  table[x=Eb/N0, y=FER] {Figures/Rayleigh_no_CSI/RLC/OSD/RLC_128_32_OSD_EW_0_0_20_MQ_1024_SIM_mHW_32_RAYLEIGH_NO_CSI.res}; \label{gp:RAYLEIGH_NO_CSI_OSD_EW_perf_5}
       \addplot[mark=triangle, Paired-6, dashed]  table[x=Eb/N0, y=FER] {Figures/Rayleigh_no_CSI/RLC/OSD/RLC_128_32_OSD_HW_0_0_20_MQ_1024_SIM_mHW_32_RAYLEIGH_NO_CSI.res} ;\label{gp:RAYLEIGH_NO_CSI_OSD_HW_perf_5}
         \addplot[mark=otimes, Paired-9, dashed]  table[x=Eb/N0, y=FER] {Figures/Rayleigh_no_CSI/RLC/OSD/RLC_128_32_OSD_LW_0_0_20_MQ_1024_SIM_mHW_32_RAYLEIGH_NO_CSI.res} ;\label{gp:RAYLEIGH_NO_CSI_OSD_LW_perf_5}
     \addplot[mark=-, cyan , dashed]  table[x=Eb/N0, y=FER] {Figures/Rayleigh_no_CSI/RLC/OSD/RLC_128_32_OSD_ILW_0_0_20_MQ_1024_SIM_mHW_32_RAYLEIGH_NO_CSI.res};\label{gp:RAYLEIGH_NO_CSI_OSD_ILW_perf_5}
     \addplot[mark=none, black , dashed]  table[x=Eb/N0, y=FER] {Figures/Rayleigh_no_CSI/RLC/OSD/RLC_128_32_OSD_ML_0_0_20_MQ_1024_SIM_mHW_32_RAYLEIGH_NO_CSI.res};\label{gp:RAYLEIGH_NO_CSI_OSD_ML_perf_5}    
    \end{groupplot}
  \path (current bounding box.north west) -- coordinate (legendpos)
      (current bounding box.north east);
    \matrix[
    matrix of nodes,
    anchor=south,
    draw,
    inner sep=0.2em,
    draw
    ]at(legendpos) 
    {
    \ref{gp:RAYLEIGH_NO_CSI_GRAND_EW_perf}&  EW &[1pt] 
    \ref{gp:RAYLEIGH_NO_CSI_GRAND_LW_perf}&   LW &[1pt] \ref{gp:RAYLEIGH_NO_CSI_GRAND_HW_perf}&   HW &[1pt] \ref{gp:RAYLEIGH_NO_CSI_GRAND_ILW_perf}&   ILW &[1pt] \ref{gp:RAYLEIGH_NO_CSI_GRAND_ML_perf}&   ML &[1pt] \ref{gp:RAYLEIGH_NO_CSI_GRAND_LUT_perf_4}&   LUT &[1pt]  \\
      };
  \end{tikzpicture}
  \caption{Comparison of decoding performance of different error patterns with GRAND, OSD and POSD on a Rayleigh fading channel with no perfect knowledge of channel state information. { Straight lines represent simulations run with $10^6$, $10^5$ and $10^4$ maximum number of queries with GRAND, POSD and OSD respectively. Dashed lines represent simulations run with a maximum number of queries of $10^3$.}}\label{fig:performance_RAYLEIGH_NO_CSI}
\end{figure}

\subsubsection{Decoding Performance Comparison}

 Fig. \ref{fig:performance_RAYLEIGH_NO_CSI} shows the performance of \ac{GRAND}, \ac{POSD} and \ac{OSD} with the proposed \ac{EW}, and the existing \ac{LW}, \ac{HW}, \ac{ILW} and \ac{ML} \ac{TEP}s. We discuss the performance of each of these decoders with different \ac{TEP}s and varying codes.

\textit{\ac{GRAND}:}
We simulate GRAND on RLC [127,85] and 5G \ac{CA}-Polar code [128,105+11]. For a maximum number of queries of $10^6$, we can observe that the \ac{EW}, \ac{ML} and LW error patterns result in almost the same decoding performance. We can also observe that EW error patterns have a $0.8 \rightarrow1$ dB gain compared to ILW error patterns at target FER $10^{-5}$. \ac{HW}
\ac{TEP}s do not reach the target FER $10^{-5}$ within the simulated range of \ac{SNR}s.
{For $MQ=10^3$ queries, the  EW, LW, ILW, and HW TEPs exhibit the same ranking (in terms of which performs better than the other) as that observed with $MQ=10^6$. LUT TEPs perform similarly to the EW TEPs with $MQ=10^3$, with RLC [127,85] and CA-Polar [128,105+11].}

 \textit{\ac{POSD}:}
We simulate POSD on BCH [127,64] and RLC code [128,32]. For a maximum number of queries of $10^5$, we can observe that the \ac{EW} and LW error patterns result in almost the same decoding performance at target FER $10^{-4}$ with a $0\rightarrow0.4$ dB degradation in performance compared to ML TEPs. Additionally, we can observe a $0.6 \rightarrow 0.8$ dB gain using EW TEPs compared to \ac{ILW} error patterns at target FER $10^{-4}$. \ac{HW}
\ac{TEP}s also achieve poor decoding performance and they achieve at least a $5.5$ dB loss in performance at FER $10^{-4}$ compared to EW TEPs. {For $MQ=10^3$ queries, the  EW, LW, and LUT TEPs exhibit the same decoding performance. We can observe that the ML TEPs outperform the EW, LW, and LUT TEPs by at most $0.2$ dB at target FER $10^{-4}$. Meanwhile ILW TEPs exhibit a performance loss of $1$ dB to $1.5$ dB compared to the ML TEPs.}

 \textit{\ac{OSD}:}
We simulate OSD on BCH [127,64] and RLC code [128,32]. For a maximum number of queries of $10^4$, we can observe that the \ac{EW}, \ac{HW}, \ac{ML}, \ac{ILW} \ac{TEP}s result in the same performance. \ac{LW}
\ac{TEP}s result in a performance loss of $1.4 \rightarrow1.9$ dB compared to the EW error patterns at target FER $10^{-5}$.  {For a maximum number of queries of $10^3$ we observe that the LUT, HW, ML and EW TEP performance is approximately the same. At target FER $10^{-5}$ we observe that the ILW TEPs attain a $0.4$ dB to $1$ dB loss compared to the LUT, HW, EW and ML.}

\begin{figure}[t]
\begin{tikzpicture}
\begin{axis}[
    xlabel={{$MQ$}},ylabel={ Percent Overlap ($\%$)},
    ymin=0, width=\columnwidth,height=6cm,
    ybar,
    xmode=log,
    bar width=10pt, 
    xtick=data, 
    enlarge x limits=0.15, 
    legend entries={},
    legend style={draw=none, fill=none}
]
\addplot[fill=magenta, opacity=0.6] table[
    x index=0,  
    y index=6,  
    header=true] {Figures/Comp_additive_fading/fading_comp.tex};\label{plot:osd_fadingncsi_awgn}
\addplot[fill=cyan, opacity=1] table[
    x index=0,  
    y index=2,  
    header=true] {Figures/Comp_additive_fading/fading_comp.tex};\label{plot:grand_fadingncsi_awgn}
    \addplot[fill=teal, opacity=1] table[
    x index=0,  
    y index=5,  
    header=true] {Figures/Comp_additive_fading/fading_comp.tex};\label{plot:osd_fading_awgn}
\addplot[fill=Paired-6, opacity=1] table[
    x index=0,  
    y index=1,  
    header=true] {Figures/Comp_additive_fading/fading_comp.tex};\label{plot:grand_fading_awgn}
\end{axis}
    \matrix[
        matrix of nodes,
        anchor=north,
        draw,
        inner sep=0.2em
    ] at ([yshift=40pt,xshift=15pt]current bounding box.north)
    { 
\ref{plot:osd_fadingncsi_awgn}&  OSD NCSI  &[1pt] \ref{plot:grand_fadingncsi_awgn}&  GRAND/POSD NCSI &[1pt] \ref{plot:osd_fading_awgn}&   OSD CSI &[1pt]  
\ref{plot:grand_fading_awgn} &  GRAND/POSD CSI &[1pt]    \\
    };
\end{tikzpicture}
\caption{Percentage of overlap of the EW error patterns generated on an AWGN channel with different decoders and types of fading channels. EW error patterns are generated for code [127,64].}\label{fig:percent_intersection_teps_grand_fading}
\end{figure}
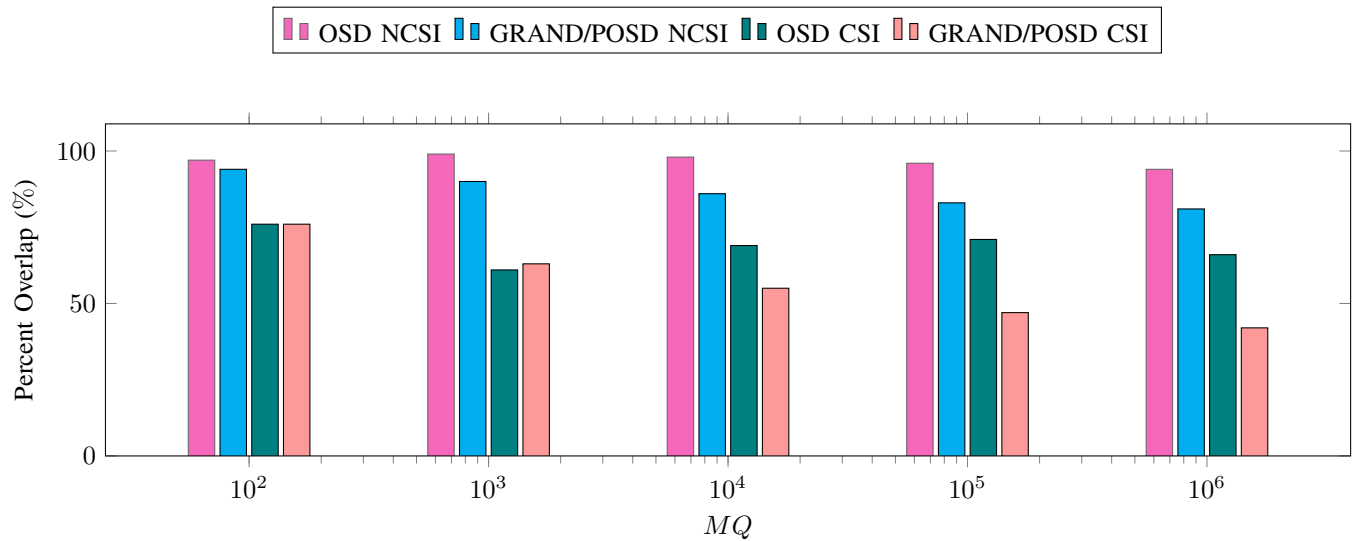

\subsection{Overlap in Error Patterns with AWGN}

Fig. \ref{fig:percent_intersection_teps_grand_fading} shows the overlap in EW TEPs generated for the AWGN model and the Rayleigh fading models. We can see that the overlap between the EW TEPs produced with Rayleigh fading channel (NCSI) and the AWGN channel is more than $75\% \rightarrow 95 \%$ across all $MQ$. This is due to the fact that the additive noise impacting the Rayleigh fading channel with no CSI is the same as that of the AWGN channel. The Rayleigh fading with ideal CSI has less common TEPs with the AWGN channel with around $40\% \rightarrow 75 \%$ overlapping TEPs with AWGN channel. This is due to the fact that the additive noise is modified before decoding, as it is divided by the fading coefficient when the fading coefficient is known. This results in a lower overlap of TEPs than the case of the NCSI fading channel.

\subsection{Conclusion}
Since GRAND, POSD and \ac{OSD} try to remove the effect of additive noise on the received signal, the error patterns used with each of these decoders with an additive Gaussian noise should not change irrespective whether there is a fading multiplicative gain. We observe that when the noise profile is the same, the EW error patterns are very similar. However, when the additive noise profile is modified, the EW TEPs change to accommodate for a different additive noise profile. We also observe that the EW TEPs can accommodate for Rayleigh fading channels with slight to no degradation in decoding performance compared to the other pre-generated TEPs.

\section{Conclusion}

In this work, we introduced two systematic frameworks that use the channel \acp{pdf} to generate channel-adaptive error patterns catered for each of GRAND, POSD and OSD. By ranking test error patterns based on their expected weighted Hamming distance, we provide a systematic alternative to Monte-Carlo simulations which are often unsuitable for high \ac{SNR} scenarios  {or hard to simulate channels. Our theoretical framework systematically allows the inclusion of rare, high-weight error patterns that are notoriously difficult to capture via empirical Monte-Carlo methods at high \ac{SNR}s. This makes the proposed method highly suitable for Ultra-Reliable Low-Latency Communication (URLLC) applications. Simulation results confirm that our proposed error patterns generally match or outperform the current state-of-the-art pre-generated error pattern on memoryless channels, including AWGN, Rayleigh fading and mixture of Gaussian channels. The strong performance of \ac{EW} patterns on all the different variations of channels and different decoder configurations demonstrate that GRAND, POSD and OSD decoders can be calibrated entirely offline for specific, non-ideal real-world noise environments without requiring Monte-Carlo simulations or complex analytical work. Future work will look to extend our proposed framework to accommodate higher-order modulation schemes and to explore TEP generation for channels with memory.}

\bibliography{references.bib}
\bibliographystyle{ieeetr}
\appendices
\section{Equivalence of the generalized weights to the weighted Hamming distance}\label{appdx:1}
We note that the notations used in this section are defined in Section \ref{sec:preliminaries}. Definition \ref{sec:whd} is restated as it is relevant for the proof. The weighted Hamming distance is defined as: 
\begin{equation}
w_H (\mathbf{x},\mathbf{L} )= \sum_{i=1} ^N    \left(\mathbf{c}[i] \oplus \theta(\mathbf{L}[i])\right)\times |\mathbf{L}[i]|.
\end{equation}
While \cite{battail_we_1993} has the same algorithm as \cite{ma_guessing_2024}, the metric used is the generalized distance defined as:
\begin{equation} \eta(\mathbf{y},\mathbf{c})=\sum_{j=1}^nv(y[j],c[j]),
\end{equation}
where:
\begin{equation}
   v(y,c)= ln \left(\frac{P(y\hspace{0.1cm}|\hspace{0.1cm}\theta(L))}{P(y \hspace{0.1cm}|\hspace{0.1cm}c)} \right).
\end{equation}
For BPSK with bits 0 and 1 equiprobable, we will prove that the weighted Hamming distance is equivalent to the generalized distance metric through enumerating all possible $c$ and $\theta(L)$ :
\begin{enumerate}
    \item When the assumed codeword bit is 0, and the received LLR is positive ( $\theta(L)\oplus c=0$), the following equality holds:
    \begin{equation}
        v(y,0)=\ln \left( \frac{P(y|0)}{P(y|0)} \right)=0.
    \end{equation}
      \item When the assumed codeword bit is 0, and the received LLR is negative ($\theta(L)\oplus c=1$), the following equality holds:
    \begin{equation}
        v(y,0)=\ln \left( \frac{P(y|1)}{P(y|0)} \right)=-L = |L|.
    \end{equation}
 \item When the assumed codeword bit is 1, and the received LLR is positive ( $\theta(L)\oplus c=1$), the following equality holds:
    \begin{equation}
        v(y,1)=\ln \left( \frac{P(y|0)}{P(y|1)} \right)=L = |L|.
    \end{equation}
      \item When the assumed codeword bit is 1, and the received LLR is negative ( $\theta(L)\oplus c=0$), the following equality holds:
    \begin{equation}
        v(y,1)=\ln \left( \frac{P(y|1)}{P(y|1)} \right)=0 .
    \end{equation}
\end{enumerate}
Hence the value $v(y,c)=\left(c \oplus \theta(L)\right)\times |L|$ and $\eta(\mathbf{y},\mathbf{c})=w_H(\mathbf{c},\mathbf{L})$ for the BPSK case.
\end{document}